UNIVERSITÀ DEGLI STUDI DI PAVIA
FACOLTÀ DI INGEGNERIA

DIPARTIMENTO DI INGEGNERIA INDUSTRIALE E DELL'INFORMAZIONE

CORSO DI LAUREA MAGISTRALE IN INGEGNERIA ELETTRONICA

TESI DI LAUREA

# Experimental Investigation of Programmed State Stability in OxRAM Resistive Memories

## Indagine Sperimentale sulla Stabilità dello Stato Programmato in Memorie Resistive OxRAM

Candidato: **Georgi Gorine**

Relatore: **Prof. Guido Torelli**
Correlatore: **Dott. Alessandro Cabrini**
Supervisore: **Dott. Andrea Fantini**

**A.A. 2013/14**

**Thesis defended at University of Pavia on April 28th, 2015.**

Examining Committee:

**Prof. Danilo Manstretta**

**Prof. Piero Malcovati**

**Prof. Lodovico Ratti**

**Prof Luca Tartara**

**Prof. Guido Torelli**

# Table of Contents

# Abstract

Various technologies and materials have been used so far for engineering memory devices. In the last 50 years, the family of semiconductor memories has greatly expanded, with new solutions able to achieve higher density and speed, lower power consumption, and more functionalities, yet lowering overall costs. Nevertheless, limitations of each memory type have also become more apparent. In particular, the mainstream technology in the semiconductor market, Flash memory, is expected to soon reach a point where storage capacity and performance cannot be further improved. Several emerging memory technologies seek to overcome these limitations. One of these emerging technologies is the OxRAM resistive memory. Oxide-based Random Access Memory (OxRAM), is part of the larger family of Resistive RAM (RRAM) memories. Differently from Flash, RRAM memories retains information, not by storing electrons, but instead by having logic values associated to resistance states. A reversible switching mechanism, that cause a structural modification in the material, is able to commute the memory cell from a low resistance state (LRS) to a high resistance state (HRS). In the specific case of OxRAM memory this effect is achieved by an electrically controlled movement and atomic reorganization of electrically conductive defects (oxygen vacancies) inside a dielectric. To achieve switching capabilities, a conductive filament between the two electrodes is first created by means of one-time forming step (a high voltage that induces a soft breakdown of the oxide).

Generally OxRAM cells consist of a transition metal oxide (typically $HfO_2$, $Ta_2O_5$, $TiO_2$) sandwiched between two metal electrodes (typically TiN, TaN, W but also Pt, Ir). In order to research solution relevant from industrial perspective this thesis, performed in imec research institute (Leuven, Belgium), will mainly consider $HfO_2$ and $TaO_2$ based materials whose adoption is widespread in high-k metal gate and MIMCAP technology.

This thesis describes the experimental investigation performed on OxRAM memories during a 6-months internship project at imec. From previous studies, HfO-based OxRAM memories showed good endurance properties (with more than $10^9$ write cycles), low operating current (as low as 10 µA), and very fast switching (programming pulses of 100 ns). However, it has been observed that, under low programming current condition, data stability becomes challenging. Therefore, this programming state stability needed further investigation.

After introducing the main emergent semiconductor memory technologies, this thesis addresses the problem of state stability, first, by presenting the equipment and algorithms used to study OxRAM state loss over short time intervals. Secondly, various experiments and tests are performed searching for a key parameter in order to control memory retention. The four parameters analyzed are: programming pulse width, fall time, temperature, and OxRAM cell material composition. Each parameter is analyzed considering the measured data distributions, according to three metrics: global behavior, local behavior and correlation. The obtained results for each experiment are provided and discussed. This work demonstrates that resistance evolution over time is affected by both a determinist drift (logarithmically dependent on time) and a random stochastic fluctuation component. It also demonstrates that this behavior is intrinsic in the nature of the device itself and does not depend on processing issues or variation in the programming conditions. Thirdly, a provisional extension to imec`s Hourglass model for OxRAM memory is developed, in order to describe the resistance relaxation leading to program instability. The model has been updated with the collected data and is able to model both the random stochastic component, described with the Random Walk model, and the deterministic process, described by a resistance drift over time. This experimental study proved the intrinsic nature of the program instability phenomena in OxRAM devices, which in turns leads to the ineffectiveness of verify algorithm for low current operation.

# Estratto

Diverse tecnologie e materiali sono stati impiegati finora per la realizzazione di dispositivi di memoria. Negli ultimi 50 anni, la famiglia di memorie a semiconduttore si è notevolmente ampliata, con nuove soluzioni in grado di offire maggiori densità e velocità, bassi consumi di energia, e più funzionalità, pur con una riduzione dei costi complessivi. Tuttavia, le limitazioni di ogni tipo di memoria sono diventate più evidenti. In particolare la memoria Flash, tecnologia leader nel mercato dei semiconduttori, dovrebbe raggiungere presto un punto in cui la capacità e le prestazioni non possono essere ulteriormente migliorate. Diverse tecnologie di memoria emergenti cercano di superare questi limiti, tra queste troviamo la memoria resistiva OxRAM. Le Oxide Random Access Memory (OxRAM), fanno parte della grande famiglia di memorie RAM non volatili resisitive (RRAM). A differenza della memoria Flash però, la memoria RRAM conserva informazioni non grazie all'immagazzinamento di elettroni, ma associando i valori logici (0,1) al valore di resistenza della cella. Questo è possibile grazie ad un meccanismo di commutazione reversibile che provoca una modifica strutturale del materiale inducendo un cambiamento della cella da uno stato di bassa resistenza (LRS) ad uno stato di alta resistenza (HRS). Nel caso specifico della memoria OxRAM questo effetto si ottiene con un movimento controllato elettricamente di riorganizzazione atomica di difetti elettricamente conduttivi (vacanze di ossigeno) all'interno di un dielettrico. Per innescare questa possibilità di commutazione della cella, viene creato in primo luogo, un filamento conduttivo tra i due elettrodi mediante la fase di "forming" o creazione (un'alta tensione che induce una rottura dielettrica controllata dell'ossido).

Generalmente le celle OxRAM sono realizzate con un ossido di metallo di transizione (tipicamente $HfO_2$, $Ta_2O_5$, $TiO_2$) inserito tra due elettrodi di metallo (tipicamente TiN, TaN, W, ma anche Pt, Ir). Per soluzioni rilevanti da un punto di vista industrial, questa tesi, realizzata presso l'istituto di ricerca imec (Leuven, Belgio), prenderà in considerazione principalmente OxRAM a base di ossidi di Afnio ($HfO_2$) e Tantalio ($TaO_2$).

Questa tesi descrive l'indagine sperimentale effettuata su memorie OxRAM nel corso di un tirocinio di 6 mesi presso imec. Da studi precedenti, le memorie OxRAM a base di HfO, hanno mostrato buone proprietà di resistenza all'usura (con più di $10^9$ cicli di scrittura), bassa corrente operativa (a partire da 10 µA), e tempi di commutazione molto veloci (impulsi di programmazione di 100 ns). Tuttavia, è stato osservato che, in condizioni di bassa corrente di programmazione, la stabilità dei dati programmatic non è più garantita e pertanto, sono necessarie ulteriori indagini.

Dopo aver introdotto le principali tecnologie di memoria a semiconduttore emergente, questa tesi affronta il problema della stabilità, in primo luogo, presentando la strumentazione e gli algoritmi utilizzati per valutare la perdita dello stato programmato delle OxRAM su brevi intervalli di tempo. In secondo luogo, eseguendo vari esperimenti e test, viene ricercato un parametro chiave che influisce sulla stabilità permettendone un controllo. I quattro parametri analizzati sono: larghezza dell'impulso di programmazione, tempo di discesa dell'impulso di programmazione, temperatura, e composizione del materiale della cella di memoria OxRAM. Ogni parametro viene analizzato considerando le distribuzioni dei dati misurati, in base a tre parametri: comportamento globale, comportamento locale e correlazione. I risultati ottenuti per ciascun esperimento sono forniti e discussi. Si dimostra che l'evoluzione della resistenza nel tempo è influenzata, sia da un drift deterministico (dipendente dal logaritmico del tempo), sia da una component di fluttuazione stocastica casuale. Si dimostra inoltre, che questo comportamento è intrinseco nella natura del dispositivo stesso e non dipende da problema di processo o da variazione nelle condizioni di programmazione. In terzo luogo, viene fornita un estensione provvisoria al modello di imec per le memorie OxRAM "Hourglass", in modo da includere una descrizione del rilassamento della resistenza che porta a problem di stabilità. Il modello è stato aggiornato grazie ai dati raccolti ed è in grado di modellare sia la componente casuale stocastica, descritta con il modello "Random Walk", sia il processo deterministico, descritto da un drift della resistenza nel tempo. Questo studio sperimentale ha dimostrato la natura intrinseca dei fenomeni di instabilità nella programmazione di dispositivi OxRAM, che a sua volta porta alla inefficacia di un algoritmo di verifica a basse correnti operative.

# Thesis Outline

The present thesis is organized as follows.

**Chapter 1** provides a review of memory devices and explains the relevance of this thesis in broad research field pertaining to semiconductor memories. First, memory devices for storing digital data are introduced. After providing a brief overview of the semiconductor and memory devices market size, a taxonomy of memory devices is given according to their market readiness Furthermore, this chapter explores the problems of the currently dominant Flash memory, introduces the concept of the "memory gap" and the drivers for the search of the Next-Generation Non-volatile Memory. Finally, a number of non-volatile memory devices are described in detail.

The research goals of this thesis are illustrated in **Chapter 2**. Before directly addressing the research questions, this chapter provides the methodological framework of the research: the followed experimental plan, the programming choices made, the measurement setup and the tools used for collecting the data, as well the software developed for performing the needed tests.

The experimental part is reported in **Chapter 3**. After a description of the point of reference conditions used for all the measurements, this chapter continues by presenting all the results obtained from the different test for program stability with both Unverified Single-Pulse and Program-and-Verify algorithms. Moreover, for each test carried out, the physical insight, the measured data and the conclusions developed, are presented.

**Chapter 4** concludes the thesis by providing a preliminary behavioral model of the relaxation in both unverified and verified programming based on the data presented in Chapter 3. At first, the analytical Hourglass model currently used in imec is presented. Then, an explanation of the random switching behavior of an RRAM memory cell is given in terms of a random walk stochastic process. Lastly, after fitting the experimental data with different laws, the oscillations component is extracted.

# Chapter 1
# Review of Semiconductor Memory Devices

## 1.1 The Semiconductor Market

Exactly 50 years ago (in April 1965), in what later became a worldwide known article [1] , Gordon E. Moore stated that the number of transistors per unit area in an integrated circuit would double every 18/24 months. This prediction, known today as “Moore`s Law”, has been one of the driving principles in the semiconductor industry, leading to the great revolution of modern electronics.

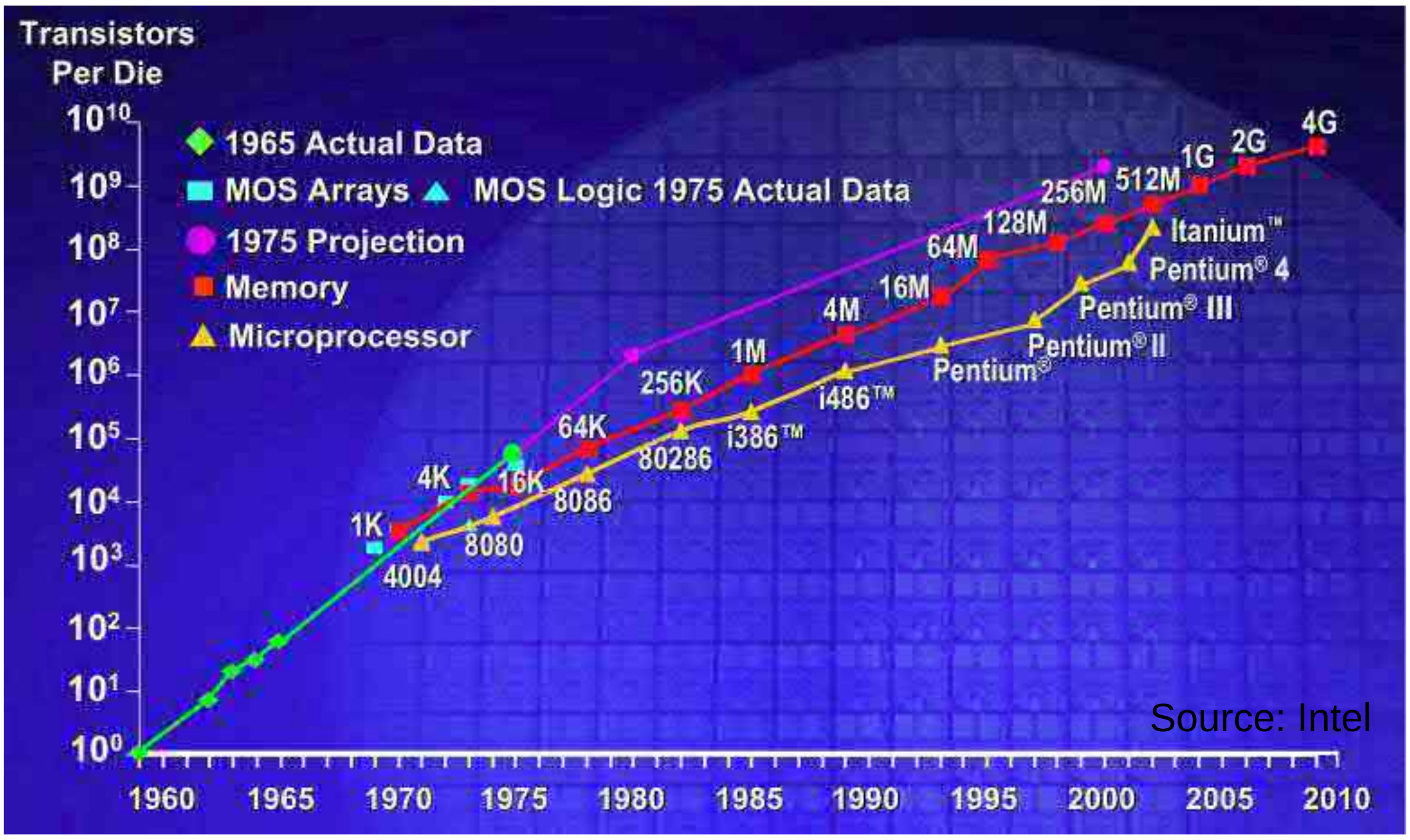

*Figure 1.1: Moore`s Law confirmed by Intel CPUs and DRAM memory developments*

Every year, engineers managed to reduce transistor size and overcome the technological difficulties connected to such size reduction. Semiconductor companies were thus able to introduce on the market, new upgraded devices with better performances and lower costs. The performance of Central Processing Units (CPUs) and memories continued to improve for decades confirming that Moore`s prediction is an actual law (Figure 1.1).

Steady advances in manufacturing technology (particularly in lithography) have allowed for a continuous reduction in transistor size. In 1974, R.H. Dennard [2] proposed a transistor scaling methodology (Figure 1.2) which maintains the electric field in the device constant. Scaling not only increases packing density, but also gives improvements in transistor switching speed and power dissipation.

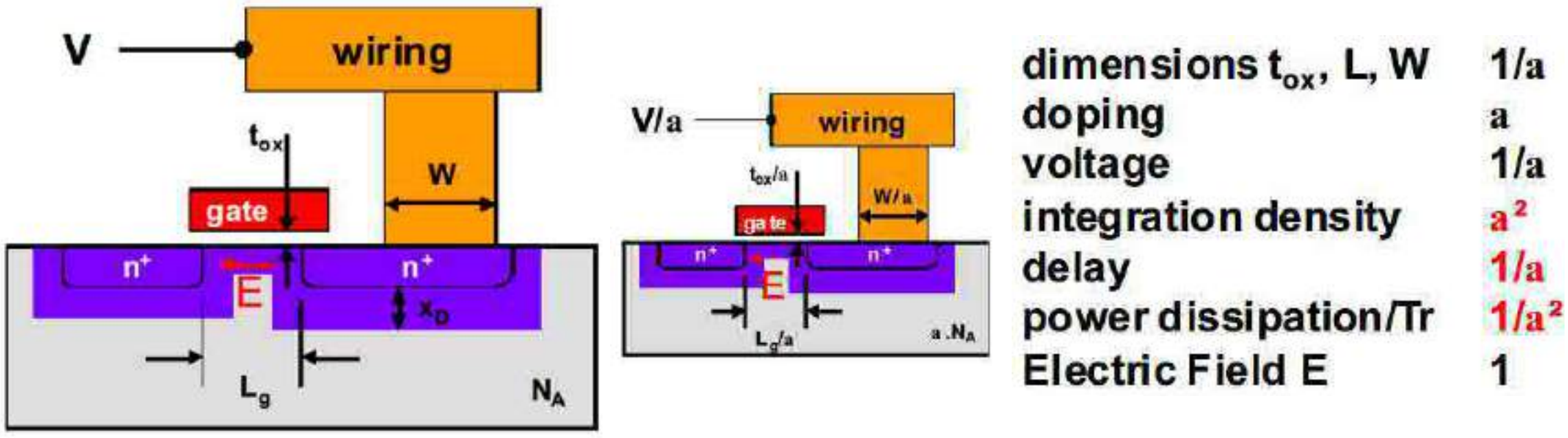

***Figure 1.2: The constant-field scaling theory predicts increased speed and lower power consumption of MOS transistors when the critical dimensions are scaled down. [3]***

As summarized by the "Virtuous Circle" in Figure 1.3, the semiconductor market experienced a positive feedback loop catalyzed by the transistor size scaling and the constant evolution of fabrication technology. Transistor Scaling leads to an increase in the number of transistor per die thus allowing lower unity cost and better performance, which in turn leads to a Market Growth with consequent Investments in research for the next technology node.

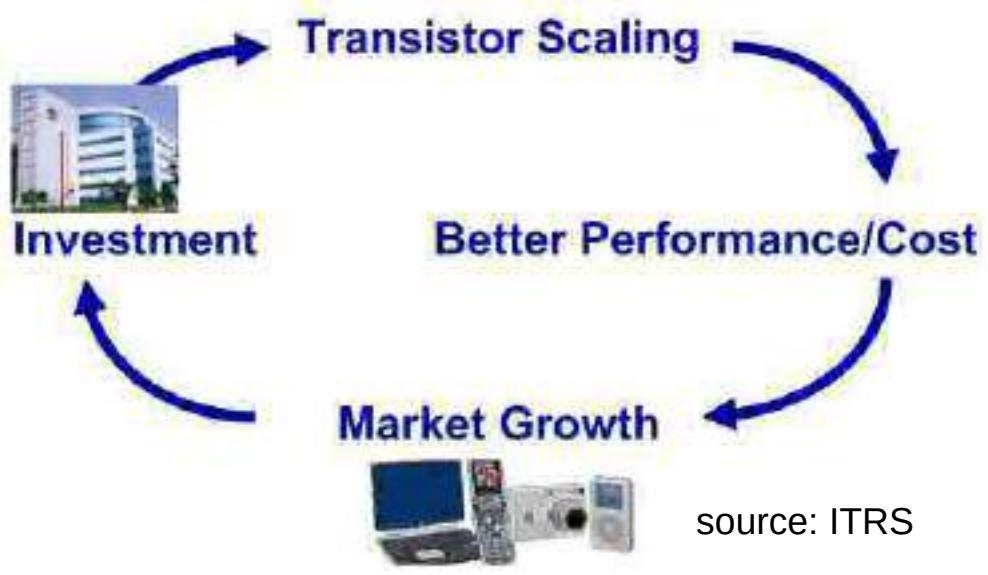

***Figure 1.3: The virtuous circle of the semiconductor industry.***

The planar process developed by J. Hoerni [4] at Fairchild Semiconductor in 1960 also contributed to the great development of the semiconductors industry. The planar process is the manufacturing technique used in the semiconductors industry to build entire integrated circuits on a single layer of silicon (wafer). The planar process is still used nowadays, but instead of producing small wafers of 25 mm diameter, the great evolution of semiconductor industry led to produce, with acceptable yield, wafers up to 300 mm (and in the near future also up to 450 mm) with many more devices per wafer. The increase in the size of the wafer results in price reduction per single chip and therefore allows the production of billions of chips for countless consumer applications.

The results of decades of successful investments in R&D in the semiconductors field are summarized in Figure 1.4.

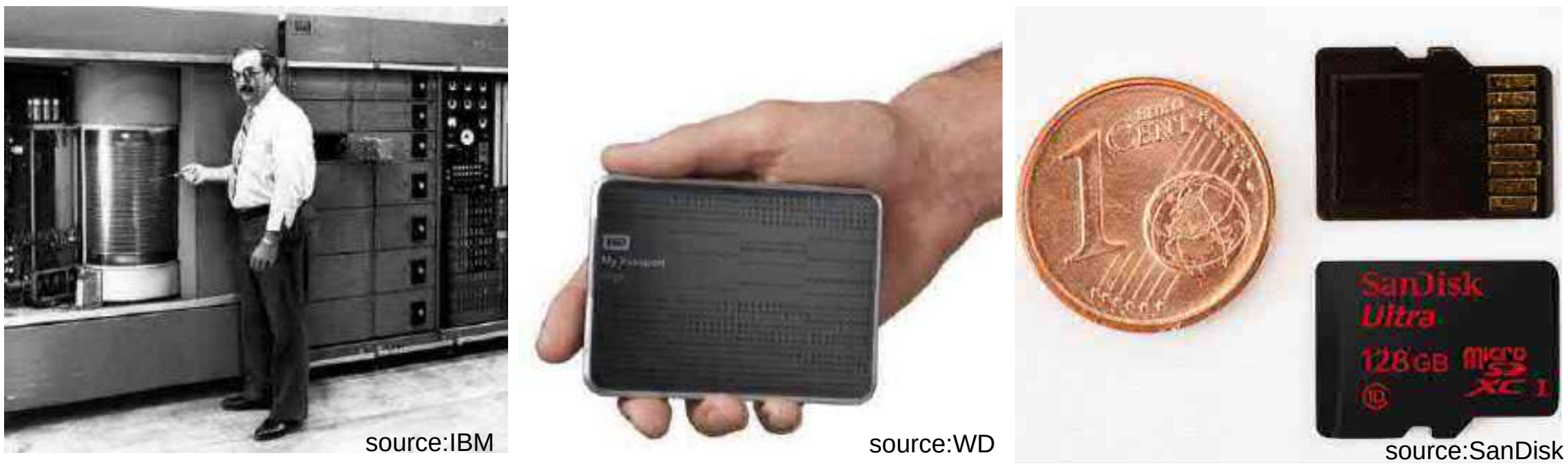

(a) (b) (c)

***Figure 1.4: A 1956 IBM305 (a), a 2012 Western Digital hard disk drive (b,) a 2014 Micro SD (c)***

In 1956, an expensive and very large IBM305 RAMAC computer with its one-meter high hard disk drive (HDD) made with 50 magnetic disks of 24" had an overall storage capacity of 5 MB (Figure 1.4a). In 2012 a hand-sized Western Digital`s HDD was made of only 4 disks of 2.5" and had a storage capacity of 2TB (Figure 1.4b). Finally, in 2014, a SanDisk`s coin-size Micro SD is able to store as many as 128 GB (Figure 1.4c). These pictures show how dramatically memory technology evolved in the last 50 years, in terms of size, speed and cost.

The evolution of the semiconductor industry can be further exemplified by analyzing the growth of the semiconductor market size in the past decades (Figure 1.5). . As stated by the Semiconductor Industry Association [5], focusing on the U.S. economy, "The U.S. semiconductor industry is a uniquely important contributor to the U.S. economy. Thanks to rapid technological development, the industry's contribution to the U.S. economy grew 265 percent from 1987 to 2011, **more than that of any other major U.S. manufacturing industry**".

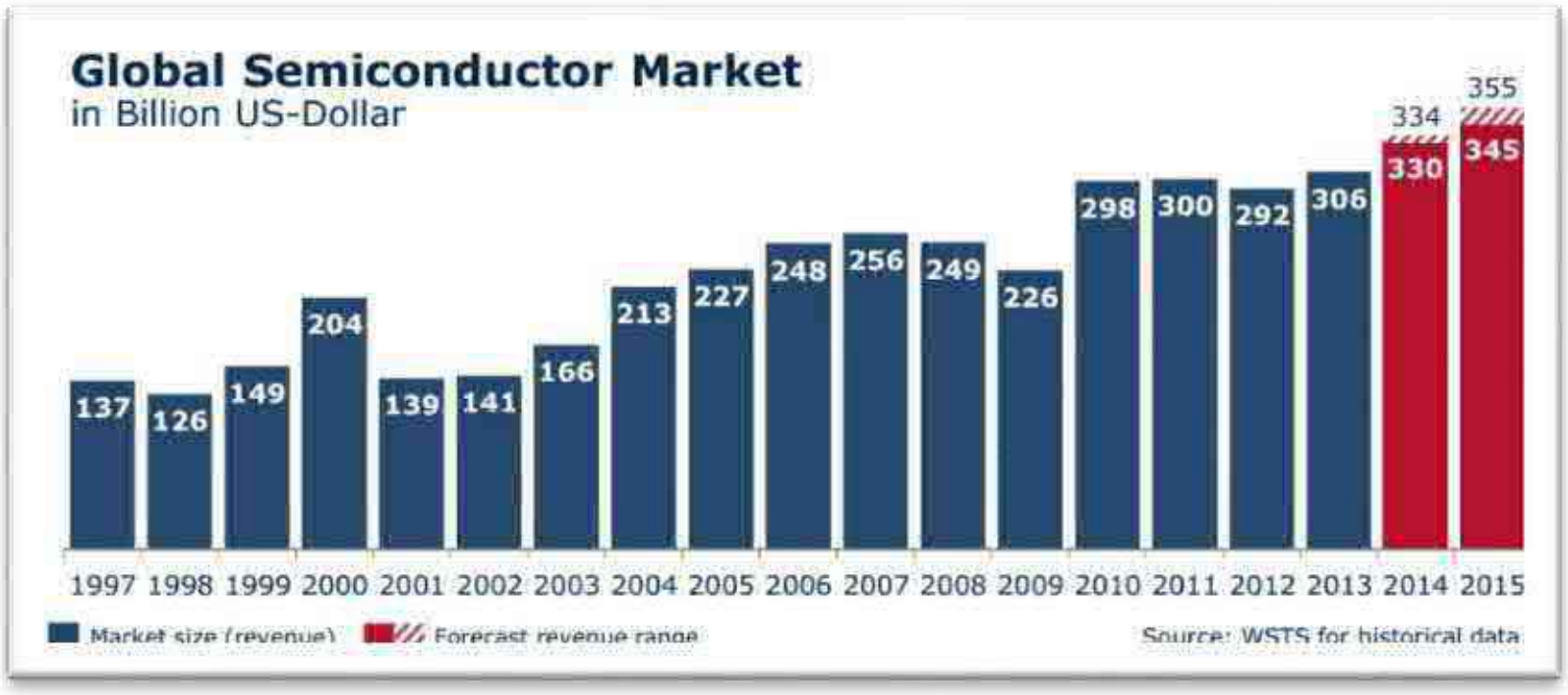

***Figure 1.5: Global semiconductor market size according to the World Semiconductor Trade Statistics[6].***

## 1.2 The Memory Market

The evolution of CPUs would not have been possible without a similar evolution in the memory field. CPUs and memory devices improved alongside and both followed Moore`s Law. With the advent of the era of digital photo cameras, smartphones and tablets, memories have become even more important. As shown in Figure 1.6, consumer electronics devices commercialized in the past few years have resulted in an exponential increase of the annual volume of shipped GBytes of NAND Flash memories.

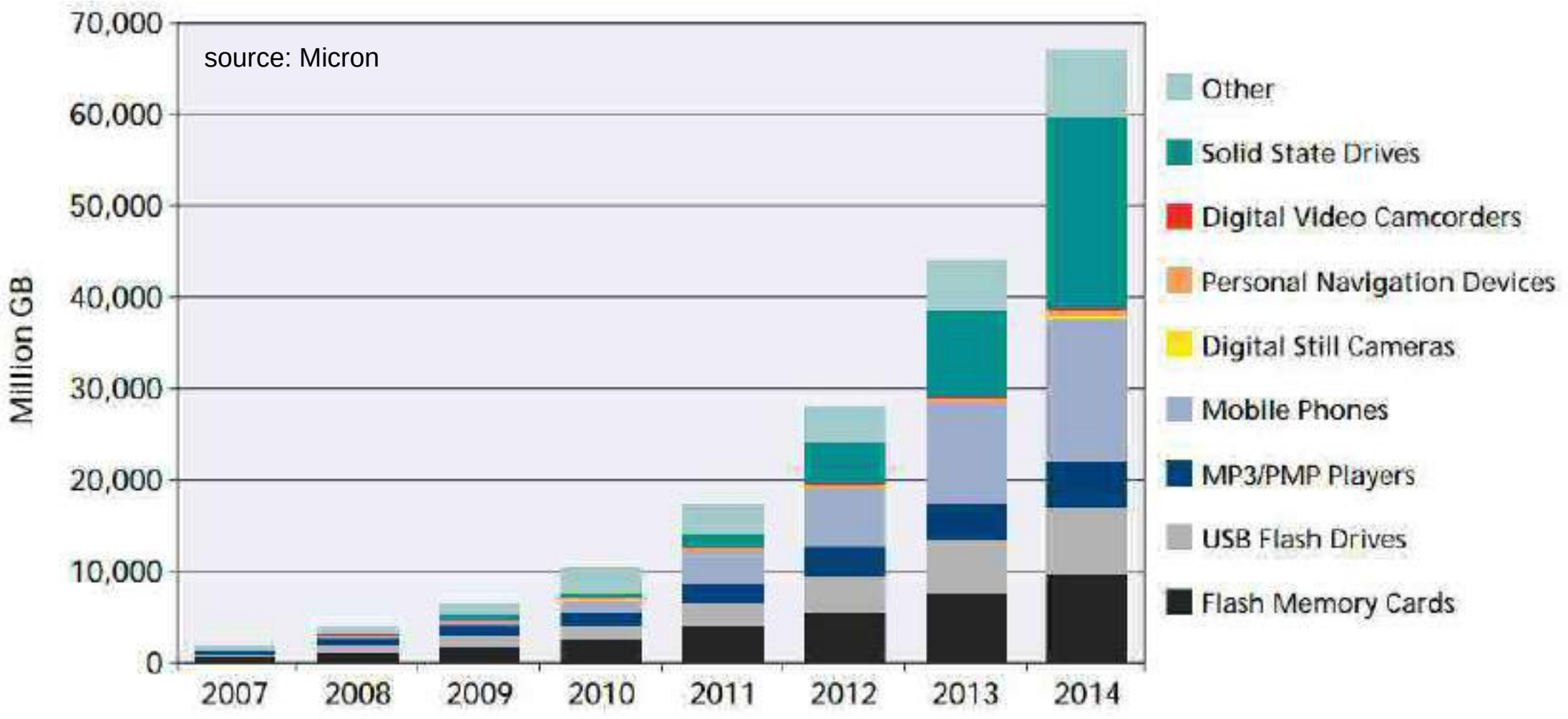

***Figure 1.6: Major markets driving non-volatile NAND flash memories.***

The memory market alone accounts for 24% of the overall semiconductor market size (Table 1.1.1). Moreover, in the past two years the memory market showed the highest year-on-year growth rate.

| Autumn 2014 | Amounts in US$M | | | | Year on Year Growth in % | | | |
|---|---|---|---|---|---|---|---|---|
| | 2013 | 2014 | 2015 | 2016 | 2013 | 2014 | 2015 | 2016 |
| Discrete Semiconductors | 18,201 | 20,441 | 21,347 | 21,980 | -4.9 | 12.3 | 4.4 | 3.0 |
| Optoelectronics | 27,571 | 29,498 | 30,958 | 31,983 | 5.3 | 7.0 | 4.9 | 3.3 |
| Sensors | 8,036 | 8,627 | 9,151 | 9,624 | 0.3 | 7.4 | 6.1 | 5.2 |
| Integrated Circuits: | 251,776 | 274,586 | 283,090 | 291,685 | 5.7 | 9.1 | 3.1 | 3.0 |
| a) Analog | 40,117 | 44,217 | 47,429 | 49,175 | 2.1 | 10.2 | 7.3 | 3.7 |
| b) Micro | 58,688 | 62,211 | 63,144 | 64,240 | -2.6 | 6.0 | 1.5 | 1.7 |
| c) **Memory** | **67,043** | **78,611** | **81,029** | **84,343** | **17.6** | **17.3** | **3.1** | **4.1** |
| d) Logic | 85,928 | 89,547 | 91,488 | 93,927 | 5.2 | 4.2 | 2.2 | 2.7 |
| **Total Products - $M** | **305,584** | **333,151** | **344,547** | **355,272** | 4.8 | 9.0 | 3.4 | 3.1 |

***Table 1.1: WSTS forecast of Global Semiconductor Market (Autumn 2014) [6].***

## 1.3 The Memory Hierarchy

During the past years, various technologies and innovative materials have been used for engineering memory devices. The first memory devices were simple punched cards, where bits were represented by the presence or the absence of holes in predefined positions. The next generation of memory devices were first magnetic tapes and then, floppy disks and Hard Disks, where each bit was stored as the polarization of special magnetic materials. Finally, in more modern technology, bits are represented by electrons trapped in a floating electrode, and the memory cell is built using semiconductor devices such as transistors.

After 50 years of evolution of computer technology, various technologies emerged, each with its own characteristics in terms of speed, size, power consumption and price. However, there is no memory device that currently possesses the full spectrum of advantages. For this reason, a tapered "level organization" is used in today's computer architecture. By doing so, the problem of the performance bottleneck between logic and memory is reduced and the CPU can virtually access a memory as fast as the fastest and as big as the biggest one. Today, any complex electronic device, such as a PC, a smartphone or a tablet, is implemented by using either the Harvard's or Von Neumann's architectures (which are the most popular computer architectures used in any digital computer). In both architectures, the CPU accesses data and instructions from different sources through I/O serial and parallel connections. In Figure 1.7, a pyramid shows all the memory types that can be currently found in an electronic device, following the above architectures, ordered from Level 0 to Level 5.

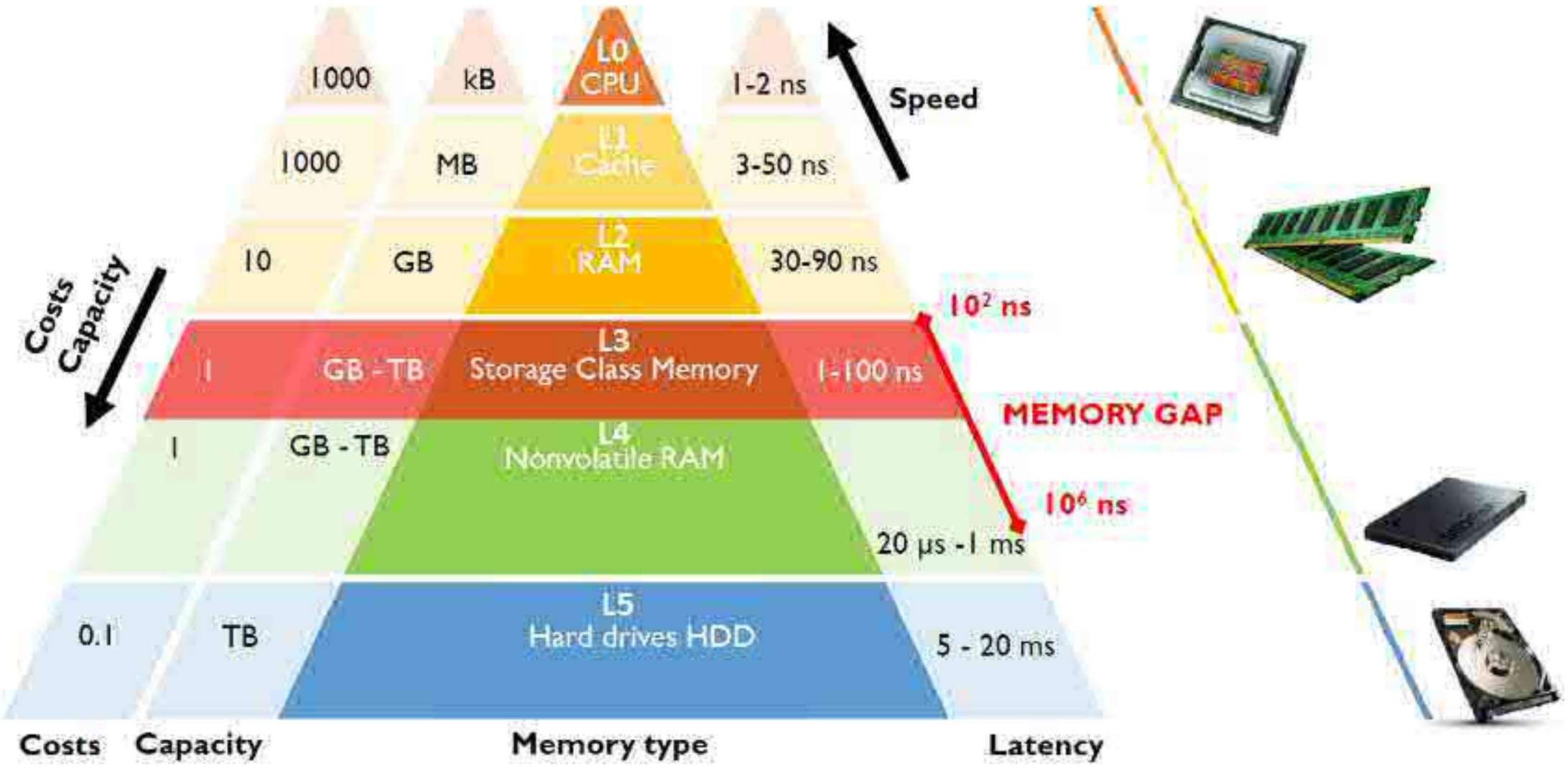

***Figure 1.7 - Pyramids of Memories: memories at the top are the fastest whereas memories at the bottom are the biggest and cheapest.***

A more detailed description of each memory Level can explain the shape of the pyramid:

- **Level 0:** this is the memory level closest to and physically embedded in the CPU. The purpose of this memory is to act as CPU's registers where the current running instructions and the data needed by the Arithmetic Logic Unit (ALU) are stored. For this reason, Level 0 memory must be as fast as possible. Static Random Access Memory (SRAM) is used as Level 0 memory. SRAMs are the fastest memories produced nowadays but their low density leads to higher cost (a single SRAM cell is built with two flip flops plus two access transistors, thus being a six-transistor -6T- cell). SRAM can retain the stored information as long as the power is on and, therefore, SRAM is a volatile memory. In commercial processors, Level 0 memory is about 1-2 KBytes.
- **Level 1:** this memory is used as cache for storing data and instructions that will be soon executed by the CPU. Also Level 1 Cache is implemented by means of SRAM cells but being further away from the ALU, there is more room for implementing more SRAM memories. Usually Level 1 cache is about 2 Mbytes.
- **Level 2:** commonly this memory is named simply as "RAM memory". This memory is located outside the CPU chip and it is interconnected with it through an I/O bus. In the past, various sockets with a different number of pins have been used in order to increase the speed of this connection. Level 2 RAM memory is usually realized by means of Dynamic Random Access Memory (DRAM). The term Dynamic is used to distinguish it from Static in order to indicate the need of this memory to be constantly refreshed in order to retain the bits stored. DRAM is much denser than SRAM because of its 1T1C structure (a single cell is made of an access transistor, T, and a charge storing capacitor, C). The lower cost per bit is traded with a higher read and write (R/W) time. Level 2 memory in modern commercially used computers is about 8 GBytes (or few GBytes).
- **Level 3:** the Storage Class Memory is a new memory concept currently at the research stage. A detailed description of this class will be given in Paragraph $1.5. As highlighted in Figure 1.7, there is a growing memory gap at this level and there is room for a new class of memories, with intermediate values of price and speed. The new Emerging Non Volatile Memories shown in Figure 1.8 are the emerging technologies, currently under development which will eventually make the Storage Class Memory a reality.
- **Level 4:** this memory was the last one to be introduced massively on the market in the last couple of years with the purpose of providing mass storage with reasonable R/W time. Level 4 memories are the Solid State Drives (SSD) built with Flash memory cells. SSDs are currently the fastest Non Volatile Memories used for mass storage. Flash memory is cheaper than DRAM, because of its higher density, however the cost advantage comes at the expense of slower R/W time. SSD nowadays are as big as 1TByte and their price is continuously decreasing.

- **Level 5:** Common Hard Drive Disks (HDD) are level 5 memories. These are the cheapest media available to store information, but their speed is limited by the intrinsic physical limitation of a mechanical head reading rotating magnetic disks. Today HDDs can store several TBytes.

## 1.4 Semiconductor Memories

Memories from Level 0 to Level 4 are all Semiconductor Memories produced on Silicon wafers, while only Level 5 memories are still mechanical devices.

As shown in Figure 1.8*a*, semiconductor memories can be categorized in two classes based on their ability to maintain stored data when the power is off:

a) **Volatile Memories** (VMs) cannot keep stored bits when the power supply is down. Static and Dynamic Random Access Memories (SRAMs and DRAMs) are typical examples of semiconductor volatile memories.

b) **Non - volatile Memories** (NVMs) are memories able to keep stored data even in the absence of power supply. EEPROMs (Electrically Erasable Programming Read Only Memories) and Flash memories belong to this class.

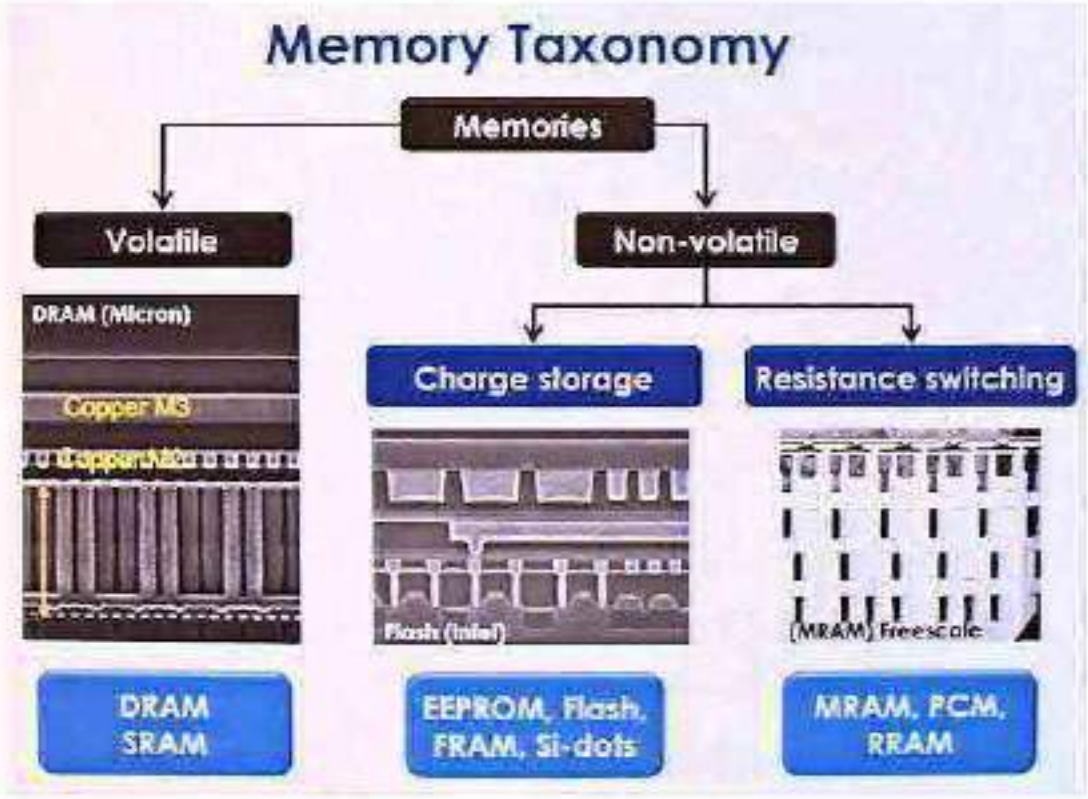

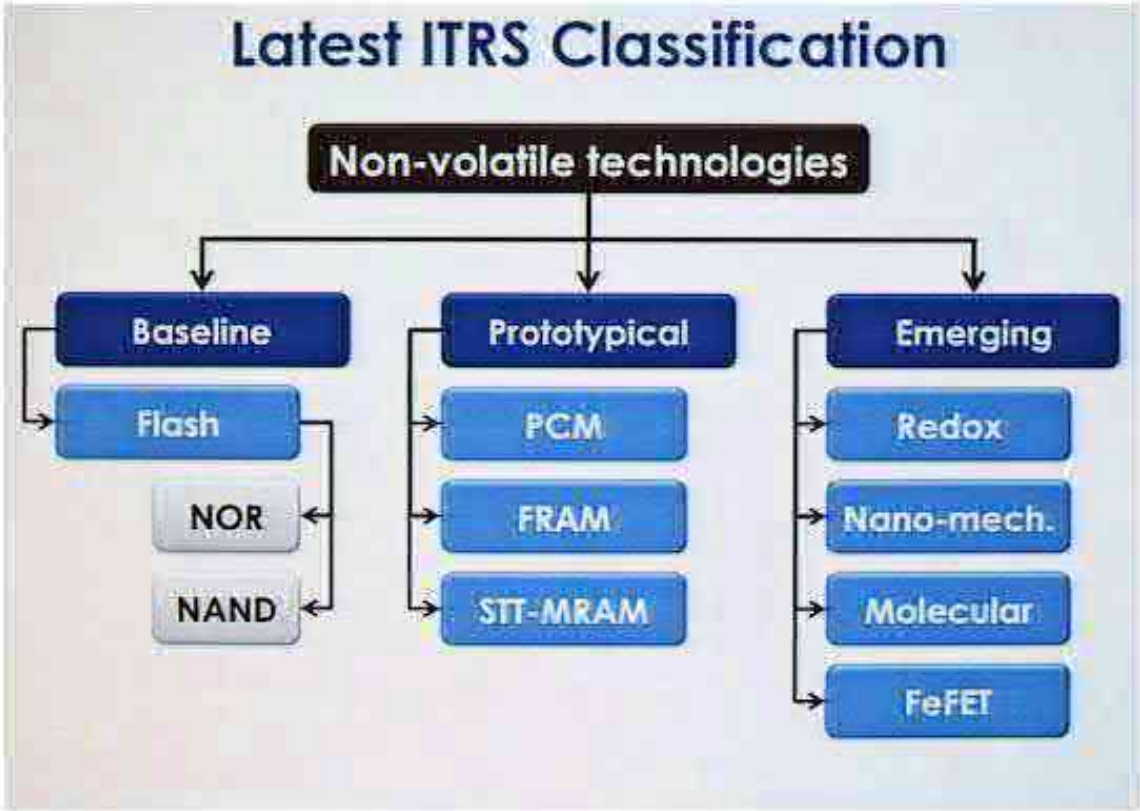

***Figure 1.8: Categories of standard semiconductor memories (a) and emerging memories (b) [7]***

Non-Volatile memories can be further subdivided in two classes, namely, *Charge storage* and *Resistance switching*, on the base of the physical principle used for storing data. Figure 1.8*b* illustrates a refinement in the taxonomy of Non Volatile Memories. The main contenders to the currently dominant Flash memory are categorized based on their technological maturity. While Flash memory is considered the *Baseline* nonvolatile memory, since it is highly mature, well optimized, and popular on the market, contenders can be subdivided in:

a. **Prototypical**: memory technologies that are at a point of maturity where they are commercially available (generally for niche applications), and have a large scientific, technological and systematic knowledge base available in the literature.
b. **Emerging:** the less mature memory technologies, which could offer significant potential benefits if various scientific and technological hurdles can be overcome.

## 1.5 Storage Class Memory

In the last 50 years, the family of semiconductor memories has greatly expanded achieving higher density, higher speeds, lower power consumption, more functionality and lower costs. At the same time, some of the limitations within each type of memory also became more apparent. Hence, there are currently several emerging technologies that aim at going beyond those limitations and potentially replacing all or most of existing memory technologies. One of these emerging technologies may become the new universal semiconductor memory.

As shown in Figure 1.7 and Figure 1.9, there is an empty window between HDDs and RAMs, called “the memory gap” that could be filled by a new type of memory. This gap has been increasing in past years, with the price per bit of HDDs decreasing in price, but not raising in speed and DRAMs becoming faster following the CPUs evolution, but not decreasing in price.

All major semiconductor companies are researching the so- called “Next Generation Memory” in order to fill the “memory gap” and eventually replace currently used memories.

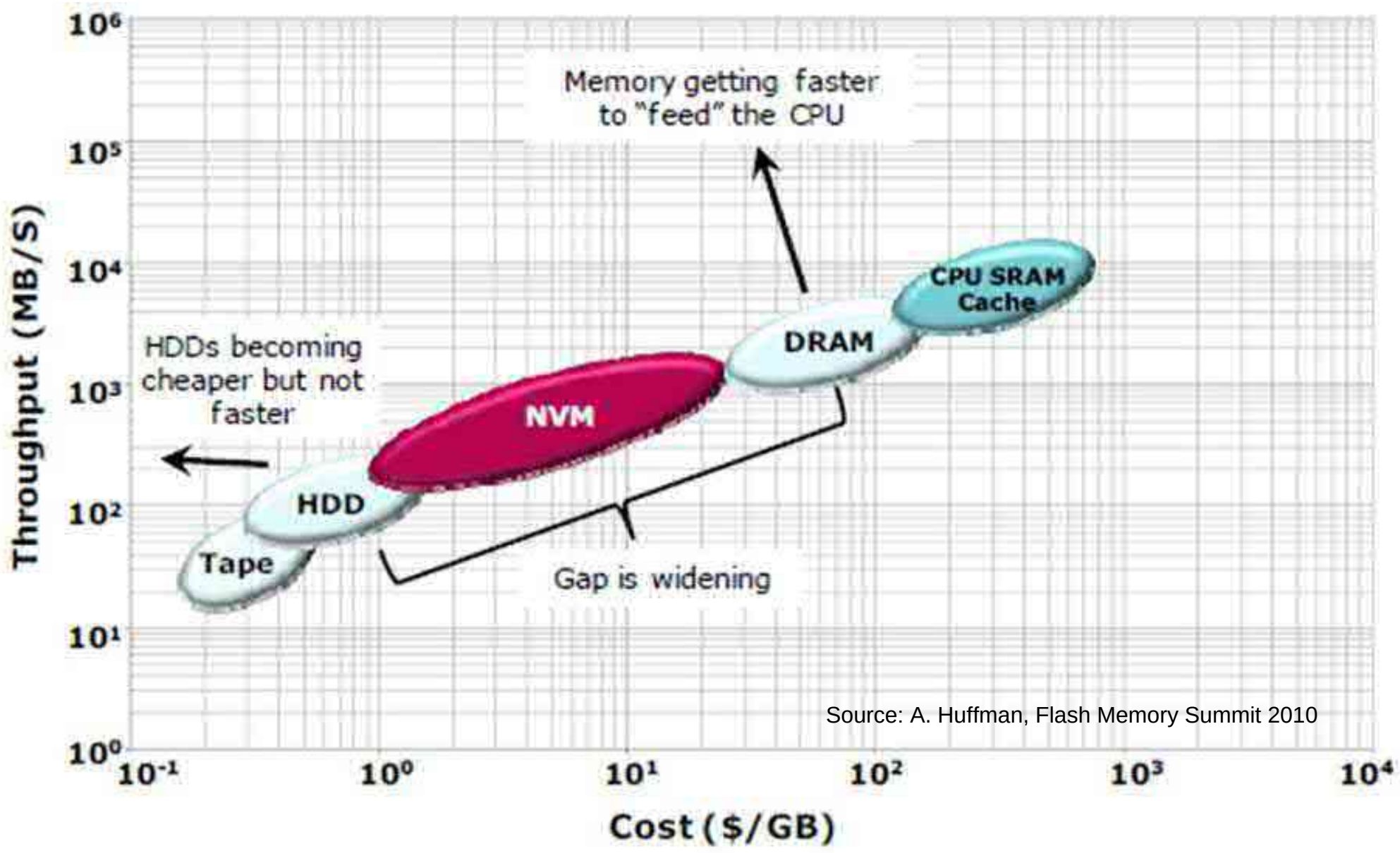

***Figure 1.9: Speed versus Cost graph highlighting the enlarging memory gap.***

It is expected that the "memory gap" would be filled by the Storage-class memory (SCM), as reported in the Level 3 of the hierarchy in Figure 1.7. Storage-class memory describes a device category that combines the benefits of solid-state memory, such as high performance and robustness, with the archival capabilities and low cost of conventional hard-disk magnetic storage. Such a device requires a nonvolatile memory technology that could be manufactured at a very low cost per bit.

In principle, the "memory gap" could be filled by two distinct SCM. These memory types can be differentiated from each other in terms of access time: we thus have a slower (and less expensive) Storage-type SCM (S-type SCM) and a faster (and more costly) Memory-type SCM (M-type SCM). Table 1.2 lists a set of target specifications for SCM devices and systems, which are compared against benchmark parameters offered by existing technologies.

| Parameter | Benchmark | | | Target | |
|---|---|---|---|---|---|
| | HDD | NAND flash | DRAM | Memory-type SCM | Storage-type SCM |
| *Read/Write latency* | 3-10 ms | ~100µs (block erase ~1 ms) | <100 ns | <200 ns | 1-5µs |
| *Endurance* **(cycles)** | unlimited | $10^3$-$10^5$ | unlimited | >$10^9$ | >$10^6$ |
| *Retention* | >10 years | ~10 years | 64 ms | >5 days | ~10 years |
| *ON power* **(W/GB)** | 0.003-0.05 | ~0.01-0.04 | 0.4 | <0.4 | <0.10 |
| *Standby power* | ~52%-69% of ON power | <10% ON power | ~25% ON power | <5% ON power | <5% ON power |
| *Areal density* | ~ $10^{11}$ bit/cm$^2$ | ~ $10^{11}$ bit/cm$^2$ | ~ $10^9$ bit/cm$^2$ | >$10^{10}$ bit/cm$^2$ | >$10^{10}$ bit/cm$^2$ |
| *Cost* **($/GB)** | ~0.1-1.0 | 2 | 10 | <10 | <3-4 |

***Table 1.2: Target device and system specifications for SCM [7].***

The benchmark parameters or necessary attributes of the new-generation SCM, as reported by the International Technology Roadmap for Semiconductors [8] in the 2013 Emergent Research Device report are:

- ***Scalability:*** scalability is the main reason that allowed the tremendous evolution of the semiconductor market. The goal of a new memory technology is to replicate this success allowing the Moore`s Law to be valid over additional decades.
- ***Areal Density:*** along with scalability, a high areal density through either MLC (Multi-level Cells) and/or 3D integration is needed to successfully beat the NAND flash memory. Areal density is usually reported in terms of bits/cm$^2$ or by multiples of the squared Minimum Feature Size (each cell size is defined as the square of the feature size $F^2$, times the cell area factor X).
- ***Fabrication Cost:*** the cost per GByte must be between the cost of NAND flash and DRAM that are the two competitors. This can be achieved by guaranteeing CMOS technological compatibility in order to reuse the great know-how and infrastructure investments for current technologies.
- ***Retention:*** data retention is the characteristic that distinguishes Volatile from Non Volatile memories and must therefore be optimized. S-type SCM must be provided with longer retention than m-type SCM.
- ***Latency:*** latency is defined as the time required to access a cell in reading (read latency) or writing (write latency). For the M-type SCM latency must be low enough to allow a synchronous operation with the logic.

- ***Power Consumption:*** with increasing speed and density, power consumption becomes of great concern. The ideal SCM must be as much energy efficient as possible to allow a longer battery autonomy for mobile devices.
- ***Endurance:*** one of the main characteristics of a memory device is the maximum number of write cycles that it can ensure. High endurance is a must for operating memories as the M-type SCM, in which the single bits are continuously switched during a program execution.
- ***Variability:*** as for any new technology device-to-device variability must be taken into account to ensure a reliable device. The reproducibility of the switching characteristic of a memory cell is often dependent on integration process and material stability.

Moreover, emerging memory solution can target a different subgroup of memory type and, depending on the choice, has to meet different requirements and specifications. The different memory types currently targeted by emerging technology developments are the following:

- **Mass Storage:** this class of memories is the classical Level 4 memory used for storing big amounts of data. Emerging memories targeting this class must be very scalable to allow a very high-density technology with the lowest price per bit possible. The main competitor in this class is the NAND flash memory.
- **Stand Alone:** this class includes all memories realized on a single chip that are ready to be used or installed in bigger systems. Stand Alone memories are realized with the best technology for the memory cell and the memory array takes most of the chip area. A Mass Storage memory is usually an array of Stand Alone chips interconnected and managed by a single common controller.
- **Embedded:** this class of memories differs from the others in that they represent a memory integrated directly on-chip along with the logic which implements complex functions (typically including a microprocessor). In embedded solutions, the technology chosen is the best for realizing the logic whereas the memory section must be easily realized with similar processes as the logic, to reduce the number of masks needed during production. The main advantage of embedded memory is the possibility to remove the slow and cumbersome inter-chip connections with the realization of a SoC, i.e. of an entire Systems on a single Chip. Key factors for a good embedded memory are low power consumption (for battery-operated applications) and high speed. Today the main competitors in this area are SRAM (used in CPUs) as well as DRAM and EEPROM (used in micro-controllers).

## 1.6 Non-volatile Memories

Solid state non-volatile memories (NVMs) can be classified in three categories based on the physical mechanisms that allows non-volatile data storage:

- **Memory based on Charges**: this is the "classical" non-volatile memory. The single bit is represented by the presence of some electrical charge. Practically, all solid-state memories commercially used nowadays are of this type.
- **Memory based on Spins**: from the old magnetic tape to the modern rotational hard disks, all these memory supports use the change of spin of atoms, to store bits. In the last years, also semiconductor-based memories have been proposed
- **Memory based on Atoms**: the information of a single bit is stored by modifying the crystal lattice of a material by an induced movement of atoms, cations or anions. Most of the new generation non-volatile solid-state memories belong to this class.

| CHARGES | SPINS | ATOMS |
|---|---|---|
| • FLASH<br>• FeRAM<br>• FeFET | • MRAM<br>• STTRAM | • PCM<br>• CBRAM<br>• OxRAM |

***Figure 1.10: Classification of main NVMs based on the storage or switching mechanisms used.***

## *1.6.1 Memory Based on Charges*

### *1.6.1.1 The FLASH Memory*

The first solid state non-volatile memory device was suggested in 1967 at Bell Labs [9] by D. Kahng and S. M. Sze, who used a MOS transistor with an additional floating gate to store data. Starting with this invention, the first memory device employing avalanche injection, the FAMOS (Floating-gate Avalanche Injection MOS), was developed in 1974 by D. Frohman [10], evolving in the following years into the EPROM (Erasable Programmable Read-Only Memory) and the EEPROM (Electrically Erasable Programmable Read-Only Memory). In 1984 F. Masuoka at Toshiba [11] invented the Flash memory. The name "Flash" has probably been suggested by a colleague of the inventor because the erasure process of the stored information reminded him of the flash of a camera.

Flash memory is by far the most popular nonvolatile memory commercially used in modern devices. From its invention in 1984, it also followed Moore`s Law trend by doubling in capacity every 24 months. In the past few years, NAND Flash memories have experienced an even greater growth being widely used in portable devices such as media players, photo cameras and mobile phones. Due to their high density, low power and high I/O performance, NAND Flash memories started to even substitute mechanical hard disks in recent years.

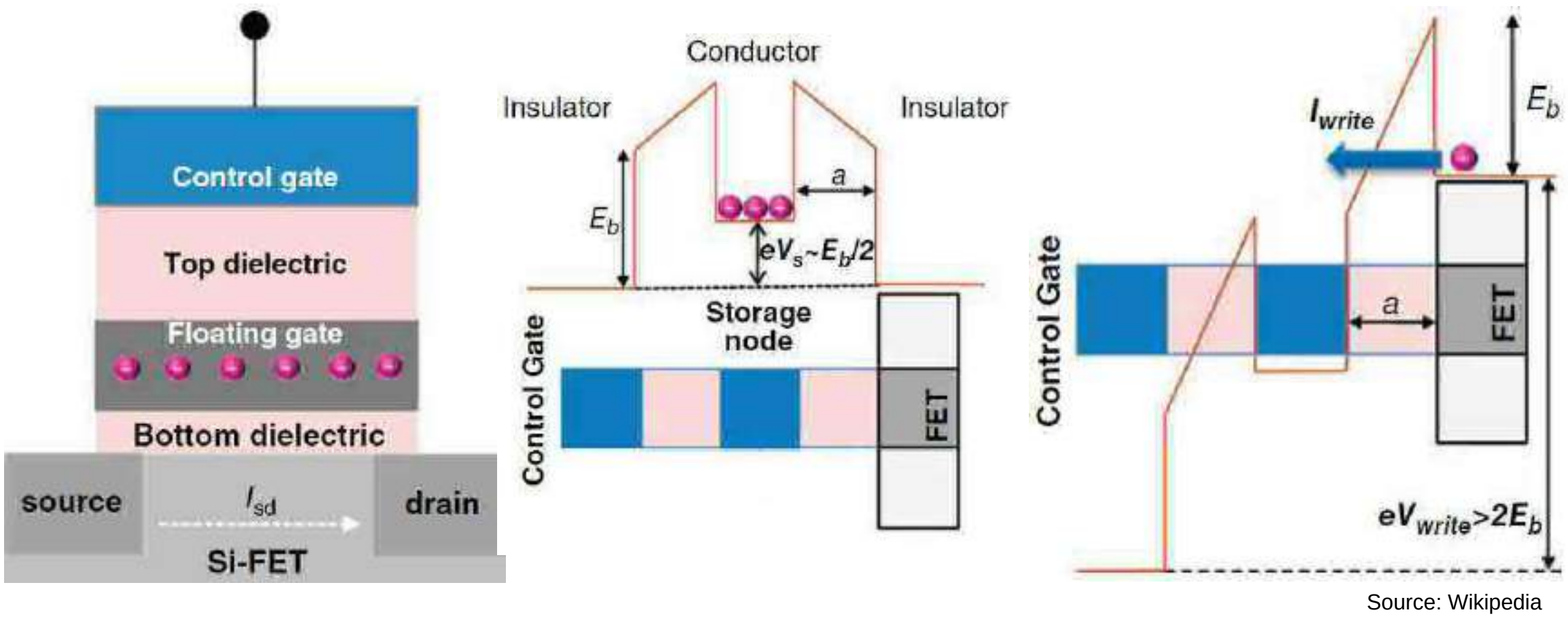

(a) (b) (c)

***Figure 1.11 - Flash memory: (a) cross section of the cell, (b), store operation (electrons remain trapped in the floating gate), and (c) F-N programming operation (electrons are injected into the floating gate).***

Flash memory is based on the ancestor floating gate transistor. As shown in Figure 1.11*a*, the single Flash memory cell is very similar to the classical MOSFET, with the addition of a floating gate located between the gate oxide and the insulating layer. The additional gate gives rise to a capacitor inside the gate oxide, on which it is possible to store extra electrons as indicated in Figure 1.11*b.* This configuration of a metallic storage layer (M) surrounded by insulators (I) is also called I-M-I structure, and is the basic feature of a floating gate device, where a large energy barrier is formed to store electrons. By applying a voltage to the upper control gate, the energy barrier is tilted, thus allowing a “tunnel current” to flow through the insulating layer, thus causing an electric charge to accumulate on the floating gate. The “tunnel current” is explained by the quantum mechanics process known as Fowler-Nordheim (F-N) Tunneling that describes the finite probability of an electron to tunnel through a very thin triangular energy barrier as shown in Figure 1.11c. This is the program operation of the Flash memory most used nowadays. Another technique used to program the cell is the Channel Hot Electron Injection (CHEI), which occurs when, by applying an adequately high voltage to the control gate and the drain terminals of the floating gate MOSFET, electrons in the channel are accelerated and, a number of them, gain enough energy to surmount the gate oxide barrier and accumulate on the floating gate. The reverse process, namely applying a high voltage to the silicon substrate and zero V (or, more frequently, a negative voltage) to the control gate and thereby extracting the accumulated charge from the floating gate, is the erase operation of the flash memory which also exploits F-N tunneling.

Figure 1.12a shows how logic levels are stored in a Flash memory cell. The electrons trapped in the floating gate during the program operation, induce a shift in the threshold voltage. The amount of charge stored in the floating gate can be controlled with good accuracy and a Multi-Level Cell (MLC) with 2 bits per cell can also be obtained (Figure 1.12b).

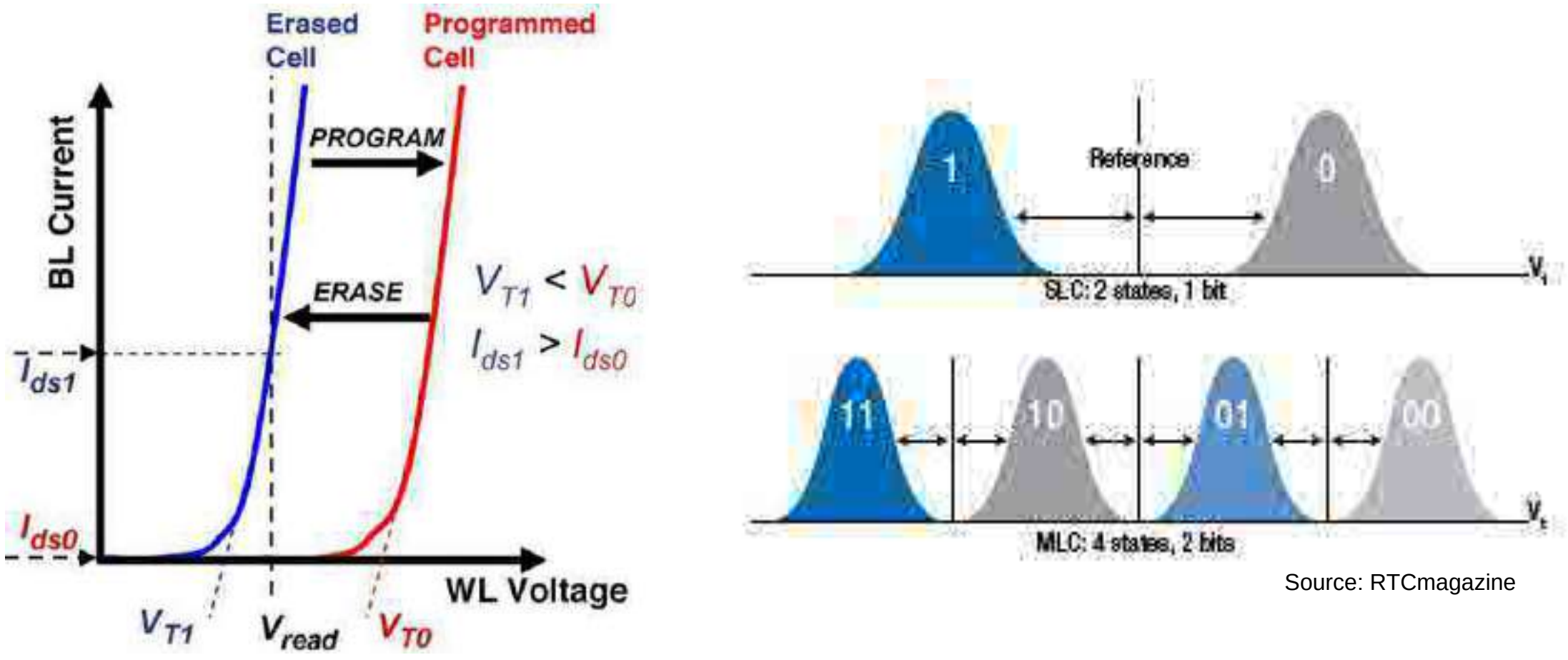

(a) (b)

***Figure 1.12: Bit logic with Vth shift (a), and Single Level Cell compared to Multi Level Cell (b)***

There are two main types of Flash memories on the market, the NOR and NAND type, which are named after the corresponding logic gates for the way the single cells are interconnected.

- In the NOR-type architecture, each cell can be accessed independently, allowing faster read access. For this reason NOR-type are mainly used in executable-code storage, for instance to store the firmware of a device.
- In the NAND-type array architecture, group of cells belonging to the same bit-line are connected in series between two bit-line select transistors, and the gates of all cells in the same row share the same word-line. This results in a much more compact layout since a much smaller number of contacts is required, but read speed is slower.

The performance comparison of NAND and NOR Flash is provided in Table 1.3.

Source: Wikipedia

| Parameter | | NAND | | NOR | |
|---|---|---|---|---|---|
| | | Current | Minimal | Current | Minimal |
| Feature size $F$ | | 16–32 nm | >10 nm | 45 nm | 25 nm |
| Access time | Write | ~100 μs | ~100 μs | ~10 μs | ~10 μs |
| | Read | ~10 μs | ~10 μs | 60–120 ns | ~60 ns |
| Retention time | | 10 yr | <10 yr | 10 yr | 10 yr |
| Write cycles | | $\sim 10^5$ | $<10^4$ | $\sim 10^5$ | $\sim 10^5$ |
| Operating voltage | Write | 15–20 | 15 | 8–10 | ~8 |
| | Read | 5 | 5 | 5 | 5 |
| Number of stored electrons | | ~50 | ~10 | ~200 | ~100 |
| Write energy (J bit$^{-1}$) | Cell level | $4 \times 10^{-16}$ | $\sim 10^{-16}$ | $2 \times 10^{-10}$ | $\sim 10^{-10}$ |
| | Array level | $10^{-11}$–$10^{-12}$ | $\sim 10^{-12}$ | $>2 \times 10^{-10}$ | $>10^{-10}$ |
| | System level | $10^{-10}$–$10^{-9}$ | $10^{-10}$–$10^{-9}$ | $\sim 10^{-9}$ | $\sim 10^{-9}$ |

***Table 1.3: Scaling and performance for NAND and NOR Flash Memory***

Flash memories on the market are currently experiencing many limitations. Scaling down the single cell size is becoming impossible due to the ensuing too high and uncontrollable leakage current. Moreover, increasing the density of states per memory cell by using a multilevel cell approach is becoming too challenging and error-prone. Solutions were proposed in terms of new high-k materials for gate dielectrics, new vertical and 3D structures and multi-stacked chips packages. However, the amount of charge stored on the floating gate remains a major bottleneck, since it cannot decrease any more. Figure 1.13 shows the total amount of electrons for different technological nodes and the number of "critical electrons", i.e. the maximum number of electrons that can leak from the Floating Gate without incurring in a substantial threshold shift and, then, losing of the stored bit.

It is interesting to note that a Floating Gate is allowed to lose at most 10% of the initially stored electrons and that, at the 20 nm node, the number of initially stored electrons is about 200. Thus, the specification for a 10-year retention is equivalent of requiring the leakage current to be less than 2 electrons per year!

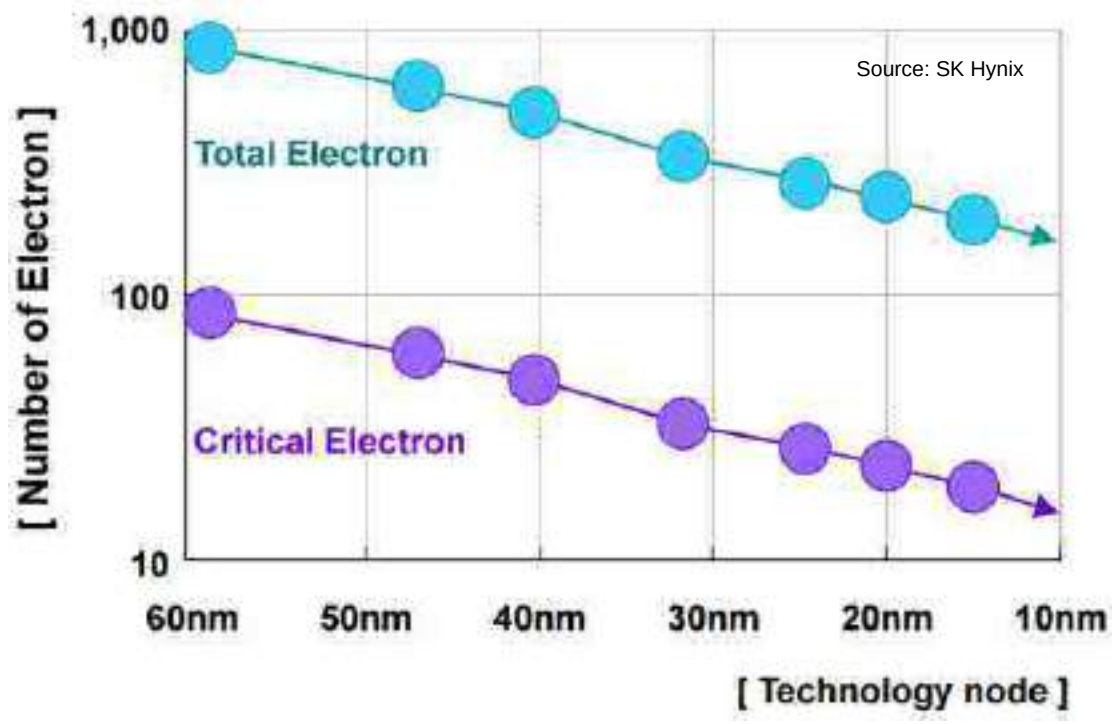

***Figure 1.13: Total number of electrons on the floating gate after programming and number of Critical electrons.***

Another disadvantage of Flash memories is the high voltage needed for programming. The Fowler-Nordheim Tunneling described before, requires several volts for the write/erase functions (~20V). For this reason, dedicated charge pump circuits are implemented and those account for most of the power consumption of the Flash memory. This issue limits, for example, the usage of Flash memories in the future Internet Of Things market, where super low power is necessary for having battery operated devices working4 for several years before discharging.

### *1.6.1.2 Ferroelectric RAM (FeRAM)*

Ferroelectric RAM (FeRAM) is an old technology first proposed by D. A. Buck in 1952 [12]. Its widespread development started in the 1980s. FeRAM is considered today a prototypical memory technology with several parts available commercially but only for niche applications (Ramtron [13]). FeRAM devices achieve nonvolatile storage by first switching and then sensing the electrical polarization state of a ferroelectric capacitor. The structure of a FeRAM cell is the same 1T1C structure as the DRAM cell, as shown in Figure 1.14, with the difference that the capacitor acts as memory element not by storing electrical charge (which then needs to be refreshed), but by switching the electrical polarization of a ferroelectric material. This is achieved by using a ferroelectric material in place of a conventional dielectric material between the plates of the capacitor. Traditionally, the most common ferroelectric materials used in electronic memories are PbZrTi (PZT) and $SrBi_2Ta_2O_9$ (SBT). These ferroelectric materials are not fully compatible with materials used in CMOS fabrication processes and must therefore be chemically or physically isolated from the underlying CMOS layers during production.

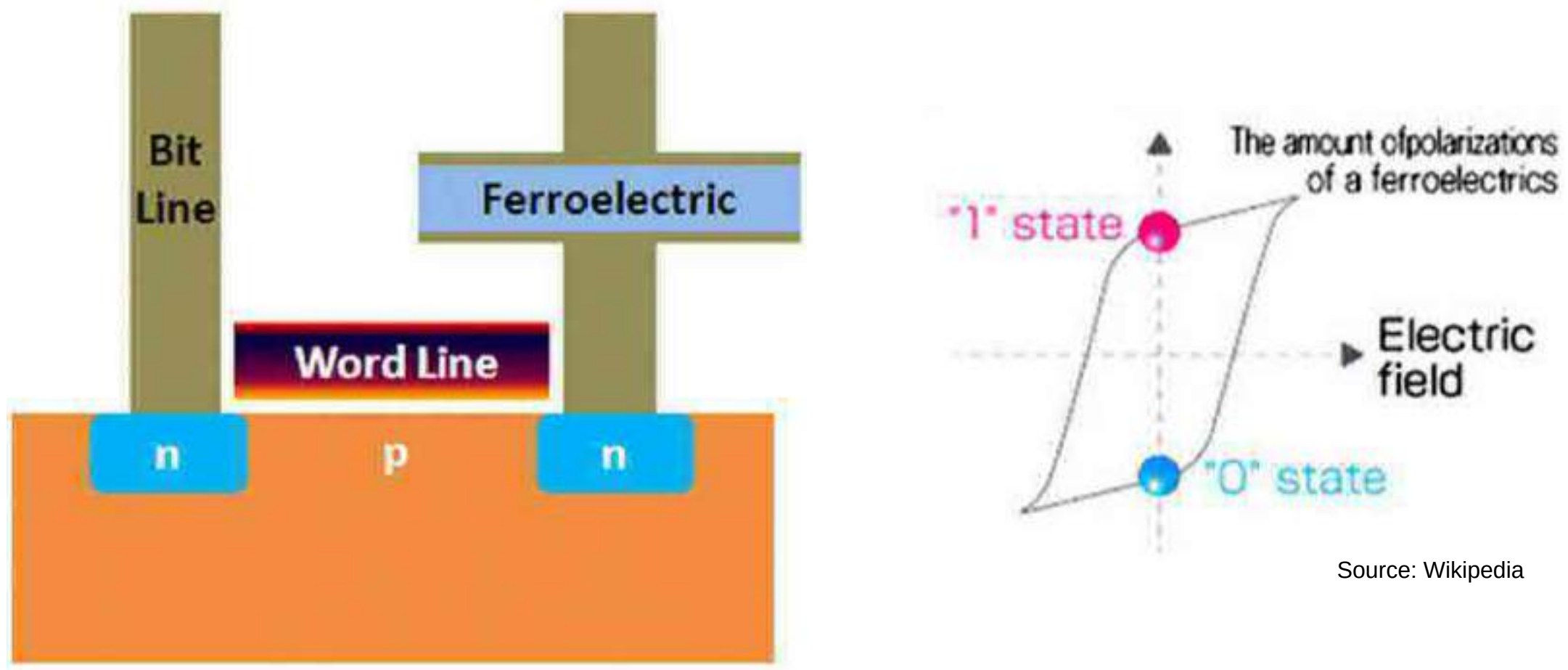

***Figure 1.14 – FeRAM: single 1T1C cell structure (left) and hysteresis loop of electric polarization vs electric field.***

Another drawback of FeRAM is the destructive readout technique, because the flow of the sensing current through the capacitor modifies its polarization. Consequently, each read access must be followed by a subsequent rewrite operation (such as in DRAMs). Initially, FeRAM was seen as a way to simply achieve all the advantages of DRAM (20ns read/write) together with long-term nonvolatile storage. However, despite years of concentrated integration efforts, FeRAM cells remain significantly larger than the DRAM cell.

### *1.6.1.3 Ferroelectric FET (FeFET)*

The ferroelectric FET (FeFET) is another memory that relies on the ferroelectric effect. FeFET was first proposed in 1974 by S. Wu [14] but is starting to become popular again thanks to the present better integration possibility and much greater knowhow on ferroelectric materials. The FeFET structure integrates a ferroelectric capacitor, similar to that in a FeRAM cell (PZT or SBT), into the gate of a MOSFET, as shown in Figure 1.15*a* and 1.15*b*.

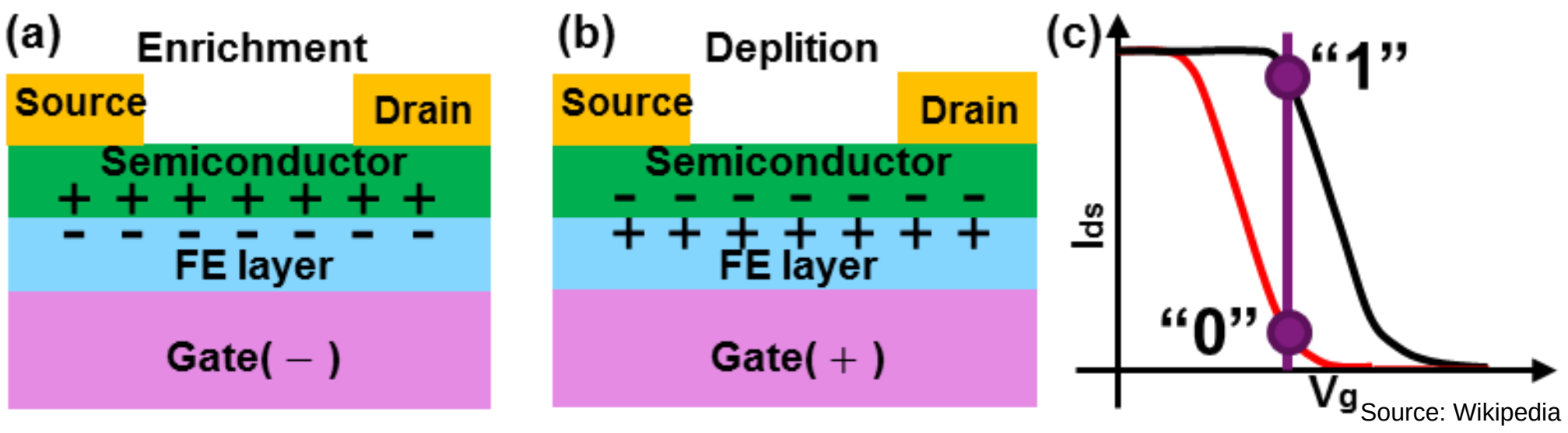

***Figure 1.15: A p-type FeFET cell (a, b) and corresponding hysteretic polarization loop (c).***

Switching is area dependent (as in a standard capacitor, the stored charge, or in this case, the ferroelectric polarization is proportional to the area of the plates) and is caused by a reversal of spontaneous electric polarization of thin ferroelectric films under an electric field with a resulting shift in the threshold voltage. The hysteresis cycle found in the Vg-Ids characteristics of the FeFET is shown in Figure 1.15*c*. The information can be read by detecting the drain current or the resistance of the channel. Also in the FeFET, the poor CMOS compatibility of the ferroelectric material, makes the integration and mass production of this technology non trivial. A solution to ease integration is to separate the ferroelectric material from the silicon substrate by means of a dielectric layer, but this approach has the drawback of a worse lifetime of the stored data due to inherent depolarizing field. However, the recent discovery of ferroelectricity in Aluminum doped Hafnium Oxide [15], a common CMOS compatible oxide (currently used in industry as High-K gate dielectric), made FeFET to gain for a second time a lot of popularity as a valid high-density CMOS compatible candidate for future memory.

### *1.6.2 Memory Based on Spins*

#### *1.6.2.1 Magnetic RAM (MRAM)*

MRAM, or magnetic RAM, is a nonvolatile RAM technology under development since the 1990s after having been first proposed by J. M. Daughton at Honeywell in 1987 [16]. MRAM is based on memory cells having two magnetic storage elements, one with a fixed magnetic polarization (pinned layer) and one with a switchable polarization (free layer). These magnetic elements are positioned on top of one another but are separated by a thin insulating tunnel barrier as shown in the cell structure in Figure 1.16. The structure is also referred to as Magnetic Tunnel Junction (MTJ). The switching is area dependent (the size of the MTJ impacts on the switching performance) and the bit is stored by changing the magnetic polarization of the MTJ. Because of the magnetic tunnel effect, if the magnetic polarizations of the two elements are parallel to one another, then the electrons will be able to tunnel (cell in the low-resistance - "ON" – state). Conversely, if the magnetic polarizations of the two elements are antiparallel, the cell resistance will be high ("OFF" state). Writing and erasing are fulfilled by forcing an adequate current through one of the two perpendicular write lines, thus inducing a magnetic field in different directions across the cell as shown in Figure 1.16. In view of power consumption and speed, MRAM wins over all other emerging memories, with a read access time of a few nanoseconds and really low current in write/erase operation. Moreover, since in MRAM no atoms are displaced during programming, and no high current flows directly through the MTJ, the endurance of such a cell is practically infinite. The only drawback lies in size and array integration, because of the intrinsic bigger dimension due to the additional word line needed for the perpendicular polarization.

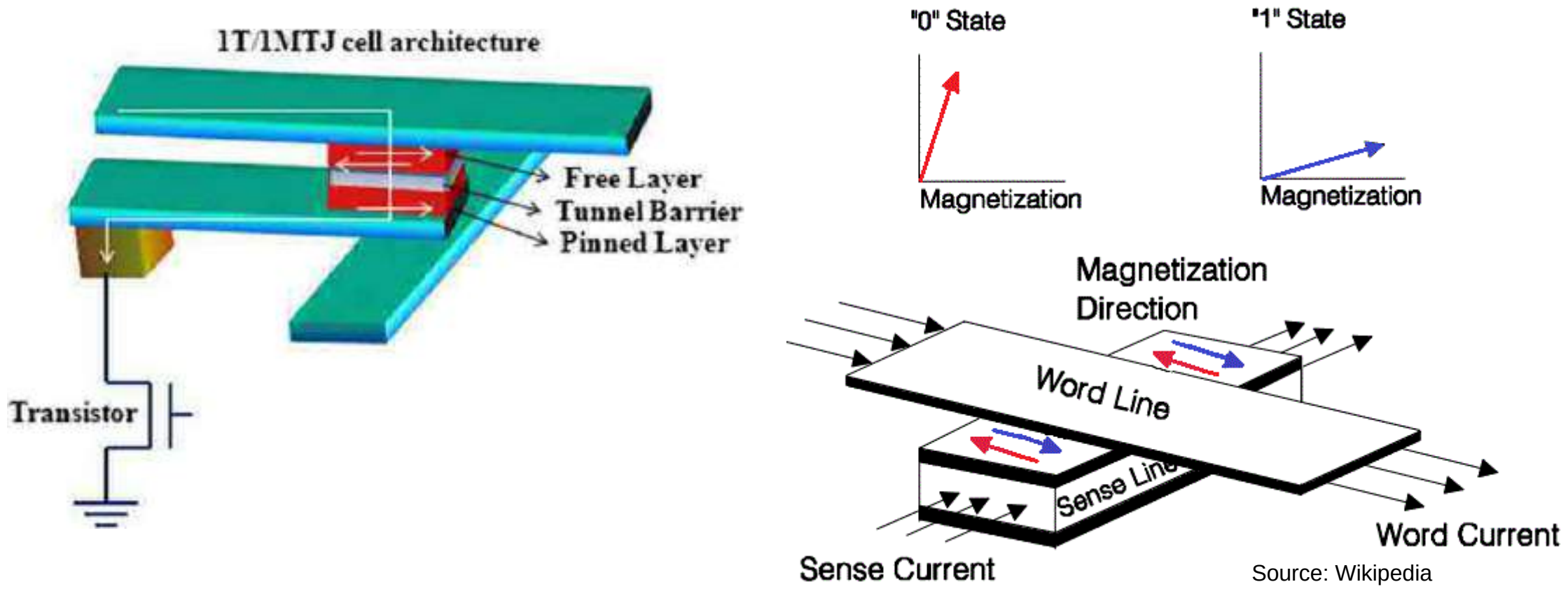

*Figure 1.16: Basic MRAM scheme and magnetization vector for different states*

### 1.6.2.2 Spin Torque Transfer RAM (STT-RAM)

The size limitation of the MRAM has been overcome by the introduction of the Spin Torque Transfer RAM (STT-RAM). The free layer of the MTJ can be switched parallel or antiparallel to the pinned layer depending on the direction of the current and the bit is stored in the spin of the electrons rather than in the magnetization like in MRAM, as shown in Figure 1.17. The different switching technique avoids the need of an extra perpendicular word line, but leads to a lower endurance of the STT-RAM because the programming current flows directly through the Magnetic Tunnel Junction, thus leading to possible breakdown. Scaling STT-RAM is beneficial in terms of power consumption during cycling, because a smaller current is needed for switching, but lowers the resistive window, thus making programming less reliable.

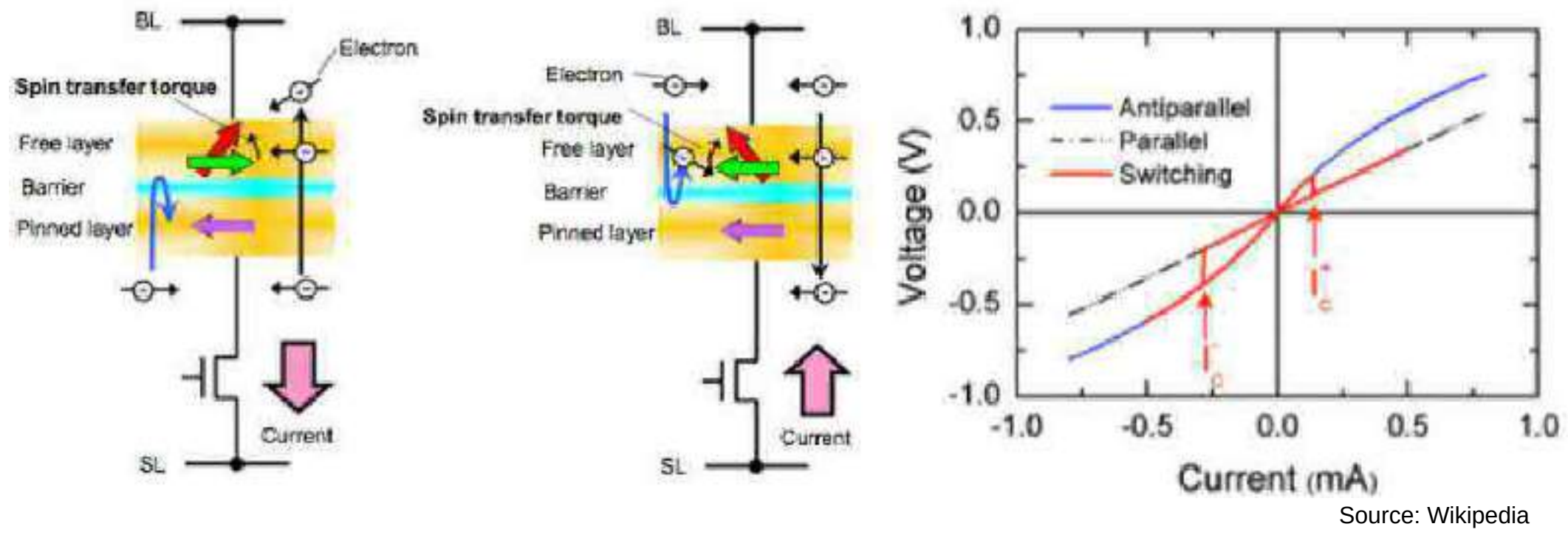

*Figure 1.17: Structure of an STT-RAM cell and example of current-induced switching.*

The smaller 1T1J cell size, the 20 ns programming/read speed, and the high endurance makes the STT-RAM a perfect candidate as Level 1-2 cache memory but, as in the case of the MRAM, the poor CMOS compatibility of the MTJ makes this technology difficult to produce and, hence, available just for niche applications.

## 1.6.3 Memory Based on Atoms

### 1.6.3.1 Phase Change Memory (PCM)

Phase Change Memory (PCM), also called Phase-change Random Access Memory (PRAM-PCRAM) is based on a class of material called chalcogenide glasses (the most commonly used compound is GeSbTe, or GST). The GST compound is a known material. In fact it has been intensively studied in 1960s by S. Ovshinsky [17] . In 1970 thanks to J. Russell [18], GST became commonly used as the data layer in rewritable compact disks and digital versatile disks (CD-RW and DVD-RW respectively) where the change in optical properties is exploited to store data.

PCM uses the resistivity difference between the amorphous and the crystalline states of the chalcogenide glass to store the logic "1" and the logic "0". As shown in Figure 1.18, the PCM consists of a top electrode, the chalcogenide phase change layer, a heater and a bottom electrode. An access transistor (not shown in Figure 1.18 for simplicity) is placed in series with the phase change element for selection purposes. The GST phase change consists of two complementary operations:

- the RESET operation, in which the chalcogenide glass is first momentarily melted by a heating resistor with a high-amplitude short electric pulse and then re-solidified, thanks to rapid quenching, into an amorphous phase with high resistivity; and
- the SET operation, in which a lower-amplitude longer pulse (10x longer) anneals the amorphous phase into a low resistance crystalline phase.

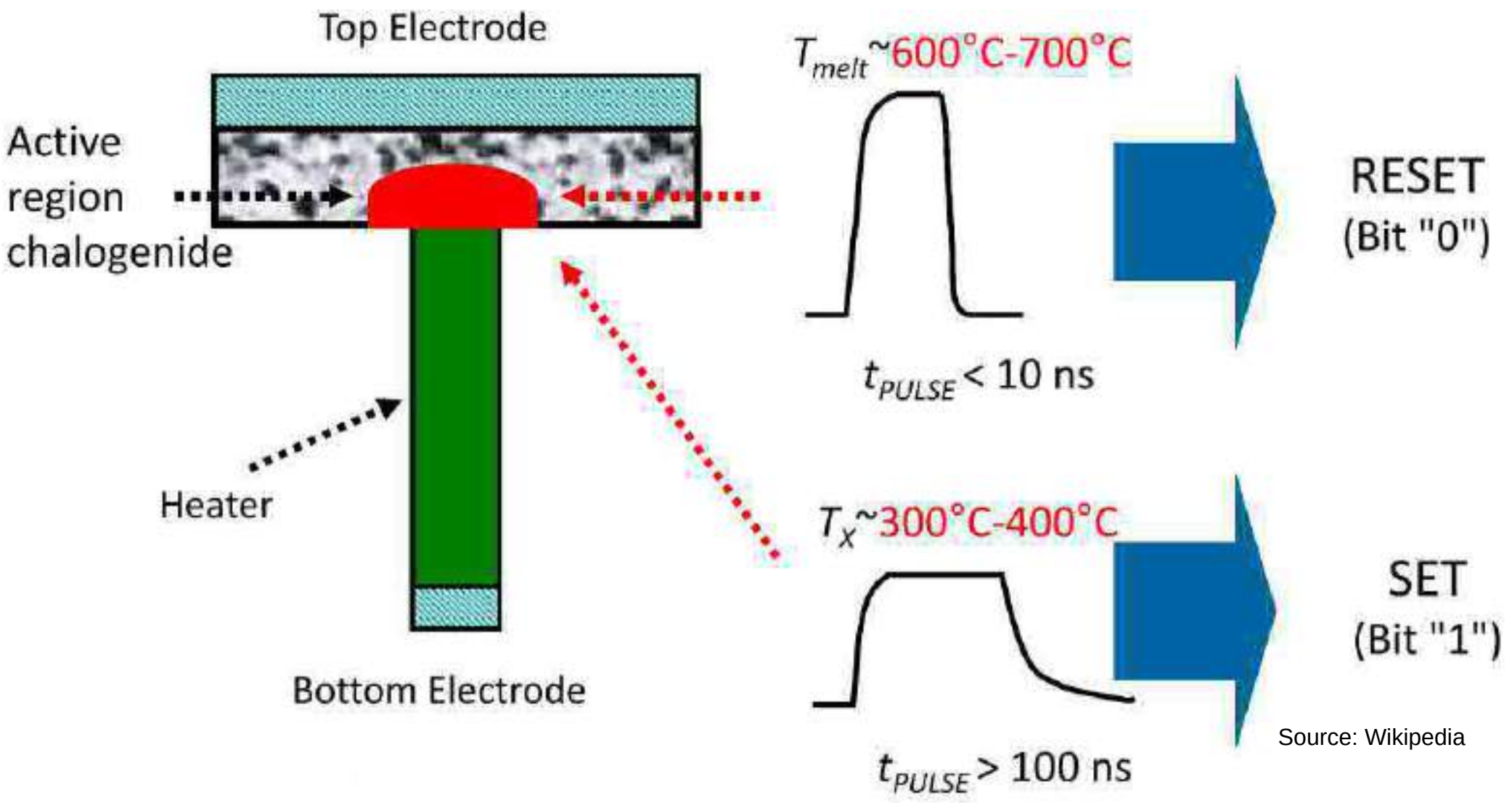

***Figure 1.18: PCM cross-section of a cell (left) and SET/RESET programming pulses (right).***

The major challenges for the PCM cell are the high current (on the order of the mA) required to RESET the GST by melting and the relatively long SET time required to allow its crystallization to occur. However, the switching mechanism is intrinsically area dependent (the GST volume next to the heater affects the switching speed) and the scaling of the cell is advantageous, leading to a reduction of the active volume and, hence, to the need for melting less material.

In terms of performance, the good switching proprieties of the GST give the PCM a high writing speed, needing just few tens of nanoseconds, which potentially makes it compatible with FLASH for the read operation, but which is several orders of magnitude faster for the write cycle. This makes it possible for PCM to function many times faster than conventional FLASH memory while using less power. In addition, PCM technology has the potential to provide inexpensive, high-speed, high-density, high-volume non-volatile storage, which is naturally prone to scaling. A possible problem affecting PCM is the high current density needed to erase the memory. However, as cell sizes decrease, the current needed will also decrease.

## 1.7 Background on Resistive RAM (RRAM)

Even though Resistive RAM is a “memory based on atoms”, since it represents the focus of the present thesis, a separate paragraph is devoted to this memory class.

Resistive RAM, or just RRAM is the common name used in the literature for devices in which the memory effect relies on the reversible transition from a High Resistive State (HRS) to a Low Resistive State (LRS) and vice versa by means of an ultra-scaled redox reaction involving movement of cations or anions. The presence of resistive switching in oxides was discovered in 1964 by P. Nielsen and N. Bashara [19], who suggested the use of such material as a memory element, but no further development was done until 2000s, when the research efforts for a post Flash era memory intensified.

In particular, four important papers marked the “rebirth” of the resistive memory device:

I. Paper by A. Beck and B.J. Bednorz, on Applied Physics Letters, in 2000, entitled *“Reproducible switching effect in thin oxide films for memory applications”* [20], is the first paper after more than 30 years of “silence”, and studies switching properties in oxides;
II. Paper by W. Zhuang *et al.*, at International Electron Devices Meeting in 2002, entitled *“Novel colossal magnetoresistive thin film nonvolatile resistance random access memory (RRAM)”* [21], is the first paper that used the term “RRAM” when showing an Al/$SiO_2$/Pt (+ other layers) based 1T1R test structure;
III. Paper by I. Baek *et al.*, at International Electron Devices Meeting in 2004, entitled *“Highly scalable non-volatile resistive memory using simple binary oxide driven by asymmetric unipolar voltage pulses”* [22], is the first advanced work that exploit simple binary oxide NiO and first introduced the term TMO-RRAM to highlight that switching occurs in a Transition Metal Oxide material;
IV. Paper by D.B. Strukov *et al.*, on Nature in 2008, entitled *“The missing memristor found”* [23], is an article that applies the Memristor concept to RRAM and was followed by great interest among circuit designers. (The memristor is the hypothetical fourth fundamental passive circuit element along with Capacitor, Inductor, and Resistor. L. Chua proposed this “missing element” in 1971 [24] actually predicting today’s resistive memories).

A RRAM memory cell is generally built by a capacitor-like MIM structure, composed of an insulating or resistive material (I) sandwiched between two metal conductors (M). In the literature there are many different types of RRAM, which use different materials both for the insulators and for the electrodes. The big variety of material combinations is because the switching behavior is not only dependent on the insulator material but also on the choice of the metal electrodes and their interfacial properties. Depending on the resulting behavior of the I-V characteristics in the first and third quadrant, the cell is called unipolar or bipolar. In unipolar switching (Figure 1.19*a*) the SET and the RESET operation can occur by successive application of electric stress of either the same or opposite polarity. On the contrary, in bipolar materials (Figure 1.19*b*), SET and RESET switching always occur with pulses of opposite polarities (with positive voltage for SET and negative voltage for RESET in Figure 1.19*a*). Usually, unipolar switching arises in symmetric MIM structures, whereas bipolar switching occurs in asymmetric stacks. However, materials were found in which both switching types coexist in one system. This is, for example, the case of Pt/$TiO_2$/Pt [25], where, depending on the value of the compliance current (or cc, is the maximum current that can flow through the RRAM, limited by an external power supply or by a built in limiting transistor), the switching changes from bipolar (cc < 0.1mA) to unipolar (cc > 1mA).

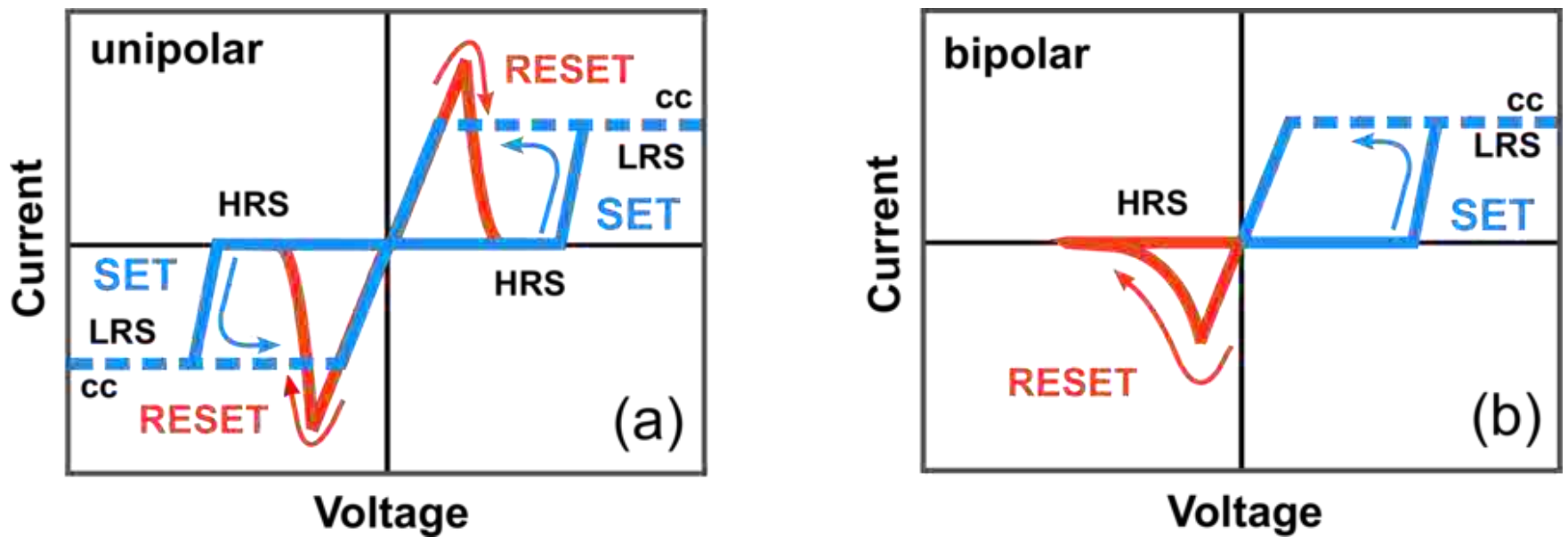

***Figure 1.19: I-V switching curves of materials showing Unipolar (a) or Bipolar (b) behaviors.***

A more detailed differentiation between different resistive switching mechanisms is shown in Figure 1.20, which has been adapted from the presentation held by D. Wouters in San Diego in 2012 at the *IEEE Semiconductor Interface Specialists Conference* [26]. In this figure the classification is based on the geometry of the memory switching mechanism. As discussed in the above paragraphs, Phase Change Memory and Magnetic RAM (but also FeRAM, FeFET, and STT-RAM) are all examples of 3D Bulk Transition materials, in which switching is volume dependent and the speed achieved and/or the energy needed for the programming are proportional to the overall size of the cell and are therefore strongly affected by scaling. Figure 1.20 also gives another level of characterization by reporting the polarity of switching. PCM is unipolar because switching depends only on the current pulse shape and not on its polarity (positive or negative), whereas MRAM is bipolar because of it asymmetric structure, in which the magneto-resistance is dependent on the alignment of the top free layer with respect to the bottom pinned layer, thus on the direction of the current.

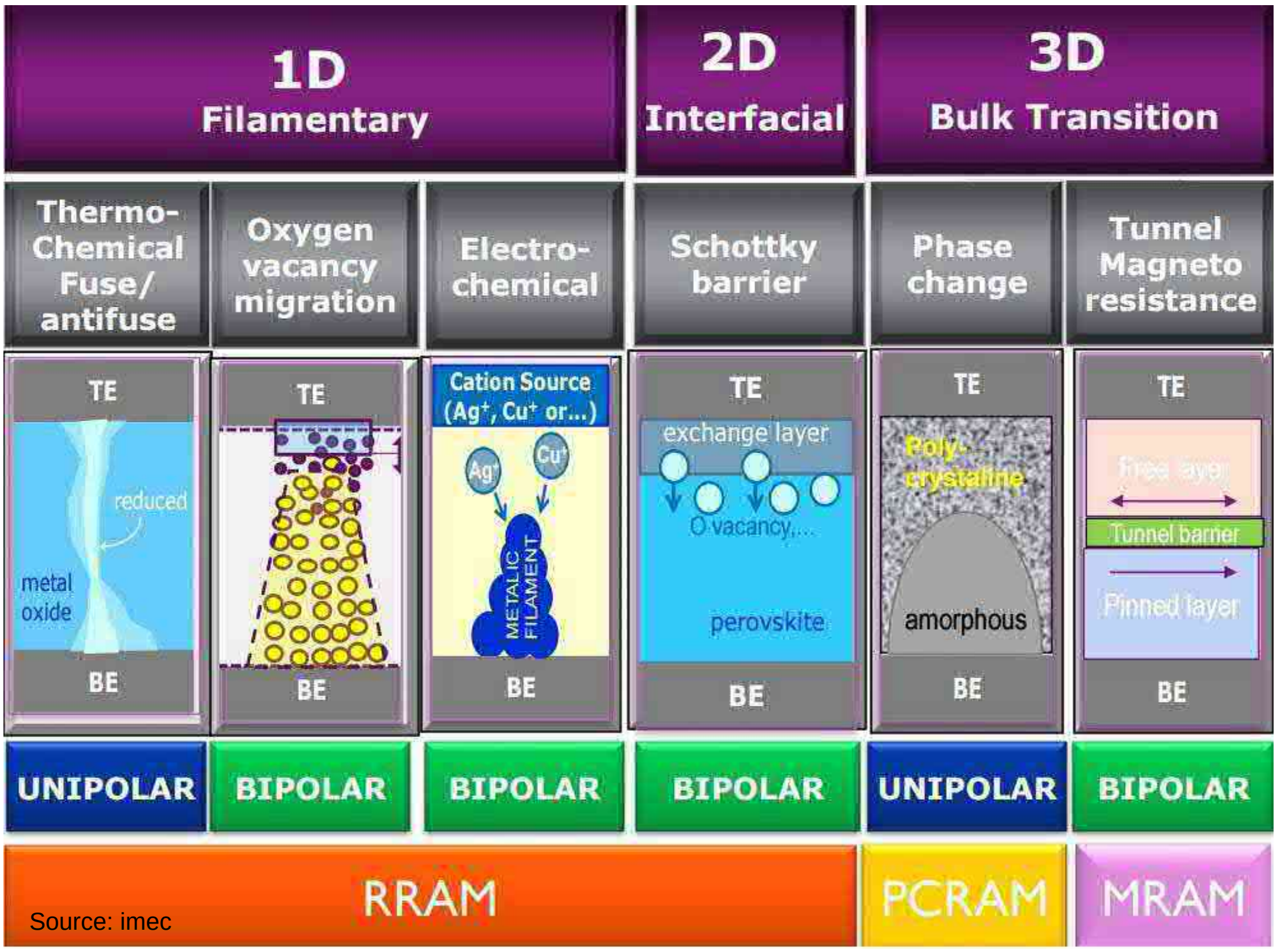

***Figure 1.20: Taxonomy based on resistance modulation geometry.***

The same characterization is applied to Resistive Memories. In Figure 1.20, three different resistance modulation geometries are listed, of which two follow the switching mechanisms reported as RRAM devices. While the 3D Bulk Transition resistance modulation has been already threated with the PCM and STT-RAM explanations, the remaining mechanisms are described in the succeeding subparagraphs in the following order:

1. Schottky barrier Memory;
2. Thermo-Chemical Memory (Fuse/Antifuse);
3. Electro-Chemical Memory; and,
4. Oxygen Vacancy migration Memory.

## *1.7.1 Schottky barrier Memory*

This type of switching is defined as a 2D interfacial resistance modulation, because it takes place at the interface. An example of such a device is the memristor presented by HP labs in 2008. Their device is a Pt/$TiO_2$/$TiO_{2-x}$/Pt structure where the switching is obtained by the movement of oxygen vacancies (Vo) along the material. When applying a positive voltage, the increase in electric field makes the oxygen drift outside the $TiO_2$ changing the shape of the Schottky barrier between the Pt electrode and the $TiO_2$ layer (Figure 1.21*a* and 1.21*b*). Therefore, thanks to the thinner barrier, electrons can more easily flow though the element and the cell results in its low resistance state (SET). When applying a negative voltage, oxygen vacancies drift backwards restoring the energy barrier to its original shape, thus switching into a high resistive state (RESET). The bipolar resistive hysteresis is shown in Figure 1.21*c*.

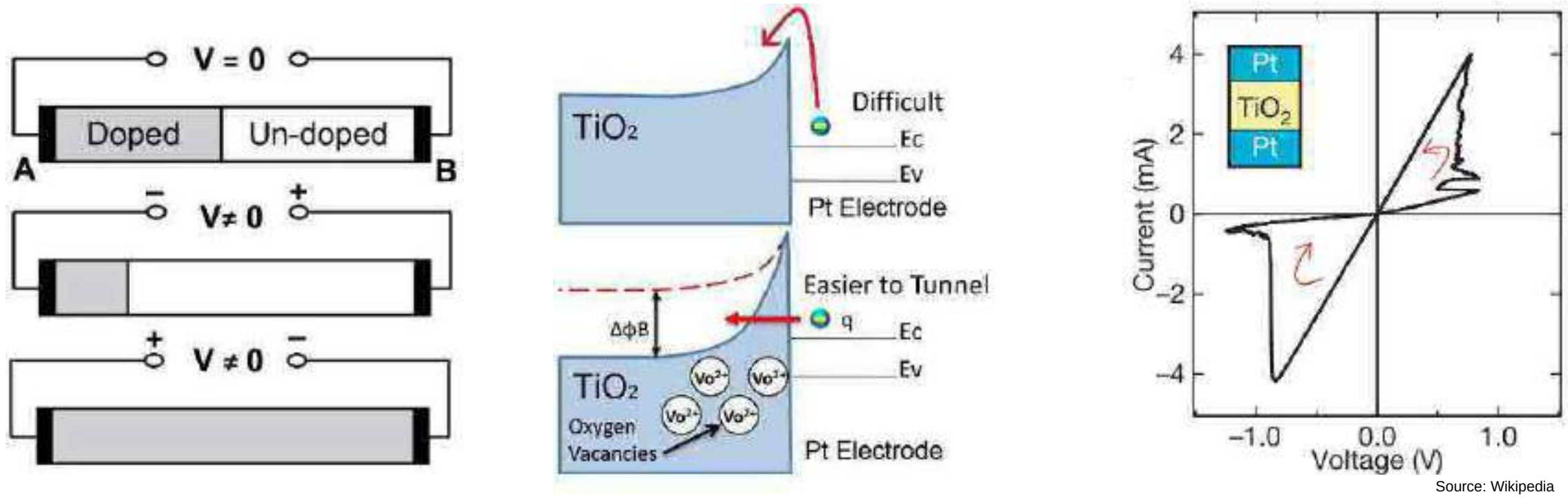

***Figure 1.21: Conduction mechanism in a memristor (a) with thinner energy barrier when a positive potential is applied to electrode A (b) and showing a bipolar resistive hysteresis (c) [27].***

## *1.7.2 Thermo-Chemical Memory (Fuse/Antifuse)*

The Thermo-Chemical Memory (TCM) or Fuse/Antifuse memory is characterized by an unipolar switching and a 1D filament conduction model. It is called thermochemical because the change in resistance takes place by both thermal reactions, in which the conductive filament (CF) modification is thermally activated, and chemical reactions, (specifically, a redox reaction takes place). An example of chemical reaction occurring during the SET process in a Pt/CuO/Pt TCM cell is:

$$2CuO \rightarrow Cu_2O + O_{(s)}$$

$$Cu_2O \rightarrow 2Cu + O_{(s)}$$

where $O_{(s)}$ is the oxygen resulting from the reduction of copper oxide and copper dioxide.

The switching in this type of cells is also called Fuse/Antifuse because of the abrupt transition between the low resistive state and the high resistive state: when applying a voltage to the cell, a sudden change in resistance is observed due to a local phase transition in the conducting filament. As shown in Figure 1.22*a* the switching is obtained by a physical disruption of the filament connecting the two electrodes. During RESET-to-SET transition, the high current induces a Joule heating and a permanent conductive path is formed through a bond breaking/reduction in the oxide, breaking it into metal (conducting) and oxygen. Similarly, during SET-to-RESET transition the initial high ON current induces again high temperature by Joule heating and a thermal oxidation dissolves the filament, returning it to its initial insulating oxide state.

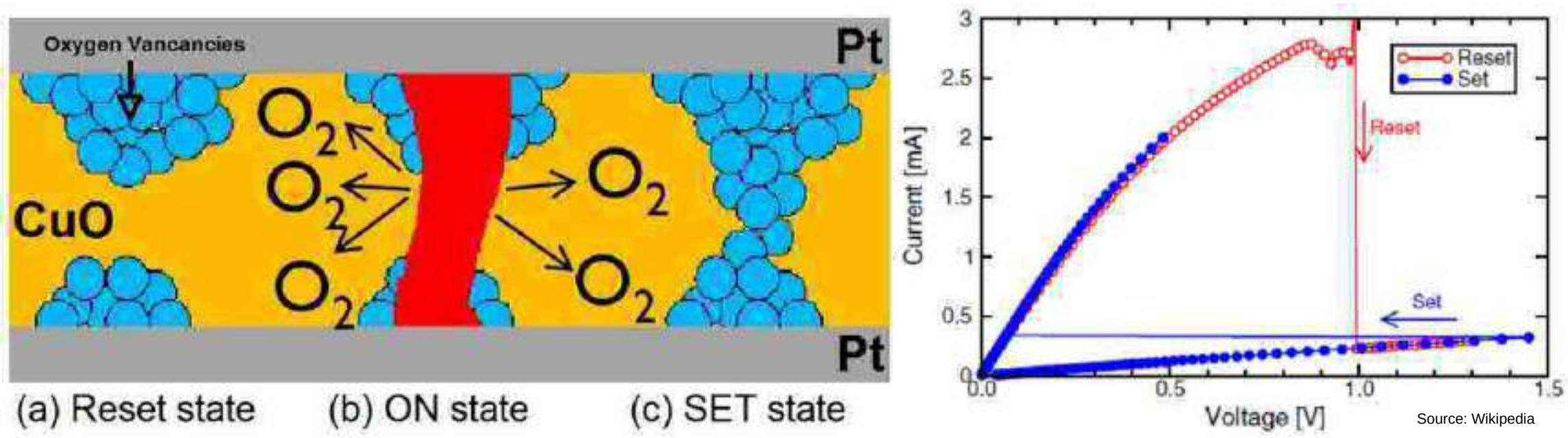

***Figure 1.22: TCM cell conducting filament switching (a) and unipolar I-V curve (b).***

The main issue of this type of memory is intrinsic in its unipolar switching as shown in Figure 1.22*b*. The state transition consists of a radial diffusion of conductive elements/defects from the CF, thus causing a loss of conductivity. As a result, the conductive elements might not be available for CF rebuilt after repetitive unipolar RESET operations, which lead to endurance limitations. In addition, the temperatures and voltages for unipolar RESET are high and this may result in an accelerated electro migration of metallic atoms. For those reasons other type of RRAM are more commonly studied because of lower power consumption and greater endurance than in the case of TCM cells.

### *1.7.3 Electro-Chemical Memory (CBRAM)*

Electro-Chemical Memory (ECM) cell, also referred to in the literature as Conductive Bridging RAM (CBRAM), is one of the most popular Resistive RAM under development. The operation principle of a CBRAM device can be described as an ultra-scaled redox reaction occurring in the solid electrolyte sandwiched in a MIM structure. The structure, shown in Figure 1.23, consists of an active electrode (e.g. Cu, Ag) and an inert electrode (e.g. Pt, W) with a solid insulating layer sandwiched in between (A variety of oxide, chalcogenide and halide materials have been used as thin films between the electrodes). Resistive switching is obtained by a voltage-induced modulation of the resistance of a conductive filament through the insulating layer, which is used as an electrolyte for cation drift. The result is the formation (SET) and the rupture (RESET) of the conductive filament, which are used to represent Boolean 1 and 0, respectively, in data storage schemes.

As shown in Figure 1.23, applying a positive voltage to the active electrode leads to the dissolution of the active metal and the deposition of a metallic (Cu) filament at the counter-electrode, which ultimately bridges the relatively insulating solid electrolyte and defines the low-resistance ON state [28].

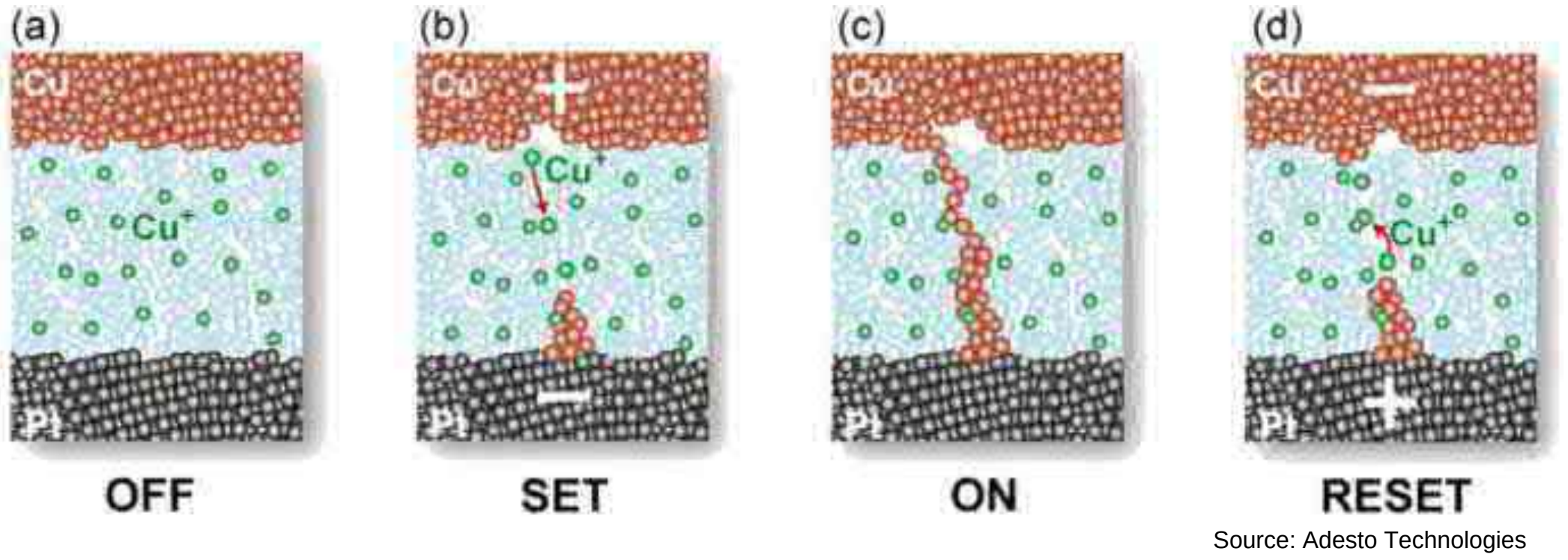

***Figure 1.23: Cross section of a CBRAM cell showing the switching process.***

More in detail, when a negative bias is applied to the inert electrode the metal ions in the electrolyte, as well as some originating from the now-positive active electrode, flow through the electrolyte and are reduced (converted to atoms) by electrons from the inert electrode. After a short period of time the ions flowing into the filament form a small metallic "nanowire" between the two electrodes. The "nanowire" dramatically reduces the resistance along that path, which can be measured to indicate that the "writing" process is complete.

Reading the cell simply requires the control transistor (not shown in Figure 1.23 for simplicity) to be switched on and a small voltage to be applied across the cell. If the nanowire is in place in that cell, the resistance will be low, thus leading to higher current (and that is read as a "0"). If there is no nanowire in the cell, the resistance is higher, which lead to a lower current (and is read as a "1"). Erasing the cell is similar to writing, but uses a positive bias on the inert electrode. The metal ions will migrate away from the filament, back into the electrolyte, and eventually to the negatively charged active electrode. This breaks the nanowire and increases the resistance again.

Due to key advantages such as scalability, low-voltage, fast-write operation, and long endurance lifetime, CBRAM technology has shown considerable progress in recent years and is considered as one of the more promising candidates for the Next Generation mainstream solid-state memory [28].

## *1.7.4 Oxygen Vacancy Migration Memory (OxRAM - RRAM)*

In oxygen vacancy migration memory or valence change memory (VCM) cells, the resistive switching effect relies on the movement of oxygen vacancies and a subsequent redox reaction on the nanoscale. This leads to a formation of a conductive filament made of oxygen vacancies, which is responsible of the resistivity change. Therefore, OxRAM cell is similar to CBRAM but instead of using a conductive filament made of displaced Cu or Ag atoms, the resistive switching is obtained by a conductive filament made of oxygen vacancies [30]. The OxRAM memory cell has a capacitor-like structure made up of a transition metal oxide sandwiched between two metal electrodes (Figure 1.24). Due to the involved ion migration the device operation is inherently bipolar.

Comparing OxRAM with scaled floating gate memories, a number of conceptual similarities are noted. In both FG and OxRAM the data is stored in a very small number of particles. In the FG memory, as previously shown, data is represented by storing electrons on the FG. Similarly, for OxRAM, data is memorized through oxygen vacancies residing in the constricted part of the filament. In the case of FG memories, electrons are confined by a potential barrier, while in case of OxRAM, the vacancies are kept in place by a diffusion.

The memory based on the above principle is named as Oxide Resistive RAM (OxRAM), and represent the kind of memory investigated in the present thesis. In the following Chapters, OxRAM will be often referred to as RRAM for simplicity.

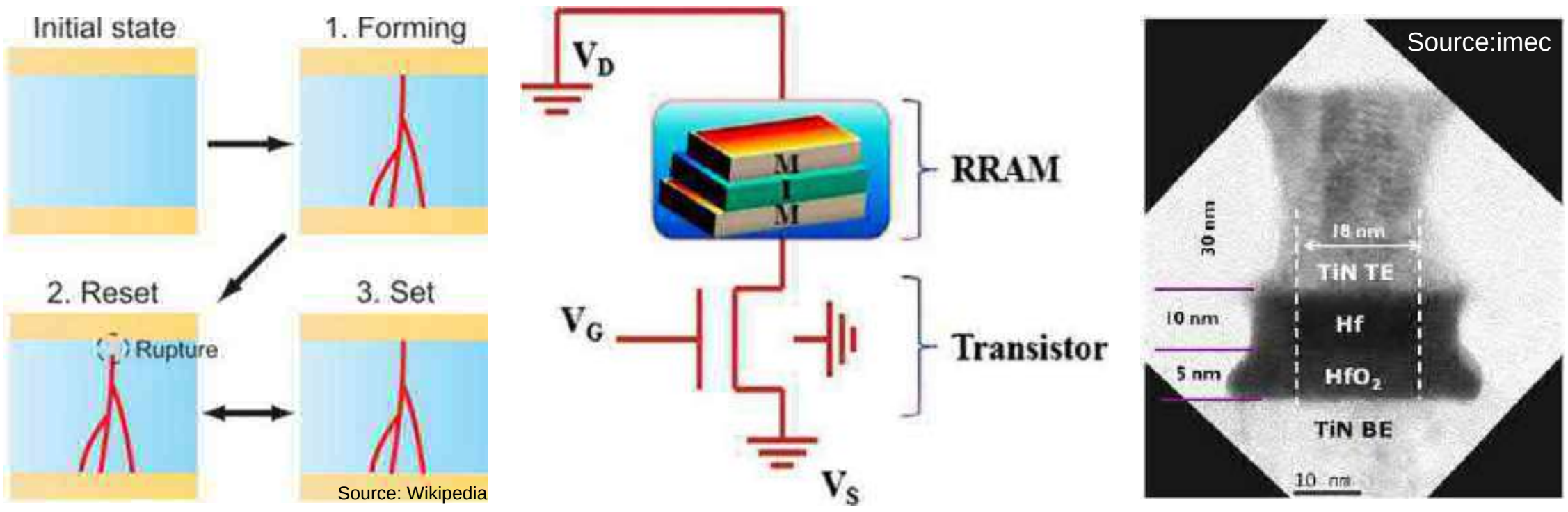

***Figure 1.24 - OxRAM: Programming cycle (left),cross section of a 1T1R cell (middle), and a SEM photo of the cross section of an HfO2-based cell, highlighting the layers composition and Top and Bottom electrodes (TE and BE) [31].***

In order to “enable” the cell, a first once-for-all forming pulse must be performed. When the RRAM cell is fabricated, it is just an MIM capacitor, as shown in Figure 1.24. The forming pulse is needed to first break the oxide and generate the initial filament of Oxygen Vacancies (red filament in Figure 1.24. The size and the number of oxygen vacancies filament depend on the magnitude of the current at which the oxide breakdown occurs, and can be controlled by a series current limiting transistor or by external compliance (better explanation of this structure is given in the next Chapter). After forming the filament, the cell is ready to act as a memory element. When an adequate voltage is applied across a RRAM cell, depending upon the voltage polarity, one or more conductive filaments made out of oxygen vacancies are either formed or ruptured. Once the conductive filaments are formed inside the metal oxide to bridge the top and bottom electrodes, the cell is in a low resistance state (LRS), and current can flow through the CFs. The larger the size of the conductive filaments, the lower the resistance. Figure 1.24 illustrates the formation of the conductive filaments in a bipolar RRAM cell. Conversely, the rupture of the conductive filaments disconnects the top electrode from the bottom electrode, resulting in a high resistance state (HRS) of the cell. When a positive current (from Top Electrode to Bottom Electrode) passes through the cell, the oxygen atoms are knocked out of the lattice and become negatively charged oxygen ions. Under this positive electric field, the oxygen ions will drift towards the anode, leaving corresponding oxygen vacancies in the metal oxide layer. The "rapture" of the filament (indicated in Figure 1.24a) is then obtained by reversing the voltage over the RRAM and inducing a negative electric field that refills the oxygen vacancies with oxygen ions, resulting in a shrinking and breaking of the conductive filament. This rapture is reversible, allowing the resistive switching. A deeper explanation of the filament behavior during switching is given in Chapter 4.

Because of its simple and highly scalable cross-point structure, with demonstrated multilevel stacking capability, this technology is very promising. It is expected that 1T1R-RRAM could meet the demand for high-speed memory technology [32].

## 1.8 Semiconductor Memories Taxonomy Summary

For the purpose of introducing the background of the research conducted in this thesis, Chapter 1 has so far listed and explained the characteristics of the main semiconductor memories, distinguished between those technologies that are currently available on the market and those at are still at an emerging state.

Figure 1.21 has been extracted from the annual report published by the International Technology Roadmap for Semiconductors association, known throughout the world as the ITRS. ITRS publishes every year, since 1998, roadmap with every semiconductor industry's future technology requirements. These future needs drive present-day strategies for world-wide research and development among manufacturers' research facilities, universities, and national labs. The objective of the roadmap is to ensure cost-effective advancements in the performance of the integrated circuit and the products that employ such devices, thereby continuing the health and success of this industry [33].

Specifically for the emerging technologies, Figure 1.21 summarizes the discussed advantages and disadvantages of each technology according to nine evaluation parameters.

In particular this thesis will discuss the investigation performed on the **variability** in retention of Metal Oxide (Bipolar Filament) OxRAM or RRAM. The "green mark" given to retention could lead to misunderstandings. Indeed the retention of a single cell, has been demonstrated to be very good, but the problem arises due to the intrinsic variability that is experienced from Device to Device and even from Cycle to Cycle. This investigation will be the main topic of the experimental part in Chapter 3 and the analysis in Chapter 4.

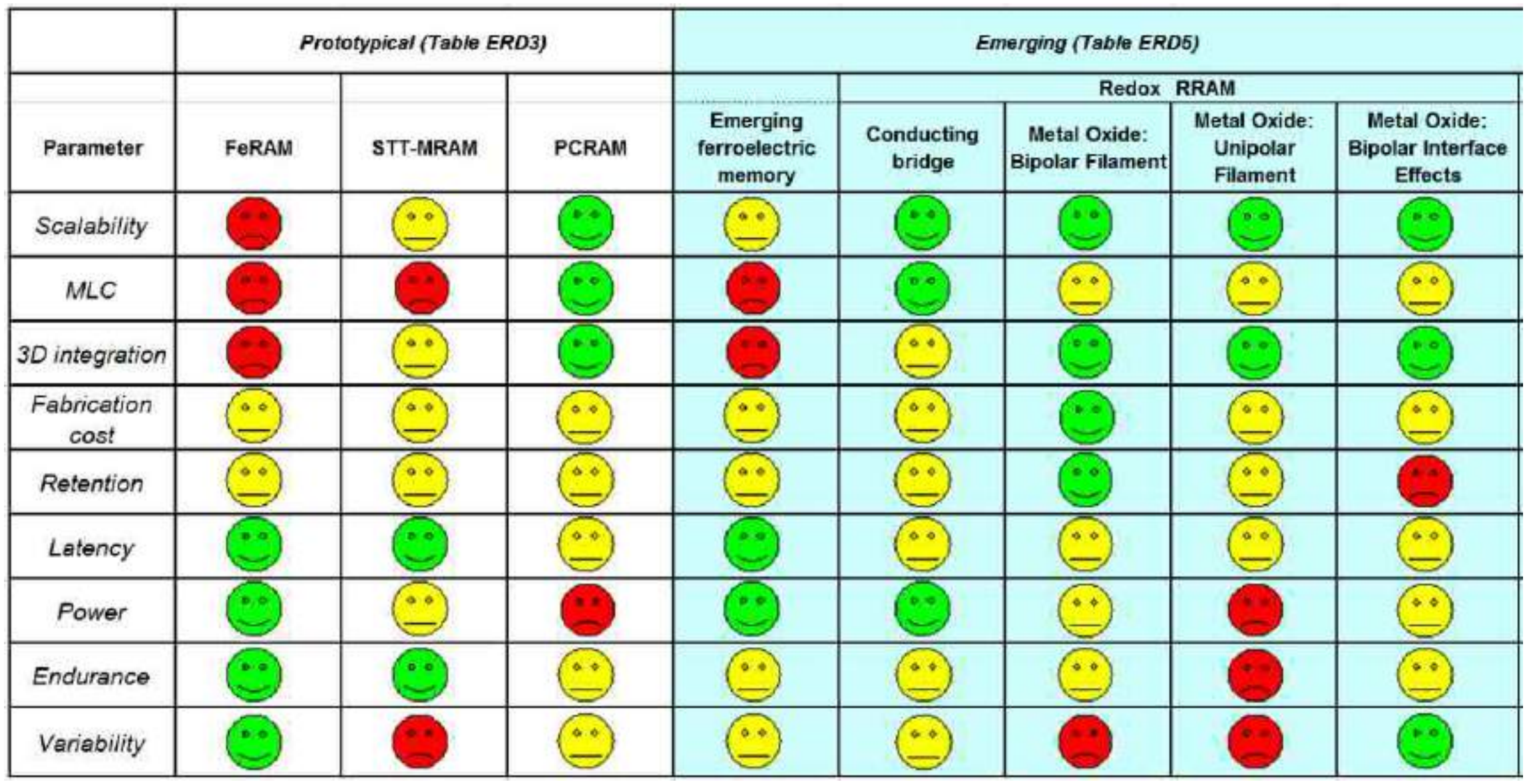

***Figure 1.25: Potential of the current prototypical and emerging memory candidates for SCM applications[8] , (Tables ERD3 and ERD5 [33]). The description of the single technology was given in the previous paragraphs as follow: FeRAM ($1.6.1.2), STT-MRAM ($1.6.2.2), PCRAM ($1.6.3.1), Emerging ferroelectric memory ($1.6.1.3), Conductive bridge ($1.7.3), Metal Oxide: Bipolar Filament (RRAM measured in this thesis $1.7.4), Metal Oxide: Unipolar Filament ($1.7.2), Metal Oxide: Bipolar Interface Effects($1.7.1).***

Figure 1.26 gives the reader few more details of the state of the art technologies that have been presented. It is obvious that in the last years a great effort has been spent by the scientific community, to improve those emerging technology, with always improving speed performances and storage capacities.

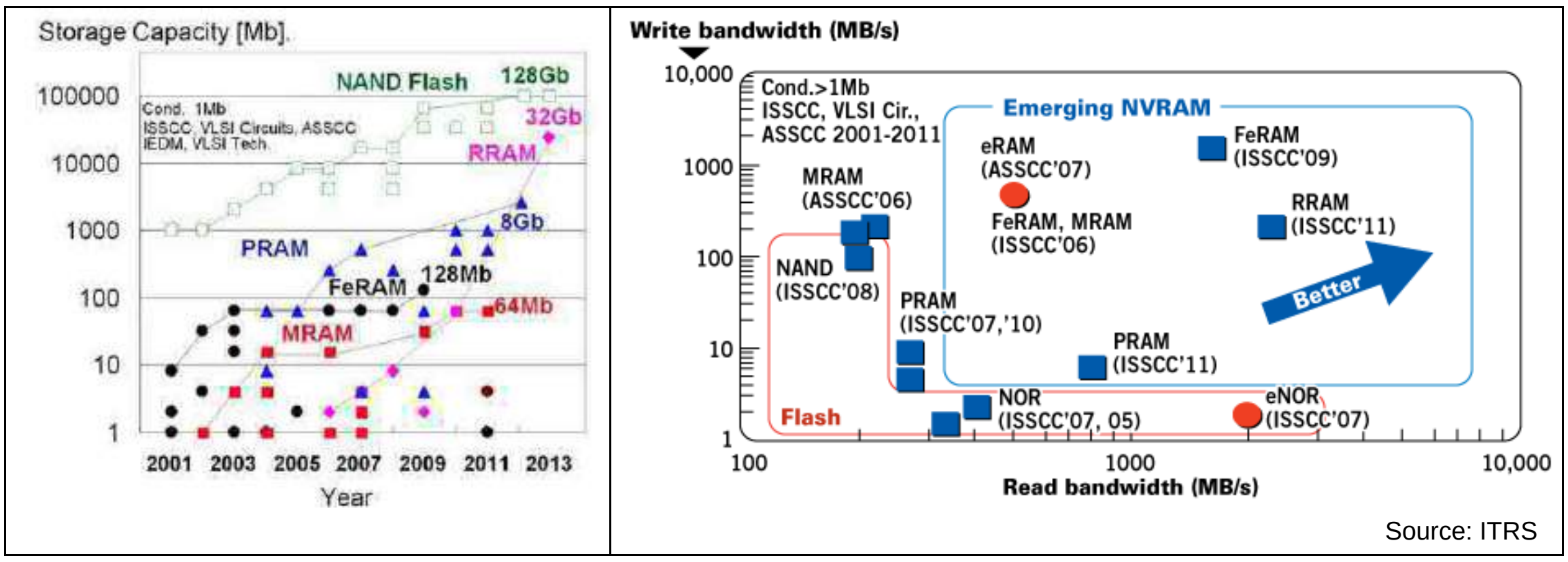

***Figure 1.26: The storage capacity evolution, during the last decade, (left), and Write-vs-Read performances (right) of the main emergent Non-Volatile technologies.***

# Chapter 2
# Experimental Set-up

This chapter presents the goals of this thesis work and the tools exploited to collect the statistical data. First, measurements procedures implemented to evaluate and characterize data retention in non-volatile memories are described. Secondly, the measurement setup and the tested RRAM cell structure are presented. At this point, the specific goals of this work will be explained. Finally, Chapter 2 describes the software developed for performing the needed tests, with a focus on data collection algorithms.

## 2.1 Test Equipment

All the data presented in this thesis were collected using the tools provided by IMEC`s AMSIMEC electrical test labs. The used setup is based on a Keithley K4200 along with a Cascade Microtech Summit 11K probe station (Figure 2.*a*) for measurements at room temperature, and a Cascade Microtech Elite 300 probe station for measurements up to 300 °C.

The Keithley K4200 is a commonly used Semiconductor Parameter Analyzer (SPA), which is chosen by test engineers to electrically characterize semiconductor devices [1]–[3]. Its most handy characteristic is the proprietary test environment KITE, which allows users to both run predefined interactive tests (called “Interactive Test Modules”, ITM) and develop their own customized and arbitrarily complex test routine (called “User Test Module”, UTM), by means of C programming language coding and proprietary Keithley libraries. Both test modules can be run as instantiated test sequences in KITE in which the user can input test parameters and retrieve measurement results both within a file and through an xy plot.

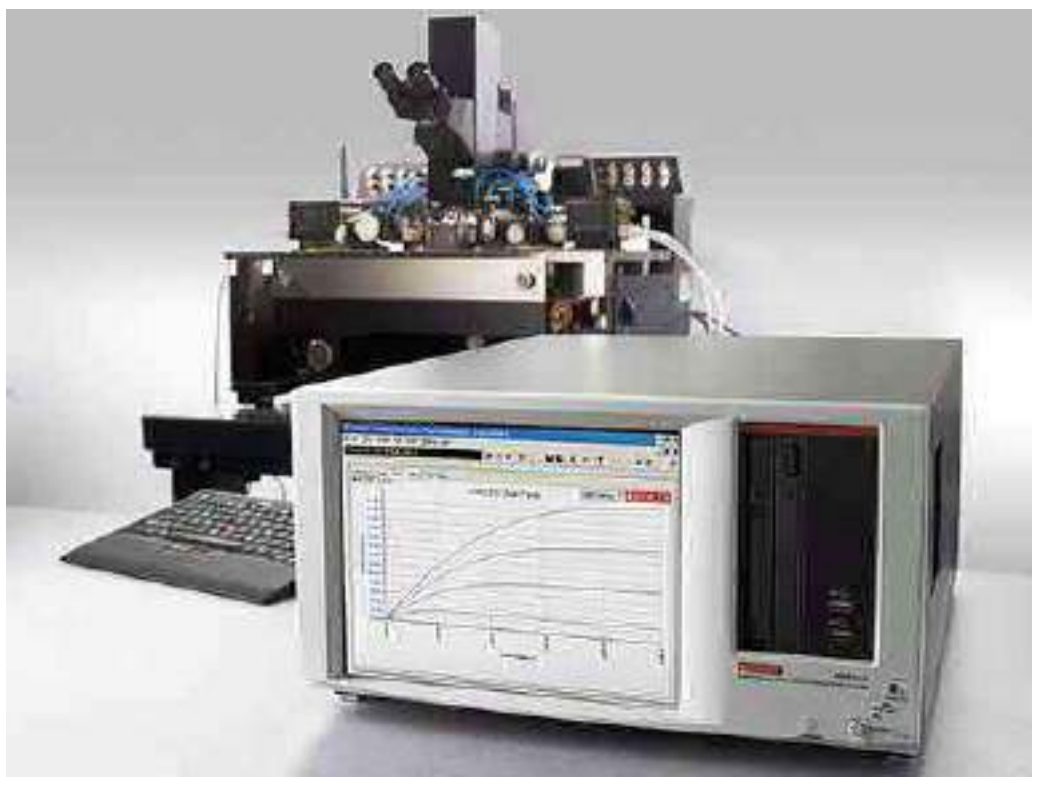

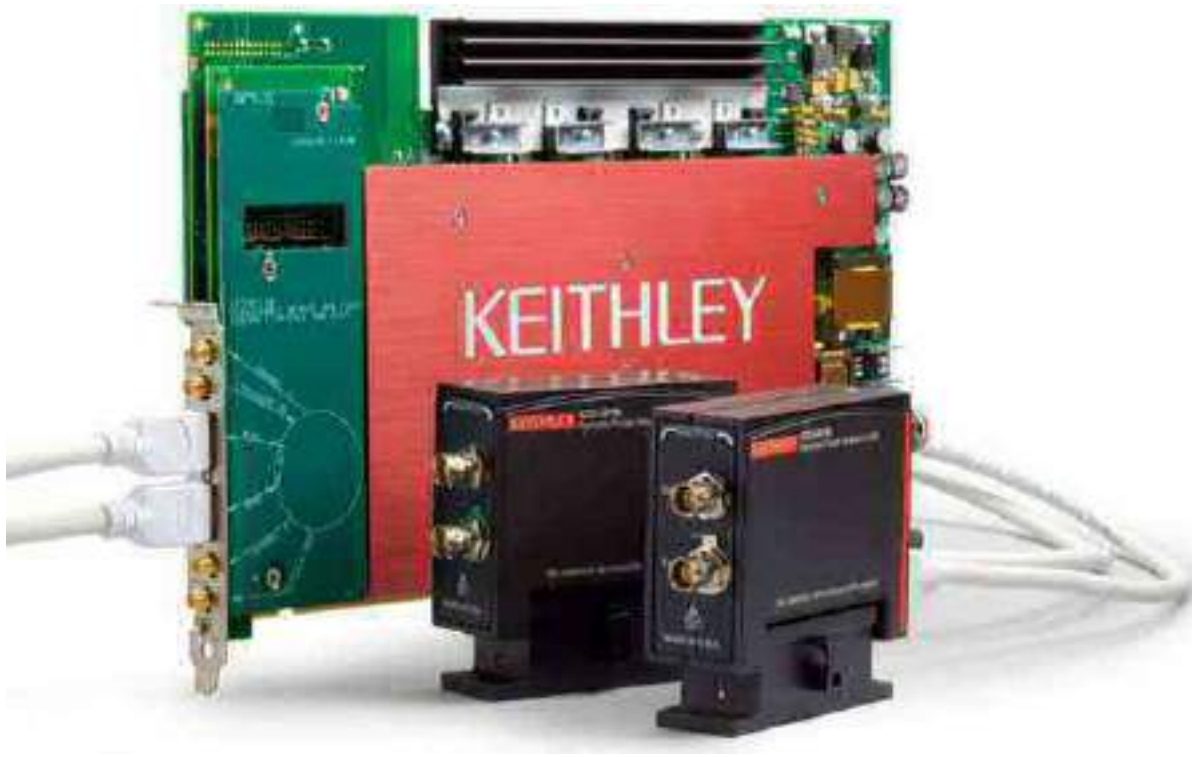

Source: Keithley

(a) (b)

***Figure 2.1:Keithley K4200 and Cascade Microtech Summit 11K (a); Keithley 4225 PMU extension card with two Remote Pulse and Measure Units(b).***

The base model “4200-SCS” is equipped with four standard high-resolution DC Sources and Measure Units (SMUs). These units allow delivering and reading currents on the order from 100fA to 100mA and voltages from a few µV to 20V [4]. However, for the purpose of data retention measurement on programmed cells, the slow operation and long measurement time of these units do not allow an accurate study of the resistance evolution in the initial time of some tens of milliseconds after programming. For these reasons, AMSIMEC`s K4200 has been equipped with an extension card denoted as “4225-PMU”, which provides two Remote Pulse and Measure Units (RPMU) (Figure 2.*b*). Those units combine and extend the functionality of a standalone pulse generator and oscilloscope in one box which is put into close contact (10 cm) with the Device

Under Test (DUT) and is directly connected to the main unit (K4200 plus 4225-PMU) by a proprietary high-speed digital link. In addition, the RPM unit allows a drastic enhancement of the signal to noise ratio in the acquired measurement results, thanks to the digitalization performed directly at the DUT side. Finally, it offers much better current sourcing and measurement characteristics, thus allowing better resolution and timing when compared with the standard SMUs.

**TYPICAL MINIMUM TIMING PARAMETERS FOR CURRENT MEASUREMENT [1]**

4225-RPM REMOTE AMPLIFIER/SWITCH (RPM optional to 4225-PMU)

| | 10V Range | | | | | |
|---|---|---|---|---|---|---|
| Current Measure Ranges | 100 nA | 1 µA | 10 µA | 100 µA | 1 mA | 10 mA |
| Recommended Minimum Pulse Width [2] | 134 µs | 20.4 µs | 8.36 µs | 1.04 µs | 370 ns | 160 ns |
| Recommended Minimum Measure Window [2] | 10 µs | 1.64 µs | 1 µs | 130 ns | 40 ns | 20 ns |
| Recommended Minimum Transition Time [3] | 1 µs | 360 ns | 360 ns | 40 ns | 30 ns | 20 ns |
| Noise [4] | 200 pA | 2 nA | 5 nA | 50 nA | 300 nA | 1.5 µA |
| Settling Time [5] | 100 µs | 15 µs | 6 µs | 750 ns | 250 ns | 100 ns |

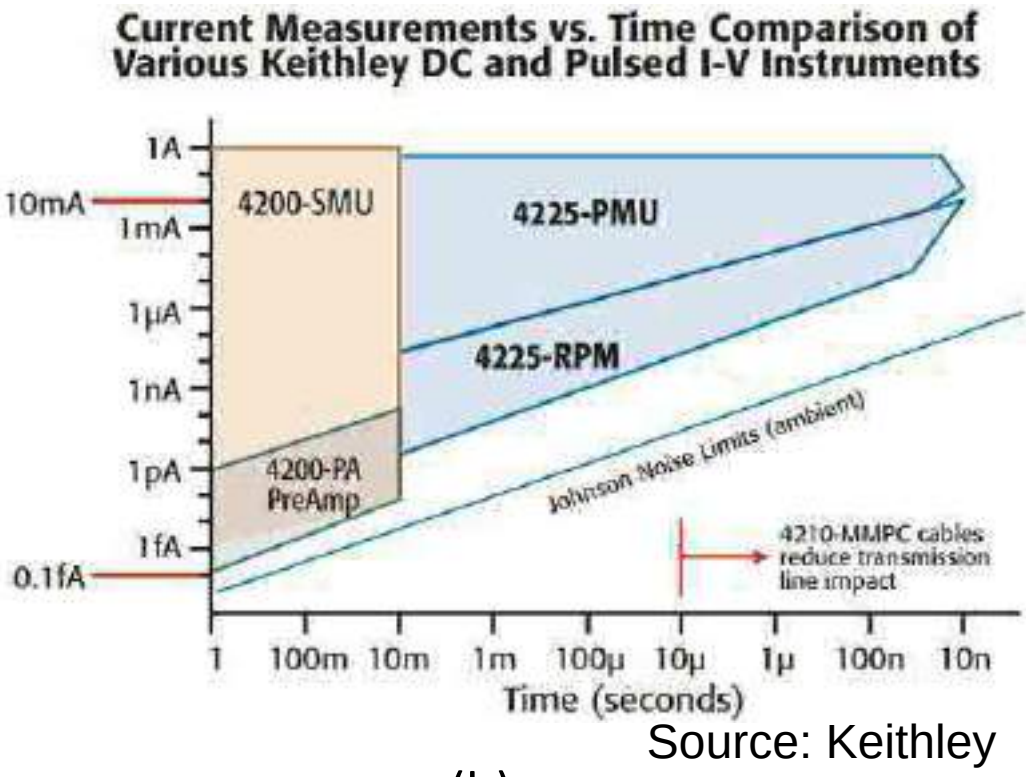

Source: Keithley

(a) (b)

***Figure 2.2: Datasheet of the 4225 RPM extension module(a) and graphical representation of Resolution-vs-Measurement Time trade off (b) [5].***

Figure 2.2b shows the difference in the achievable Current Resolution versus Measurement Time performance between the standard SMUs and the PMU card + RPM extensions. The RPMU is able to measure 10 µA currents needing only few microseconds of settling time.

## 2.2 Test Structure and Experimental Set-up

Conventional characterization strategies rely on the use of the so called "1R" (one resistor) or "1T1R" (one transistor, one resistor) structures (Figure 2.3*a* and Figure 2.3*b*), where the current limiting action during SET programming is controlled through an external load resistor and/or directly by an SPA, or by an integrated MOS transistor. While the external control suffers from capacitance effects due to the required connections between the devise and the experimental set-up, the use of 1T1R structure is in particular beneficial for RRAM cell performance and characterization because it allows for flexible current limitation. However, in order to study the intrinsic Resistive Memory Element (RME) switching properties, the actual voltage across the RME must be recalculated accounting for the voltage drop across the MOS transistor. While for the RESET-to-SET transition the programmed resistance is well controlled by trimming the current flow with the MOS transistor, in the case of the SET-to-RESET transition, the cell switching depends upon the effective voltage across the RME, which in 1T1R structures is strongly affected by the transistor non linearity.

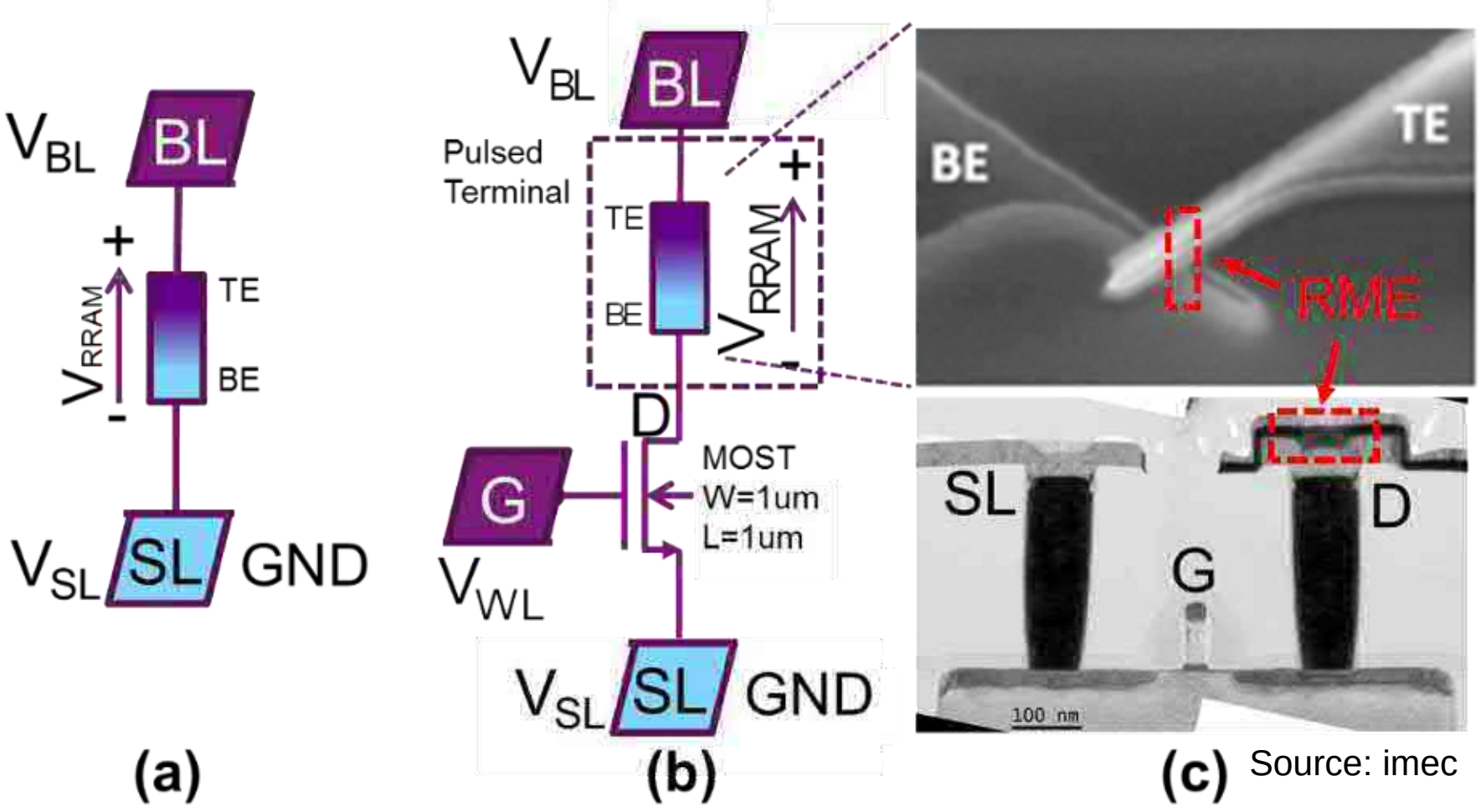

***Figure 2.3: Typical implemented 1R (a) and 1T1R (b) test structures. Top view of the crossbar structure of the RME and lateral view of the 1T1R (c). TE = top electrode; BE = bottom electrode; D = drain; G = gate; BL = Bit Line; SL = Source Line.***

As mentioned in the above the 1T1R configuration is convenient for our measurements since the current through the RME is controlled by simply adjusting the gate voltage of the series transistor with one SMU or RPMU of the K4200.

Measurements were made on wafers (Figure 2.4*a*) by using microprobes. The DUT is a three-pin structure (realized with a four-pad layout as shown in Figure 2.4*b*) that allows us to Form (initial breakdown of the RRAM oxide and formation of a conductive filament between TE and BE), Program (SET and RESET) and Read the RRAM cell. The electrical connections between the 1T1R cell and the main unit, which are depicted in Figure 2.5, are the following:

- **BL:** The Bit Line pad, or the “Drain” pad, is the pad that contacts the Top Electrode (TE) of the RRAM cell. This pin is connected to the RPMU#1 and can be driven in Pulse Mode by the PMU or in DC mode by using the SMU#1.
- **WL:** The Word Line pad, or the “Gate” pad, is the pad that contacts the gate of the MOS. This pin is connected to the RPMU#2 and can be driven in Pulse Mode by the PMU or in DC mode by using the SMU#2.
- **SL:** The Source Line pad, or “Source” pad, is the pad that contacts the source terminal of the MOS transistor. This pin is connected directly to ground through the Ground Unit (GNDU, which is an extra card found in the K4200, exclusively used as GND source).

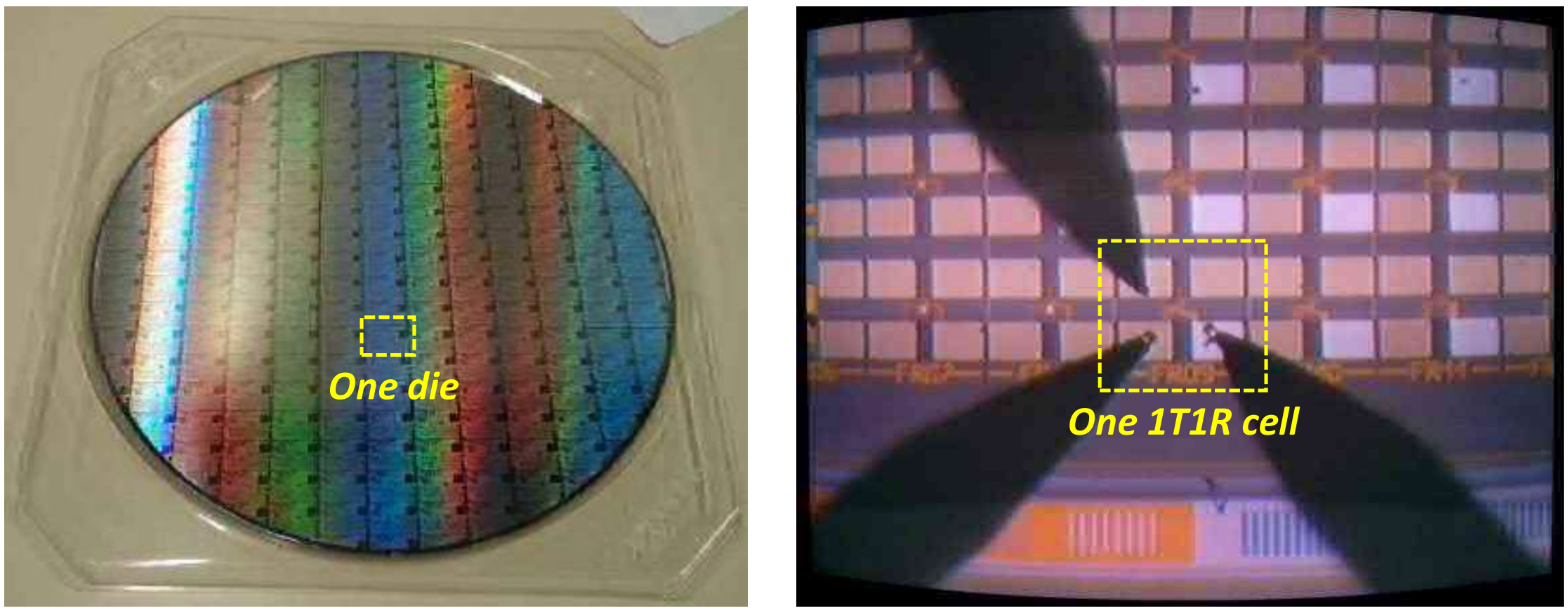

(a) (b)

***Figure 2.4: Photograph of one of the tested 300 mm wafers (a) and microphotograph of one of the RRAM cells connected to the SPA through microprobes (b).***

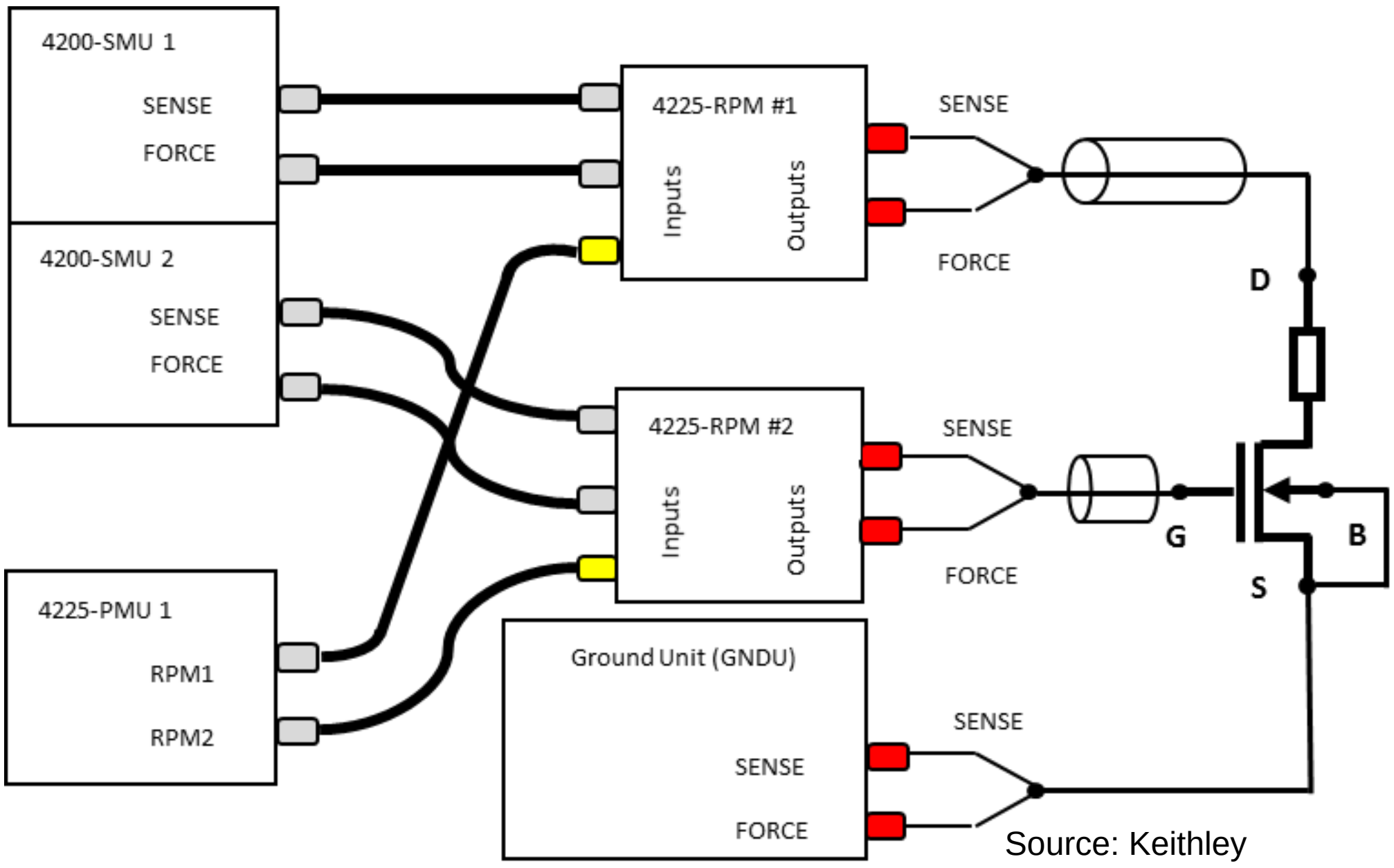

***Figure 2.5: Scheme of the electrical connections between the K4200 SMUs, the 4225 RPMUs, and the DUT. The FORCE and SENSE are separated connections commonly used in SPA to deliver significant current at an accurate voltage, separating the line delivering the "power" (FORCE) from the one carrying the "information" (SENSE)***

## 2.3 Retention Measurements

The main characteristic that distinguishes Volatile memories from the Non-Volatile memories is data retention. Data retention is the ability of a memory device to maintain information over time without the need for an external continuous power source. Some memories are used in applications where shorter retention is needed, as in the case of M-type Storage Class Memories (SCMs) requiring just few days of retention. Other memories are used in applications that require many years of retention, as in the case of S-type SCMs. A good characterization and optimization of retention is therefore mandatory when developing a new Non Volatile Memory.

As shown in Figure 2.6, which is based on tests from a presentation at IMEC [6], a considerable shift in the distributions can already be noticed 100 ms after the verified programming, which indicates unstable programming and state relaxation (see Chapter 3 for details).

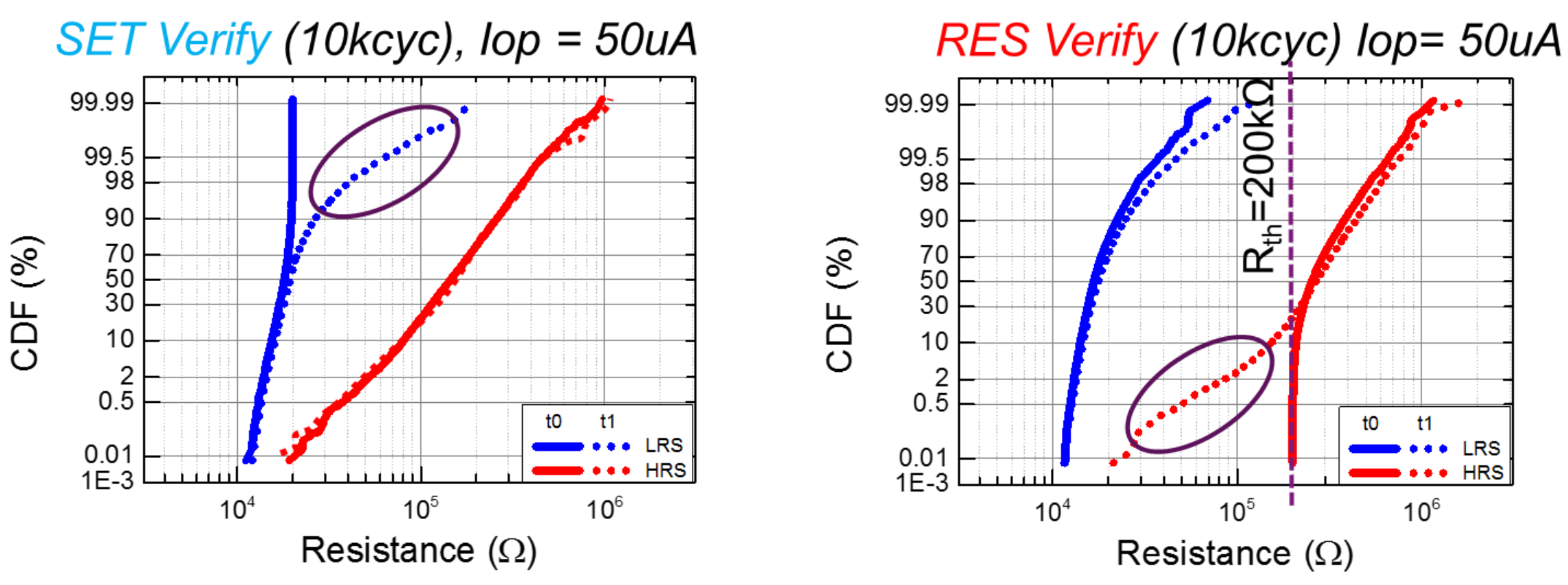

***Figure 2.6: Resistance drift (highlighted by a circle) in SET (left) and RESET(right) verified distributions after 100ms. This data was collected by switching 10K times a single cell with a programming current of 50 uA. The initial full lines are the distribution of the verified states (SET verified if R<20KΩ and RESET verified if R>200KΩ). The dotted lines are the distributions sampled after 100 ms [6].***

Data shown in Figure 2.6 were collected by using a custom UTM. This module was intended and designed mainly to perform endurance tests on RRAM cells. An example of endurance measurements obtained by using this custom UTM is shown in Figure 2.7. Testing the endurance of a memory cell means performing repetitive SET/RESET cycles until a breakdown occurs in the cell or until the resistive window (i.e. the spacing between the median values of the distributions corresponding to the SET and the RESET state) becomes too small. The module was therefore optimized to achieve the highest possible programming and sampling speed in order to perform the maximum number of cycles (either verified or unverified) in minimum time. Those test requirements are very specific for endurance measurements and are in contrast with retention characterizations, thus making the "old" UTM unsuited to the desired measurements.

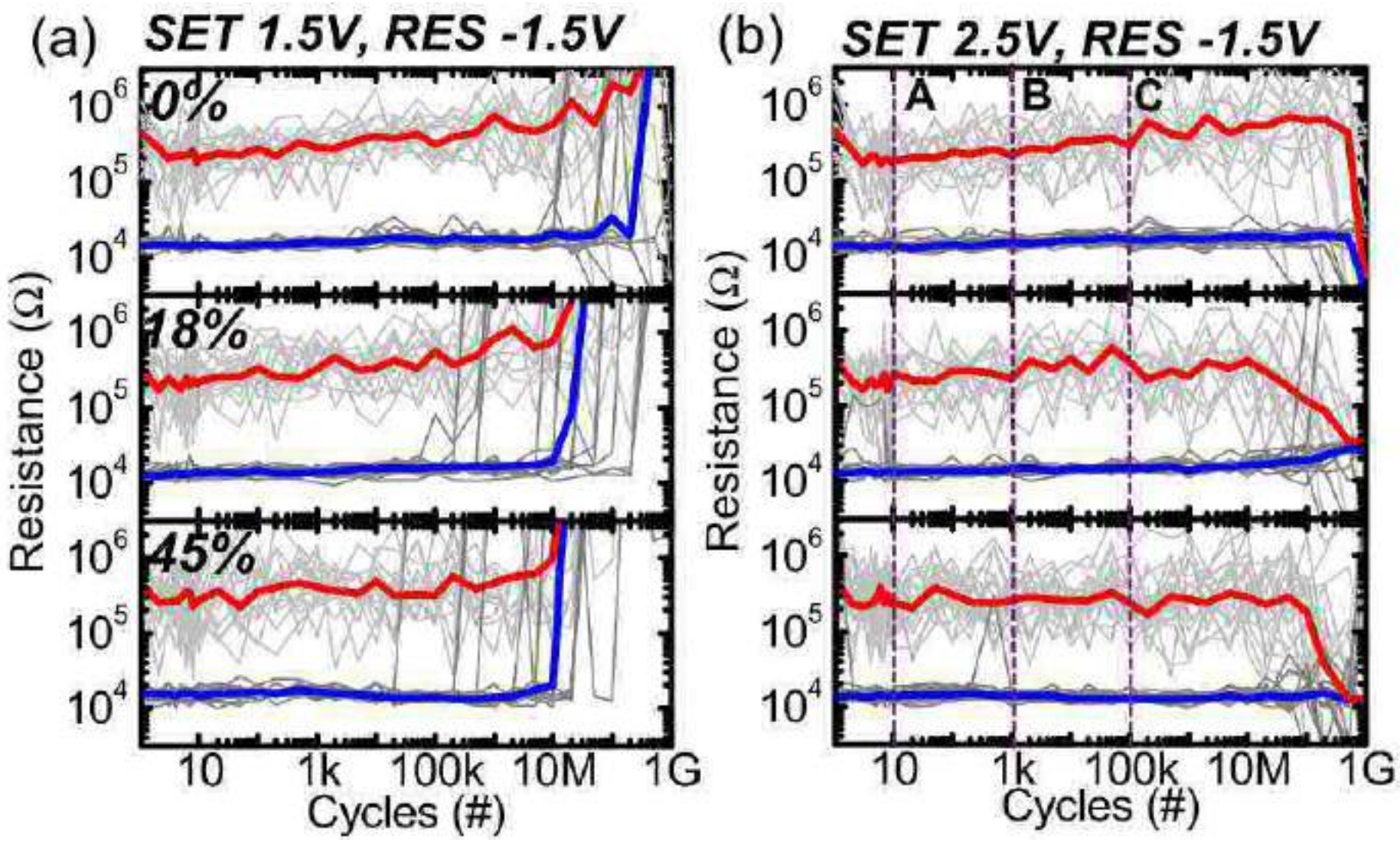

***Figure 2.7: Unverified endurance test performed over a population of 18 cells for different material compositions ($Hf_{1-x}Al_xO_y$ with x=0%, 18%, and 45%) and for different SET voltages [7]. In particular it was observed that the # of cycles before breakdown decreases with increasing x and is one decade less when using lower SET voltage.***

The main drawbacks of the above UTM for endurance testing can can be synthetized in three points:

- lack of flexibility in terms of the number of measurement pulses (only two) and resulting impossibility to track the resistance evolution over time of the programmed state;
- impractical and stringent programming sequence definition with subsequent rigidity in changing from one algorithm to another;
- unformatted output data that require some post processing before being easily imported and treated with statistical software.

To enable an easy characterization of data retention and, at the same time, overcome the above drawbacks, a new UTM module for KITE had therefore to be developed.

## 2.4 Specific Goals of the Thesis

At this point, the specific goals of this thesis work are clearer, and can be listed as follows:

- Extensive study of resistance relaxation in verified distributions, following previous IMEC`s knowledge on RRAM technology and continuing the previously conducted research [6];
- Characterization of program instability in Single-Pulse unverified distributions as a function of electrical programming parameters, temperature, and material stack;
- Development of a new UTM module for collecting the necessary data.

In order to characterize data retention, several Test Modules have been executed on the K4200 for collecting the data presented in this thesis. Both ITMs and UTMs have been joined and run in sequence in the KITE environment. The main modules are presented in the following, differentiating the AC (pulsed) measurements from DC measurements.

## 2.5 User Test Module for Retention Measurements

The specifications that needed to be implemented in the new software are summarized in the following points.

1. Arbitrary and Dynamic measurements:
    a. *Pulses of any duration and amplitude,*
    b. *Voltage amplitude or Time sweep of any pulse,*
    c. *Changeable number of measurement pulses per sequence,*
    d. *Custom sequence of pulses for each programmed state;*
2. Easy measurement set-up by loading input data from external text files;
3. Output well formatted and ready to be imported in software like JMP (pronounced *"jump"*, is a statistical software focused on exploratory analytics and visualization, developed by SAS (Statistical Analysis System) [8]) .

All the above features were successfully implemented in a new UTM module. This UTM module was effectively used to retrieve all the data presented in this thesis and is also used at present by other IMEC colleagues for their retention measurements.

The easiest way to describe the operation of the module takes into account the two steps needed to start a new measurement, namely:

1. *Arbitrary Pulse definition*: the user defines, in a standalone library file, the voltages and timing of all RESET, SET, delay, and sampling pulses that will be used in the measurement (see paragraph $2.5.1);
2. *Arbitrary Sequence and Test definition:* along with the definition total number of SET/RESET cycles to be performed for the measurement, the user defines the sequence/order in which the previously defined pulses will be applied. (see paragraph $2.5.2);

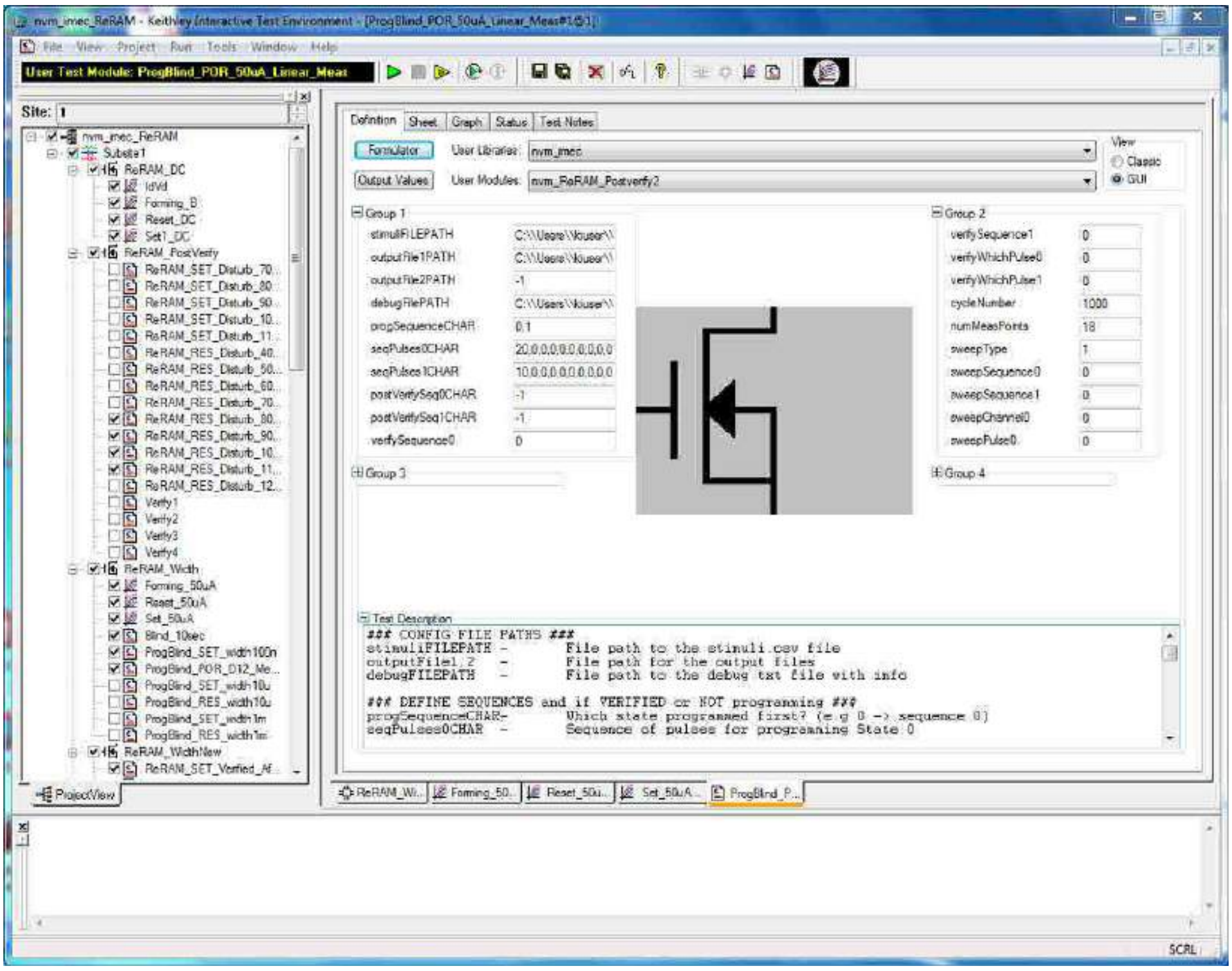

*Figure 2.8: New UTM module GUI as seen on the K4200 when loaded on KITE.*

### *2.5.1 Arbitrary Pulses*

The first step to be performed to set up a new measurement is the definition of a custom library of pulses. The new UTM has been programmed in order to accept, as an input, a CSV file (A Comma-Separated Value file stores tabular data and consists of a number of characters separated by commas) containing all the pulse definitions for a specific material or operating temperature.

Figure 2.8 shows how a Single-Pulse is simply defined by five time parameters, one for each of the five segments, plus a sixth parameter for the pulse amplitude. Each segment can vary from 20 ns to 1 s (these bounds are a constraint of the K4200) and the pulse amplitude can have positive, negative or null value (for example, a delay pulse can be implemented just by setting the amplitude to zero).

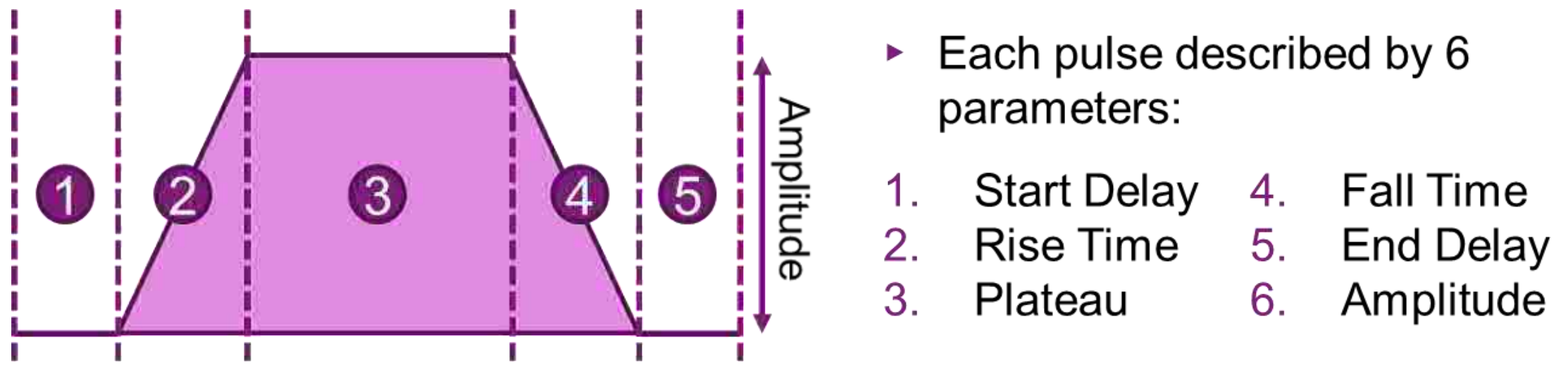

***Figure 2.9: Single-Pulse fully defined with six parameters.***

By combining different values, the user can create his/her library of pulses. For example, as shown in Figure 2.10, each RESET, SET, delay, and measurement pulse can be separately defined and addressed with a unique ID# inside the CSV library. Using the same ID# numbering along different libraries, allows a better code reusability and speed up of the measurement setup.

| ID # | TYPE of PULSE |
|---|---|
| [ 0 – 9 ] | Measure |
| [10 – 19] | SET |
| [20 – 29] | RESET |
| [30 – 39] | Disturbs |
| [80 – 89] | Log Delay |
| [90 – 99] | Linear Delay |

| ID# | Pulse Definition |
|---|---|
| 0 | MEAS |
| 10 | SET |
| 20 | RES |
| 21 | RES |
| 80 | DELAY |

***Figure 2.10: Example of library organization (left) and example of a collection of pulses with different shapes and voltage levels (right). A library is made of an arbitrary number of pulses, such as Measuring, SET, RESET, Disturbs (pulses of lower amplitude used to simulate a disturb spur during programming, and Delays (logarithmically-spaced or linearly-spaced)***

In order to help the user in building a new library and avoid any possible mistake due to a wrongly typed value, a little script with a user-friendly graphical interface was generated. The script runs in Matlab environment and allows the user to insert the various values for pulse description. The user can observe waveforms plots in a “live manner” and can save the pulse results in his/her personal pulses library.

The graphical user interface of this script is shown in Figure 2.11.

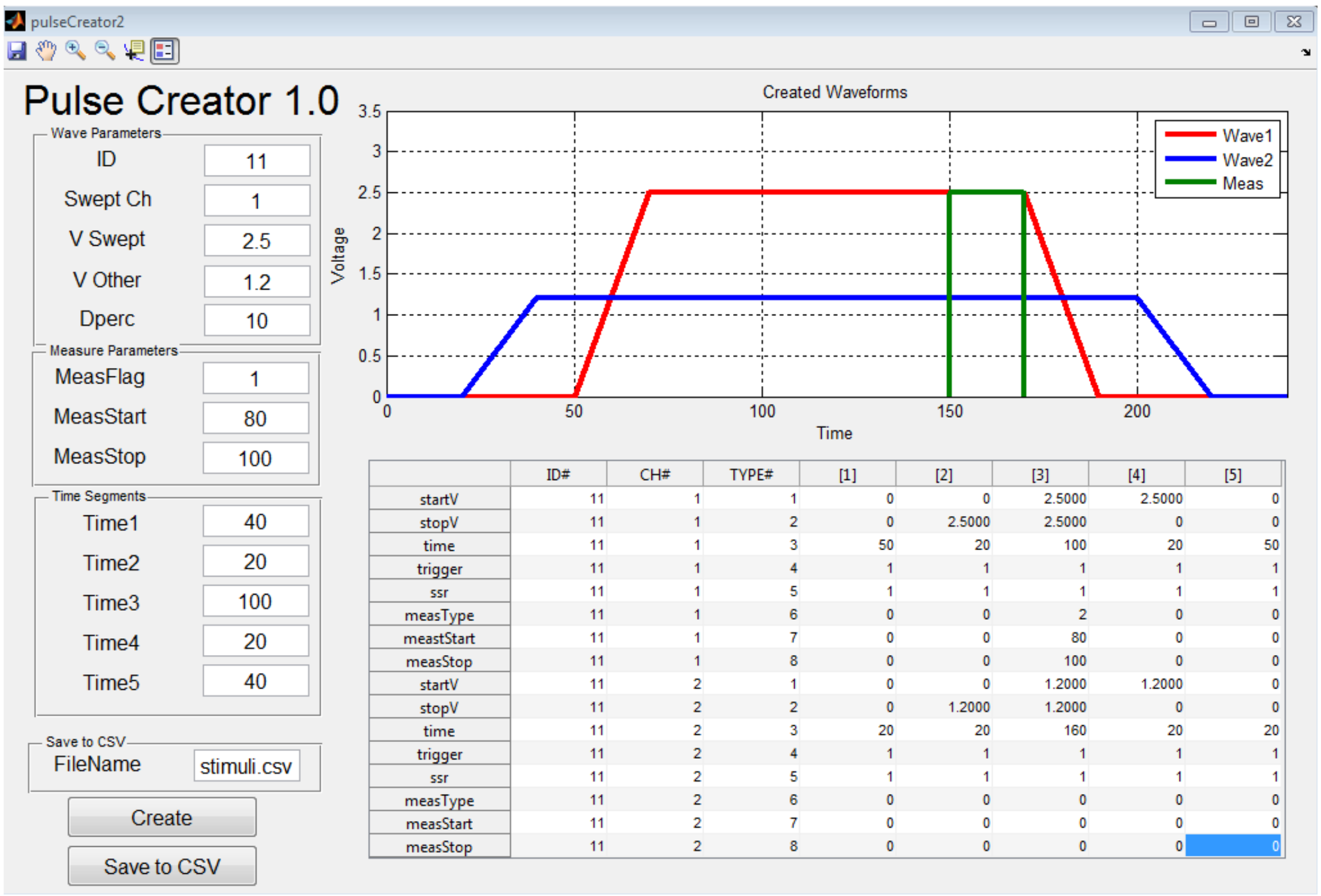

***Figure 2.11: Graphical user interface of the pulse with application running in Matlab environment.***

The flexible definition of pulses and the possibility of having different libraries for different materials are the main advantages of this new module. When setting up a new algorithm, the user does not need any more to redefine parameters in the source code of the module, but needs to refer to the pulses in the previously generated library, or generate, if needed, a new custom pulse by using the Matlab GUI and rapidly proceed with step 2 of the measurement definition (Arbitrary Sequence definition).

### *2.5.2 Arbitrary Sequence*

The second step needed to configure a new measurement is the setup of the programming sequences. Another main feature of the developed software is the possibility to input arbitrary sequences with minimum user`s effort. As already mentioned, the old software was very efficient for endurance measurements, but its rigid structure did not allow performing variations with respect to the standard pulse sequence. The newly designed UTM instead provides a fully flexible way for changing the pulse sequence and implementing a new programming algorithm. After loading a library generated during step 1, the user can insert any type of pulse combination by using ID# as a referral. Figure 2.12 shows an example of two pulse sequences for RESET and SET programming, respectively, with three measurements performed at different time after programing (using delay pulses as spacers between "Meas" pulses).

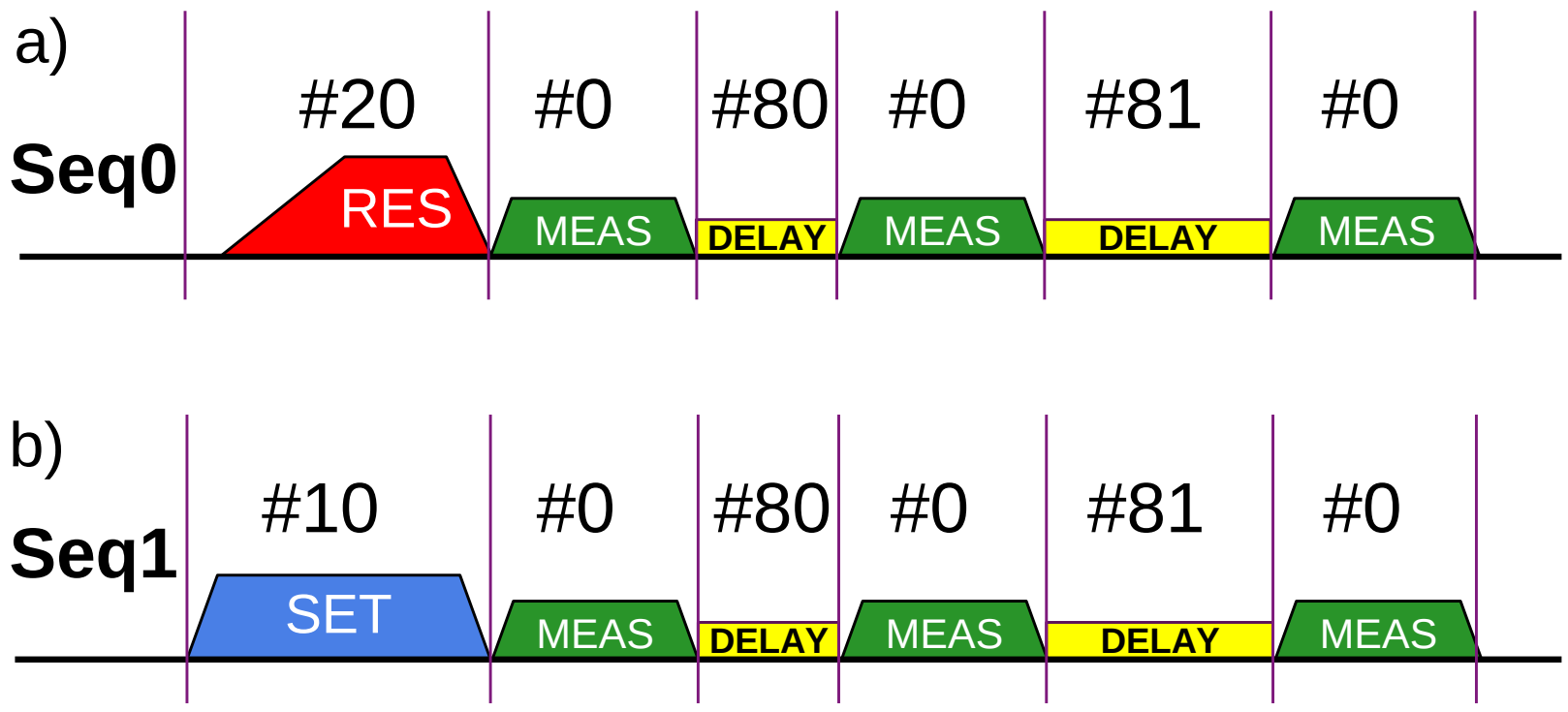

***Figure 2.12: Example of sequences generated from library in Figure 2.10, for RESET (a) and SET (b) programming. Resistance monitoring is here performed with 3 delayed measurements.***

The sequence can also be dynamically modified while the measurement is ongoing by sweeping one of the six parameters (Time or Amplitude) of a pulse. This sweep declaration is defined during the second step of the measurement setup by inserting the *sweepStart*, *sweepStep,* and *sweepStop* values. The graphical representation of two examples, namely a time sweep (a) and a voltage sweep (b) is shown in Figure 2.13.

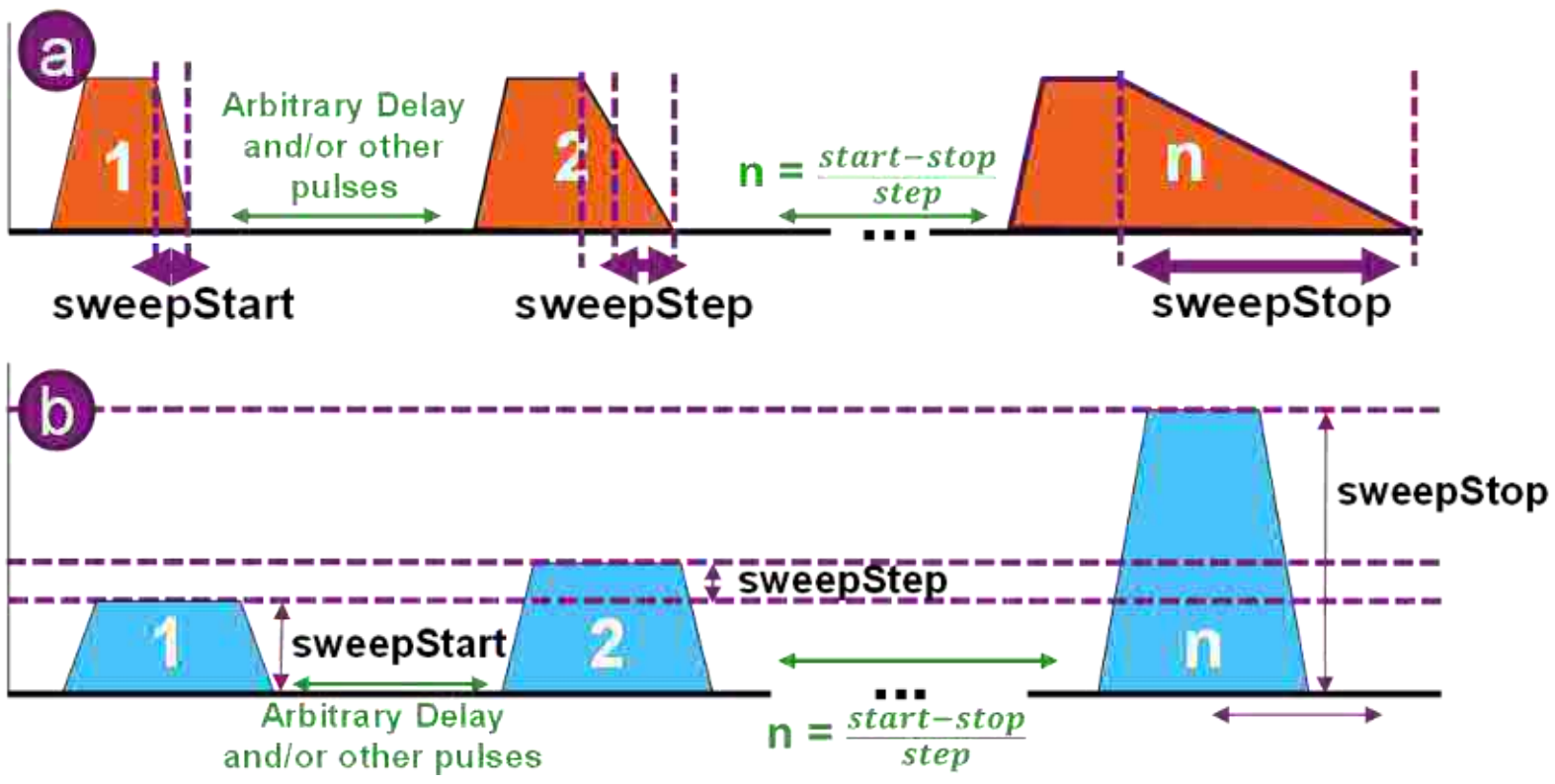

***Figure 2.13: Examples of a time sweep, with an increasing fall time (a) and an increasing amplitude (b).***

### 2.5.3 Algorithm Examples

The new test module has been employed to perform the following two basic types of measurements for retention characterization.

- *Single-Pulse Unverified Programming*, which is an algorithm used to assess program stability by analyzing unverified distributions. The filament is simply switched at each cycle with a fixed amplitude and time. An example of Single-Pulse (SP) programming sequence is shown in Figure 2.14.
- *Verified Programming*, which is an algorithm used to evaluate program stability by analyzing verified distributions. A sequence of programming pulses is applied to the cell, each followed by a read (verify) step. A read current threshold is defined and programming pulses are actually fed by the RPM until the threshold is reached. This programming algorithm is one of the most commonly used for non-volatile semiconductor memories. An example is shown in

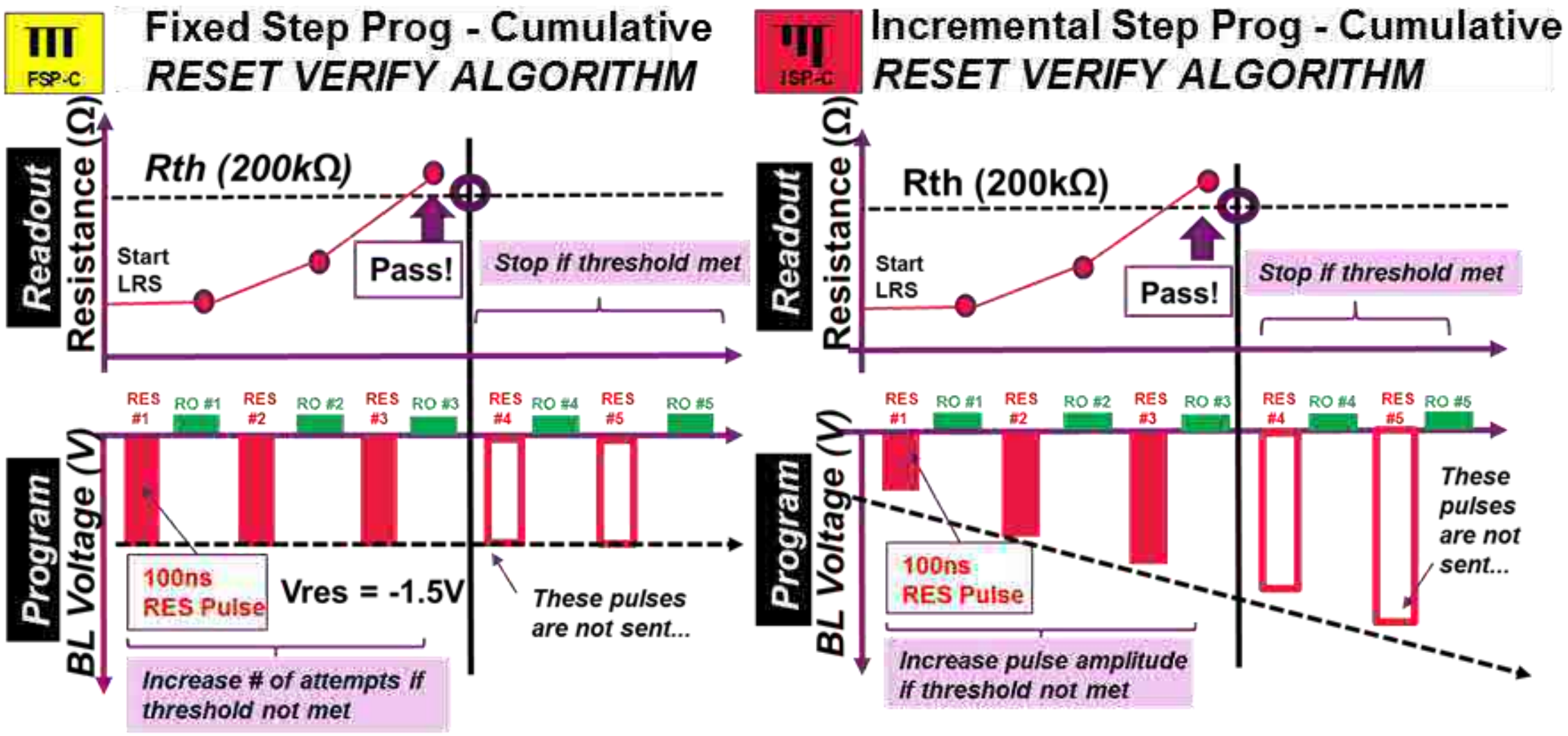

- Figure 2.15, where the sequence of resistance-tracking measurement pulses is omitted for simplicity.

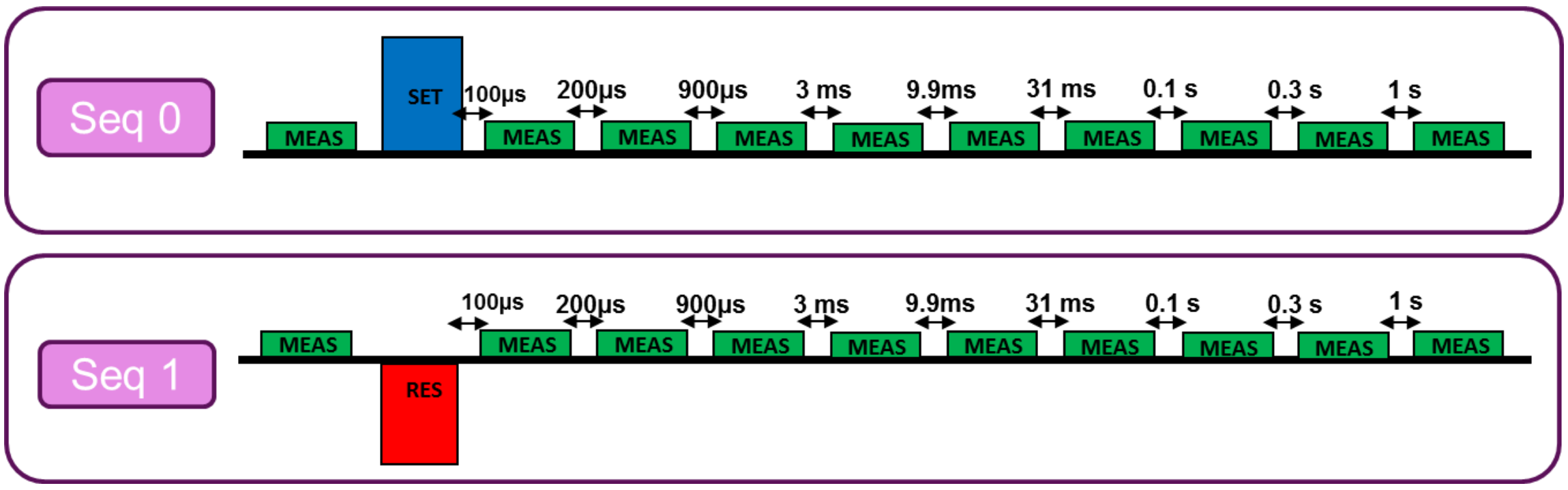

*Figure 2.14: Example of a Single-Pulse unverified programming sequence for retention measurement: after a first test measurement (whose purpose is to check the initial state of the cell), a Single-Pulse*

***programming is performed and, then, the resistance evolution over time is tracked by means of 9 log-spaced measurement pulses.***

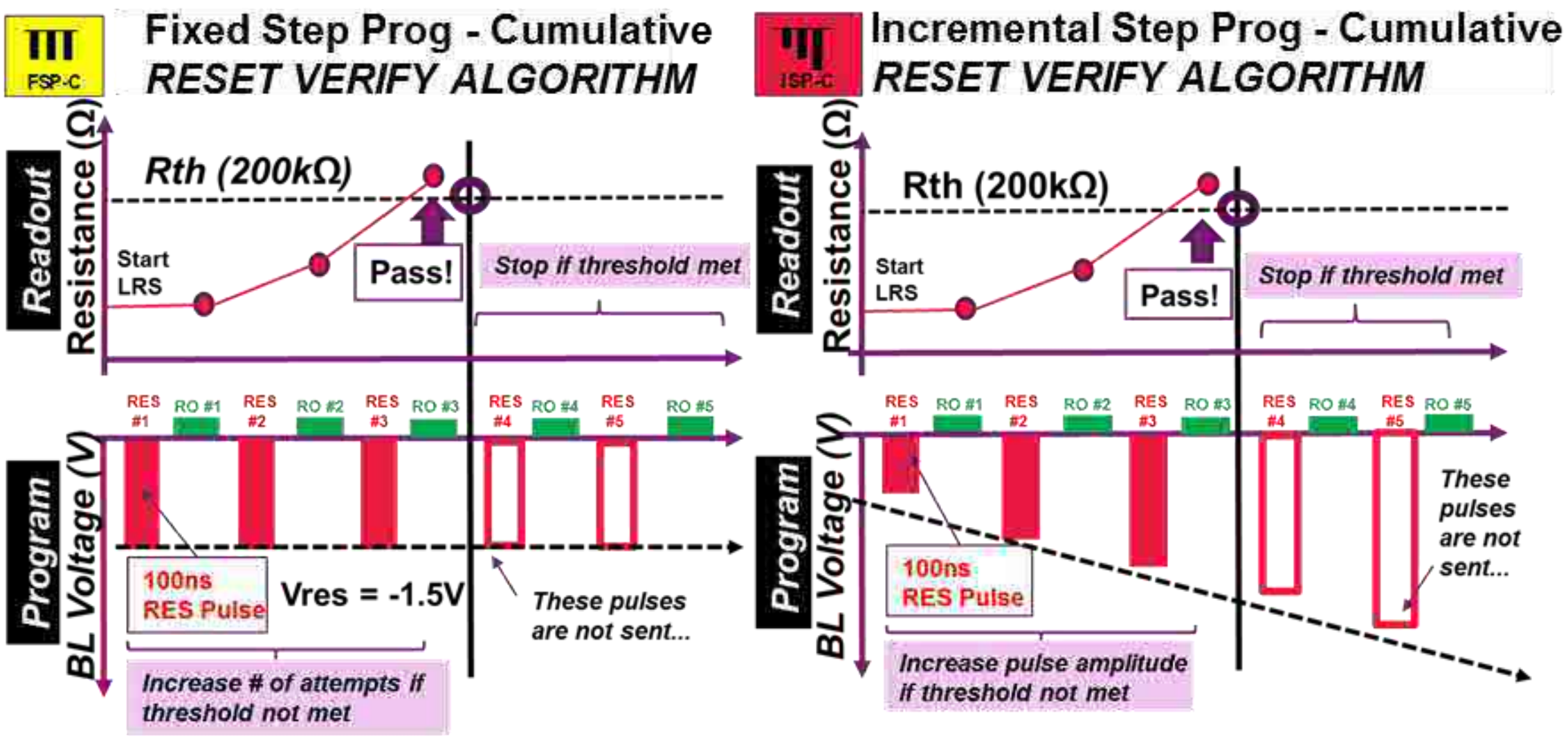

***Figure 2.15: Schemes of the two Verified Programming Algorithms that have been tested on IMEC`s RRAM. In the Fixed Single-Pulse Cumulative algorithm (left), the same amplitude is used for all RESET pulses; in the Incremental Single-Pulse Cumulative algorithm (right), an increasing amplitude is used for successive RESET pulses until the threshold value is reached.***

## 2.6 Interactive Test Modules for DC Measurements

Along with the custom UTM used for pulsed programming, several standard Interactive Test Modules (pre-loaded modules in KITE) were also used to perform DC characterization of the devices. In particular, DC modules were extensively used to perform the forming of the filament and to check the switching capabilities of a cell before continuing with the pulsed retention measurements. The main ITM modules are described below.

**Forming:** This ITM module uses the SMUs to perform a sweep on the BL while keeping the transistor gate voltage fixed for setting the current compliance. The BL voltage is swept from 0 V to 3 V while the WL voltage is kept at the voltage value (that depends on the size of the transistor of the measured 1T1R cell and on the different operating temperature) that keeps the transistor in saturation at a compliance of 50 µA. As shown in Figure 2.16*a*, the current that passes through the Resistive Memory Element remains low (ideally zero) until a voltage $V_{forming}$ is reached, after which the current increases up to the compliance value. This abrupt change is called “forming”, whereby a soft breakdown of the oxide occurs and a central filament of oxygen vacancies is produced. After the forming step, the cell is ready to be switched.

**Reset**: This ITM module uses the SMUs to perform a negative sweep on the BL while keeping the WL as high as 1.5V, to keep the transistor on with a sufficiently low on-resistance during the transition. During the sweep, the resistive element experiences the transition from the low- to the high- resistance state as shown on the left side of the Figure 2.16*b*. After voltage $V_{reset}$ is reached, the oxygen vacancies are moved by the negative electric field, thus thinning the filament and increasing its resistance.

**Set:** Similarly as in the case of the forming ITM, during a set transition, the WL is used to set the current compliance at which the set will occur. A sweep is performed on the BL and, after $V_{set}$ is reached, as shown on the right side of Figure 2.16*b*, an abrupt increase in the current through the filament indicates the transition from the high- to the low-resistance state.

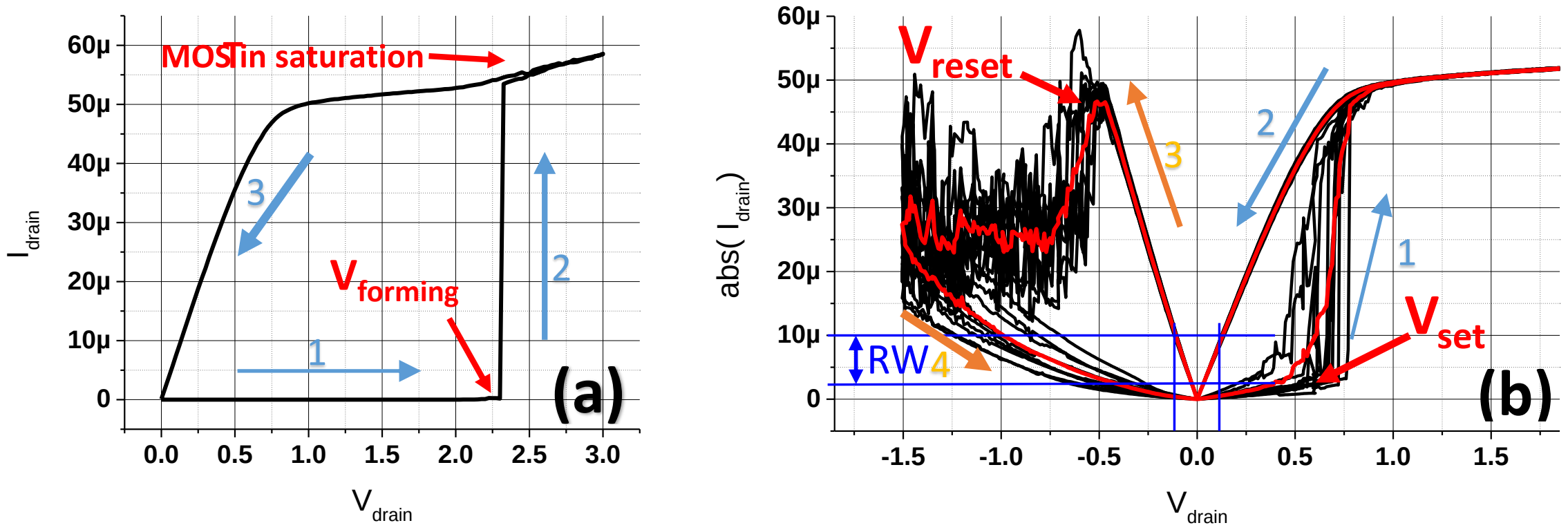

***Figure 2.16: Forming (a) and Set/Reset Cycling (b) measured on a HfO2 based RRAM cell by using DC measurement ITMs. Forming following the blue arrows: initially no current flows through the structure, but as soon as $V_{forming}$ is reached, a soft breakdown occurs creating a conductive filament, which size is limited by the saturation of the series MOST. SET and RESET occur similarly, but because the filament is already formed, lower voltages of $V_{set}$=$V_{reset}$ are required. The red line is the median of the several black cycles.***

# Chapter 3
# Measurements and Results

This chapter presents the experiments conducted for the thesis project, focusing on the algorithms used and explaining the workflow of the benchmarking.

In the Point Of Reference sub-chapter, a collection of predefined starting conditions are described, in terms of reference material, reference programming pulses, and reference retention measuring sequence. Those starting conditions build-up the foundation for the analysis of Program Stability, whose metrics are defined in the second part of this Chapter. The measurements described in the third part of the Chapter include the benchmarking performed with the Unverified Single-Pulse Programming algorithm and the Program-and-Verify Step Programming algorithms.

## 3.1 Point of Reference Measurement

The new UTM allows performing a new type of measurement for data retention analysis. The Point Of Reference (POR) is a collection of predefined starting conditions in terms of reference material, reference programming pulses, and reference retention measuring sequence. In fact, during this thesis project, a number of experiments were executed, whereby each measurement presented a single modification of the Point Of Reference conditions. This was done in order to study the effect of each modified parameter upon retention.

The POR standard measurement was defined as an unverified Single-Pulse programming performed on $HfO_2$ based 40 nm x 40 nm RRAM cells. The cell structure is shown in Figure 3..

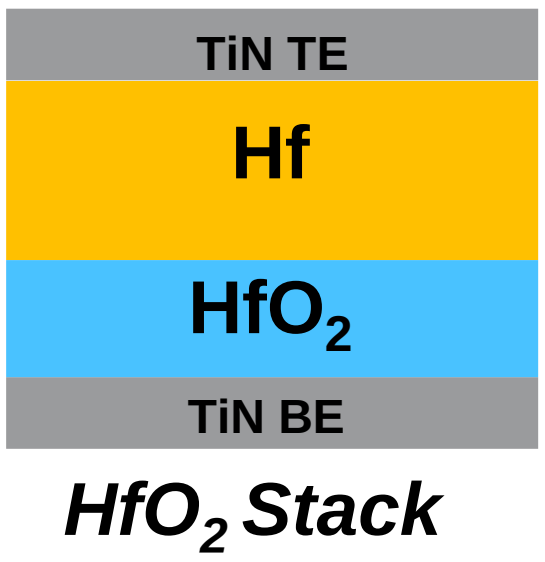

| | |
|---|---|
| **TE** | 35nm PVD TiN |
| **Capping** | 10nm PVD Hf |
| **Oxide** | 5nm ALD $HfO_2$ |
| **BE** | 35nm PVD TiN + CMP |

***Figure 3.1: $HfO_2$ based RRAM structure (composition and layer thicknesses). TE = top electrode; BE = bottom electrode. PVD = Physical Vapor Deposition; ALD = Atomic Layer Deposition; CMP = Chemical Mechanical Polishing.***

The time and voltage values of the reference Single-Pulse programming sequence for Forming, SET, RESET, and Readout, needed to obtain a resistive window of 10x at an operating current of 50 µA are given in Table 4.

| *Pulse* | *Tr/Tw/Tf* | *WL (gate)* | *BL (TE)* | *Source (BE)* |
|---|---|---|---|---|
| **Forming** | DC | Iop = 50uA (Vg = 1.15V) | 3.5 V | GND |
| **SET** | 20/100/20 ns | Iop = 50uA (Vg = 1.15V) | 2.5 V | GND |
| **RES** | 20/100/20 ns | Vg = 1.5V | -1.5 V | GND |
| **ReadOut** | 10/100/10 µs | Vg = 1.5V | 0.1 V | GND |

***Table 4: Time and voltage values used as reference programming conditions for Forming, SET, RESET, and Readout (See Figure 2.3 for details on electrical connections).***

By previous observations on imec`s RRAM devices, the resistance evolution over time was expected to follow a log(t) law. This time dependence has been demonstrated by performing measurements with different sampling rates, and the obtained results confirm that log-spaced samples can better track the resistance variation (see Appendix A for detailed measurements).

A reference readout scheme has therefore been formulated in terms of number of measurements and distance between two consecutive samples, as shown in Figure 3.2. A first readout, RD#0, is done to measure the resistance value prior to programming; then, a SET/RESET pulse is applied, followed by a sequence of nine log-spaced readout pulses. By using this reference readout scheme, the resistance evolution over four time decades can be monitored, starting 100 µs (reference readout) after the programming pulse and arriving until 1 s after this pulse (RD#9). The same monitoring scheme is repeated for RESET programming.

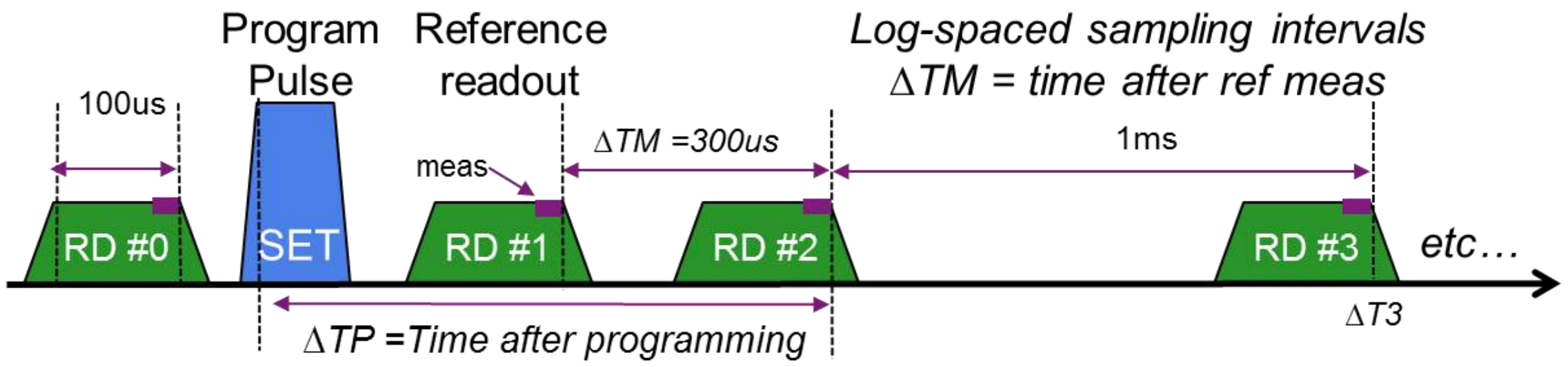

| #Readout | #0 | #1 | #2 | #3 | #4 | #5 | #6 | #7 | #8 | #9 |
|---|---|---|---|---|---|---|---|---|---|---|
| ΔTM [ms] | before | ref | 0.3 | 1 | 3 | 10 | 30 | 100 | 300 | 1000 |

***Figure 3.2: Reference readout scheme for monitoring resistance evolution over time with log-spaced readout pulses.***

Because of the intrinsic stochastic behavior of RRAM memory cells, in order to develop a robust confidence in data, an amount of statistically relevant samples must be collected. For imec`s RRAM and for the purposes of retention analysis, a number of 1000 switching cycles performed on at least 5 different cells is considered an adequate set of data points for good statistical confidence.

A reference readout scheme is also provided for the incremental Program-and-Verify algorithm. Table 5 lists the start, step, and stop values for the programming pulse amplitude, which is increased at each sub-cycle until the threshold value of the cell resistance is reached. In the case of a SET programming, the sweep is defined for the WL (gate) in order to have an increase of the operating current through the resistive element from ~10 µA to ~90 µA. On the contrary, for RESET programming, the sweep is performed on the BL (RRAM top electrode) in order to induce an increasing negative voltage from -1 V to -2 V.

| | Terminal | Start | Step | Stop | Threshold |
|---|---|---|---|---|---|
| **SET** | WL (gate) | 0.8 V | 25 mV | 1.3 V | R < 20 KΩ |
| **RES** | BL | -1.0 V | -0.1 V | -2.0 V | R > 200 KΩ |

***Table 5: Definition of voltage levels for verified programming.***

## 3.2 Analysis Metrics

The experimental plan involved gathering the required amount of statistical data for the development of a behavioral model.

For each experiment, a single modification of the POR (Point of Reference) conditions is considered. POR conditions can be found below.

Three metrics were defined as a starting point for the investigation of retention for each measurement.

### 3.2.1 Metric (A) – Global Behavior

Metric (A) involves the statistical comparison of a set of data considering the Cumulative probability Density Function (CDF), on the same plot (to appreciate the relative shifts) for each different time instant. The CDF is a useful way for representing how a particular quantity is distributed. CDF(x) represents the probability that a real-valued random variable R will be found to have a value less than or equal to x. In symbols, the CDF is defined as

$$\mathrm{CDF_R(x) = P(R \ \leq x)} \qquad \text{(Eq. 3.1)}$$

Referring, for instance, to the case of a distribution of resistances, the CDF represents the probability that a certain resistance value R takes on a value less than or equal to x. A particular value of the CDF is the Median, expressed as the value located in the middle of the probability distribution. Another value extracted from the CDF is the standard deviation, which gives a measure of the amount of variation or dispersion of a set of data values.

The global behavior of all CDFs of resistances sampled at different time instants is displayed, as shown in Figure 3.3*a* (RD1 at ref time, RD2 at ΔTM = 300 µs, RD3 at ΔTM = 1 ms etc.). Furthermore, the analysis with the Metric (A) continues by considering the evolution of the median value over time to appreciate the overall drift, as shown in Figure 3.3*b*. The standard deviation evolution (not shown in Figure 3.3*b*) is considered as well, in order to have a measure of how much the resistance distribution is spreading over time.

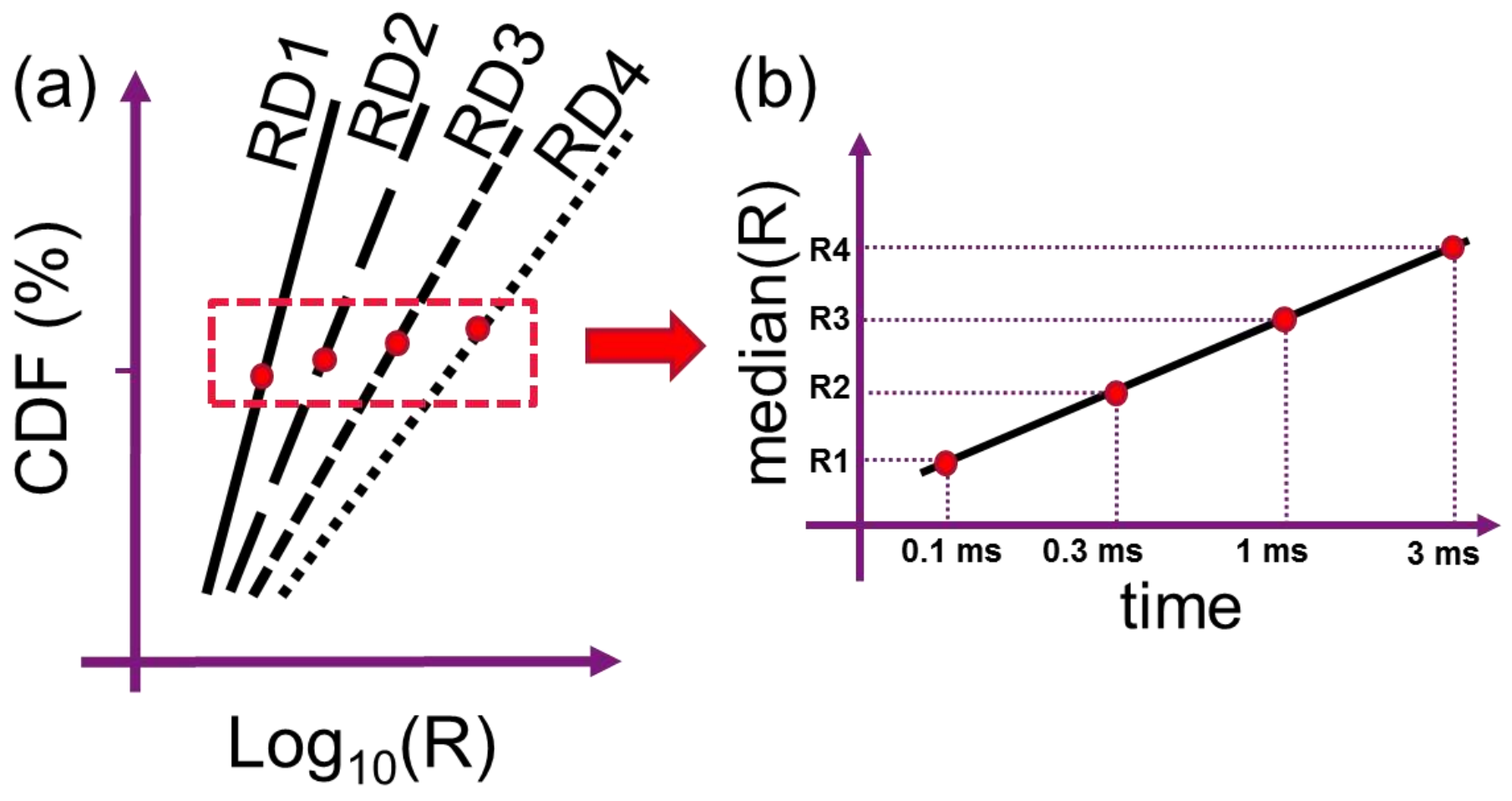

***Figure 3.3 - Metric A: Study of the Global Behavior by analyzing the CDFs at each measurement time (a) and checking the evolution trend of the Median values (b).***

### *3.2.2 Metric (B) – Local Behavior*

The analysis of the retention, after the study of the global behavior, continues by considering the local behavior. Metric (B) highlights the behavior of the tails, i.e. the top and the bottom data, which are the less probable to occur, so that their contribution is "shadowed" by the more probable data. Instead of considering all the data together, three subpopulations of the previously calculated CDFs are identified. As shown in Figure 3.4*a*, the three subpopulation have all the same amount of data (10% of the total) and are taken from the top, the middle and the bottom, corresponding to the data greater than $90^{th}$ percentile (top), around the Median (mid), and lower than $10^{th}$ percentile (bottom).

The evolution over time of the subpopulations is then studied on a separate graph, in order to analyze how different sets of data, with high, mid, or low initial resistance value behave in time, as shown in Figure 3.4*b*.

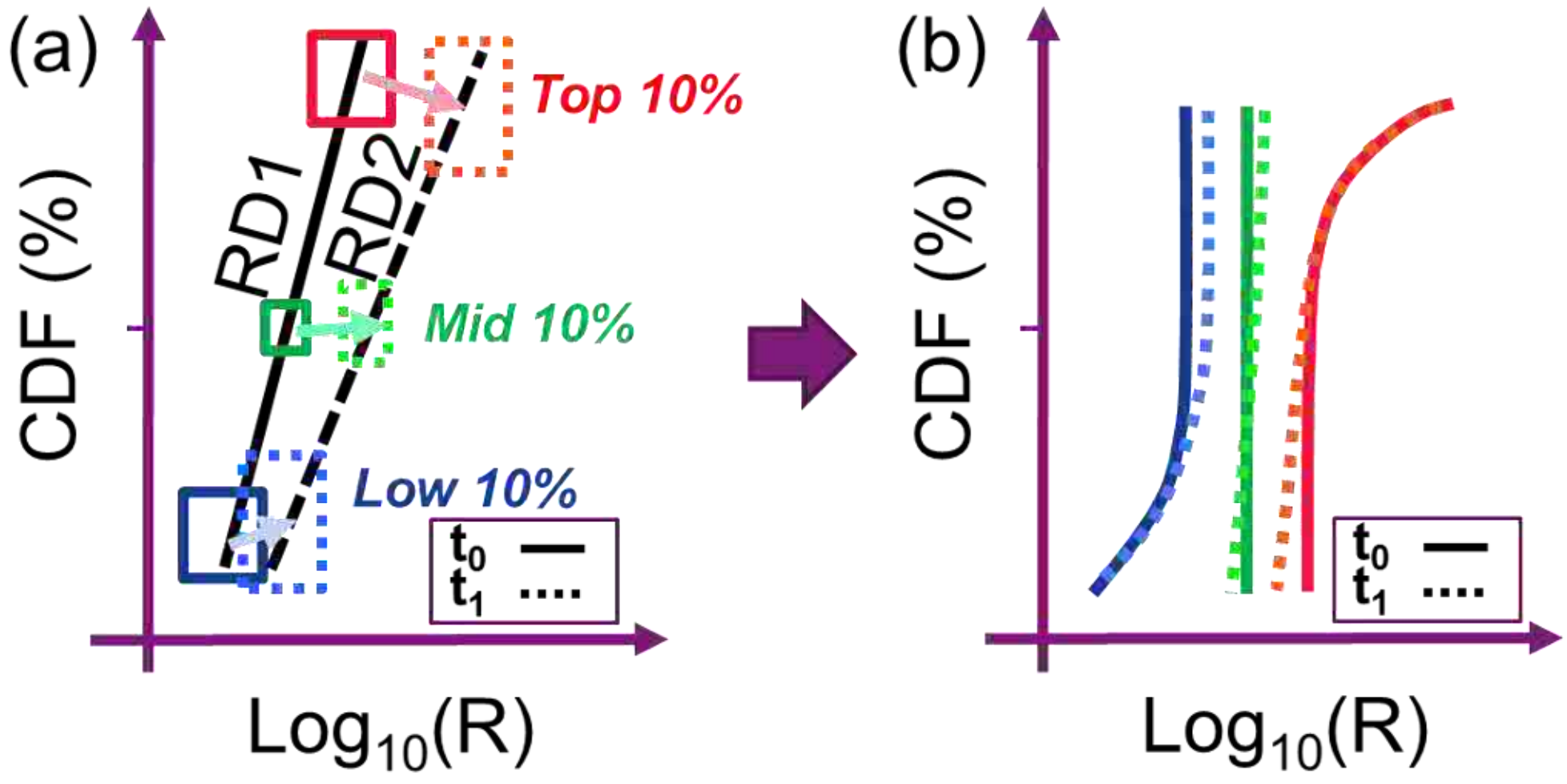

***Figure 3.4 - Metric B: Study of the Local Behavior by analyzing the CDFs of three sub-populations of the previously measured resistances. The CDF behavior of resistance values taken from the tails (both the Top (higher 10%) and the Low (lower 10%) tail) are compared with the median evolution (Mid (middle 10%)).***

### *3.2.3 Metric (C) – Correlation*

Metric C is used to synthetically describe the evolution over time of two resistance populations sampled at different moments. Observing the global CDFs and studying median and standard deviation only describes the global statistical properties of a physical process and, therefore, reveal little information on the real state stability. Another way to study state stability is by considering the data correlation between two read-outs.

A synthetic yet quantitative estimation on how memory is retained between readouts can be obtained by calculating the Pearson correlation coefficient *r* for two data vectors x,y :

$$r = r_{xy} = \frac{\sum_{i=1}^{n}(x_i - \bar{x})\,(y_i - \bar{y})}{\sqrt{\sum_{i=1}^{n}(x_i - \bar{x})^2}\,\sqrt{\sum_{i=1}^{n}(y_i - \bar{y})^2}} \qquad \text{(Eq. 3.2)}$$

where $\bar{x}$ and $\bar{y}$ are the distributions means, that are respectively subtracted from each value $x_i$ and $y_i$ of the two data vectors x and y.

If we consider two statistical populations X and Y, the same coefficient *r* can be synthetically and more intuitively expressed in terms of covariance and standard deviation (σ):

$$r = r_{xy} = \frac{cov(X,Y)}{\sigma_x \sigma_y} \qquad \text{(Eq. 3.3)}$$

A graphical representation of the different scatter plots for populations with decreasing correlation is given in Figure 3.5.

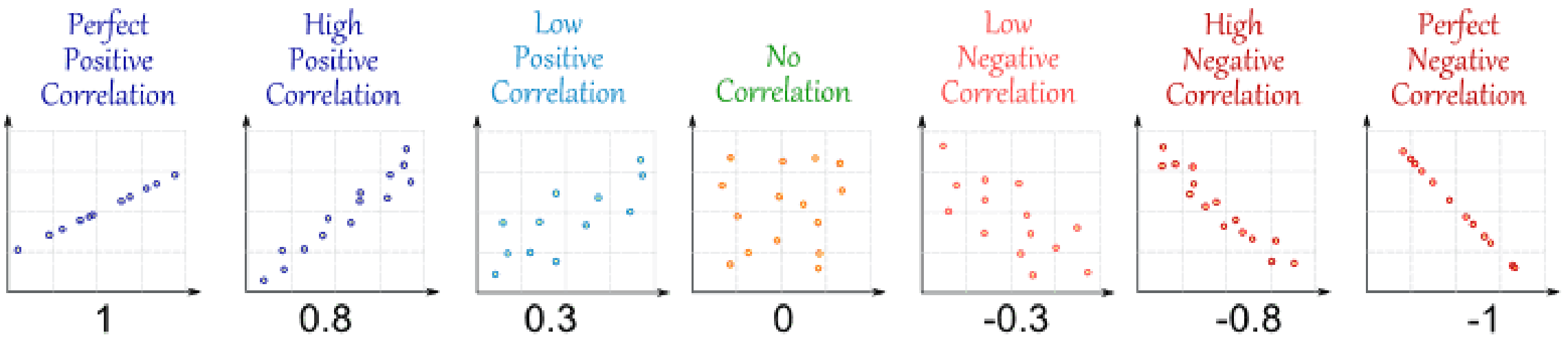

***Figure 3.5: Different scatter-plots for correlation coefficients varying from -1 to 1, of two generic data sets. When two sets of data are strongly linked, they are said to have High Correlation: when the correlation coefficient is unit, they are found to be distributed in a straight line as in the leftmost graph. For decreasing values of the correlation coefficient, the points start spreading from the bisector straight line until reaching a "ball shape", where the coefficient drops to zero. Correlation coefficient is positive when the corresponding values in the two data sets increase (and decrease) together, whereas it is negative when they have an opposite behavior (i.e. when one value in one set decreases as the corresponding value in the other set increases).***

As shown in Figure 3.6*a*, the predefined POR measurement scheme provides several readout pulses at different time instants after the Program Pulse. The resulting data can be represented on a scatter plot to globally analyze how the single readout redistributes from RD#1 to RD#2. Figure 3.6*b* shows an example of scatter-plot obtained from the measured data from which a correlation coefficient can be calculated from either Eq. 3.2 or Eq. 3.3 and displayed on the picture.

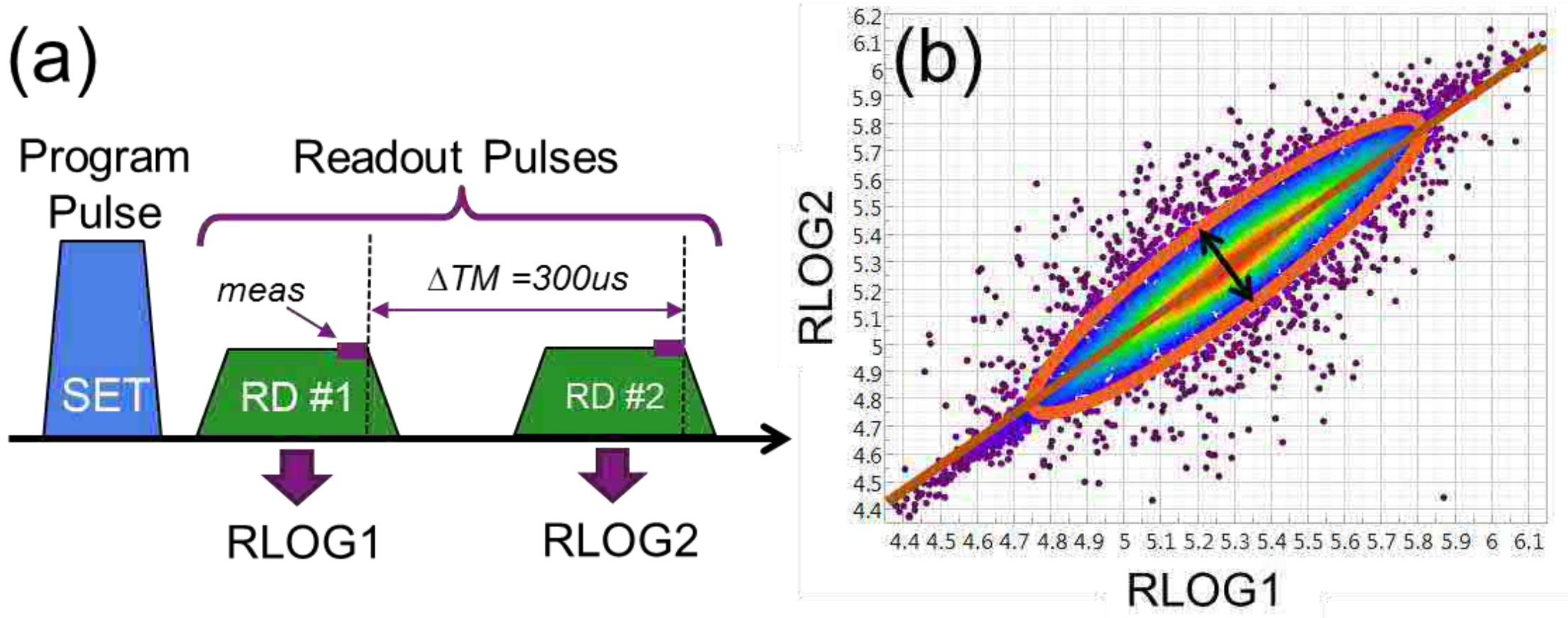

***Figure 3.6 - Metric (C): Study of the correlation between two adjacent readouts (a). The obtained resistances RLOG2 vs RLOG1 (log10 of the resistance values expressed in Ω ) are plotted on a scatter plot to calculate the correlation coefficient (b).***

## 3.3 Program Stability in Single-Pulse Switching

Figure 2.6 shows that for both logic states, in a standard program-and-verify algorithm, the value of the programmed data is quickly lost after programming. In order to understand the reason for this behavior, it was first necessary to study the stability of the states obtained by means of unverified Single-Pulse programming.

The POR conditions (listed in Paragraph 3) were used as a starting point for the successive benchmarking of different electrical or physical parameters. For each new experiment, only one single variable in the POR condition was changed, in order to evaluate its only impact on the overall switching performance, in particular on programmed state stability. A list of the benchmarked parameters is reported hereafter:

- variation of programming pulse width $T_w$;
- variation of programming pulse fall time $T_f$;
- impact of different operating temperature;
- impact of different material stack.

The type of study was scientific, meaning that even unrealistic long programming conditions were taken into consideration, in order to highlight physical trends or confirming former observed outcomes. In all experiments, the resistance evolution over time was monitored accordingly to the readout scheme shown in Figure 3.2. Next, the results for each programming condition are shown, including a comparison with the POR condition. This analysis is based on the Metrics (A), (B) and (C) discussed in Paragraph 3.2.

For better readability, only the results for the Low Resistive SET State have been reported. In most of the cases, the RESET state programming showed similar behavior. A good insight on the technology under study can therefore be obtained even by only analyzing the SET state programming.

### *3.3.1 Effect of Pulse Width variation*

The first experiment performed with the new UTM is a simple modification of the POR condition. As shown in Figure 3.7, the programming pulse width has been changed from 100 ns for Test#1 to 10 µs for Test#2, and 1 ms for Test#3. The purpose of this experiment is to assess the impact on program stability of a longer programming pulse and, hence, of a longer current flow and a larger time for the atoms to recombine.

The measurement parameters of the three performed tests were:

- DC Forming at an operating current $I_{op}$ = 50µA.
- Single-Pulse programming (following POR conditions):
    - different SET (or RES) pulse width $T_w$ (100 ns, 10 µs, 1 ms), as shown in Figure 3.7;
    - for opposite polarity restoring pulses, fixed RES (or SET) pulse (RES -1.5V for 100ns and SET 50µA for 100ns);
    - resistance monitoring using the reference readout sequence shown in Figure 3.2.
- Statistics for each test collected on 5 different cells, with 2500 SET/RESET cycles each.

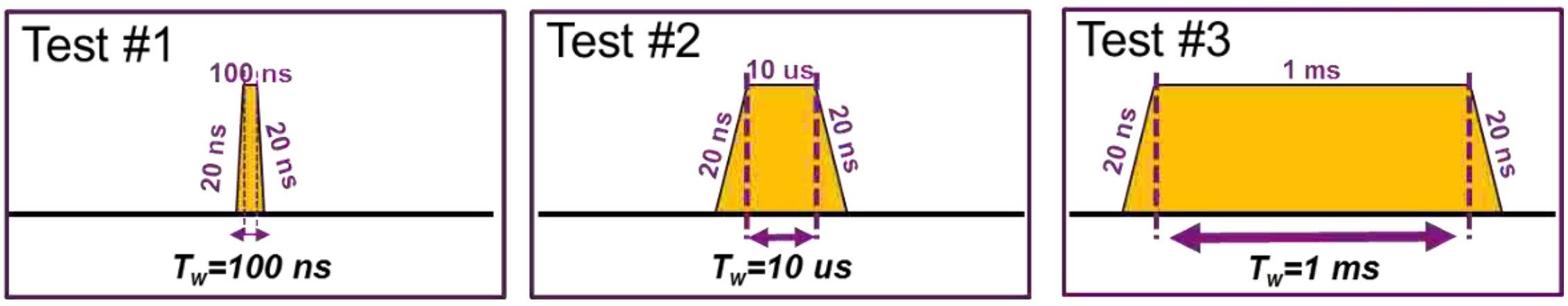

***Figure 3.7: Programming pulse timing, as changed for the three different Pulse Width tests.***

Before starting with the analysis of the collected data, all the single experiments (five experiments for 5 cells with 2500 cycles each) were compared in order to prove the hypothesis that the RRAM technology under test shows a stochastic behavior and that the outcome of a study at the array level (cycling several cells together) is statistically comparable to a study on several cells tested separately. A detailed explanation of this behavior is provided in Chapter 4. Figure 3.8 shows an example of plot (with $T_w$ = 100 ns) used to qualitatively compare the different distributions obtained for SET and RESET programming on different cells.

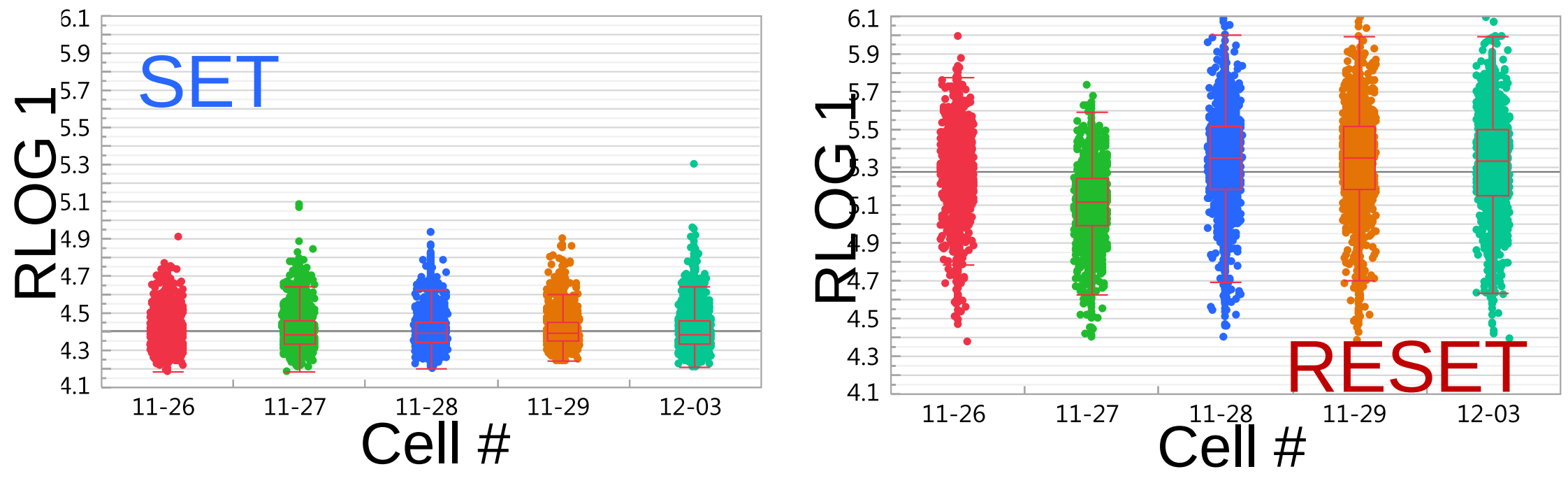

***Figure 3.8: Distributions for SET (left) and RESET (right) experiments performed on 5 different cells ($T_w$ = 100 ns). All the cells show similar median value and standard deviation: all collected data can therefore be considered.***

### 3.3.1.1 Metric A (Global Behavior) Results

Figure 3.9 shows the CDF plots for the three different programming conditions. The first plot on the left side shows, is the global behavior resulting from a measurement using POR conditions (with minimum-length programming pulse).

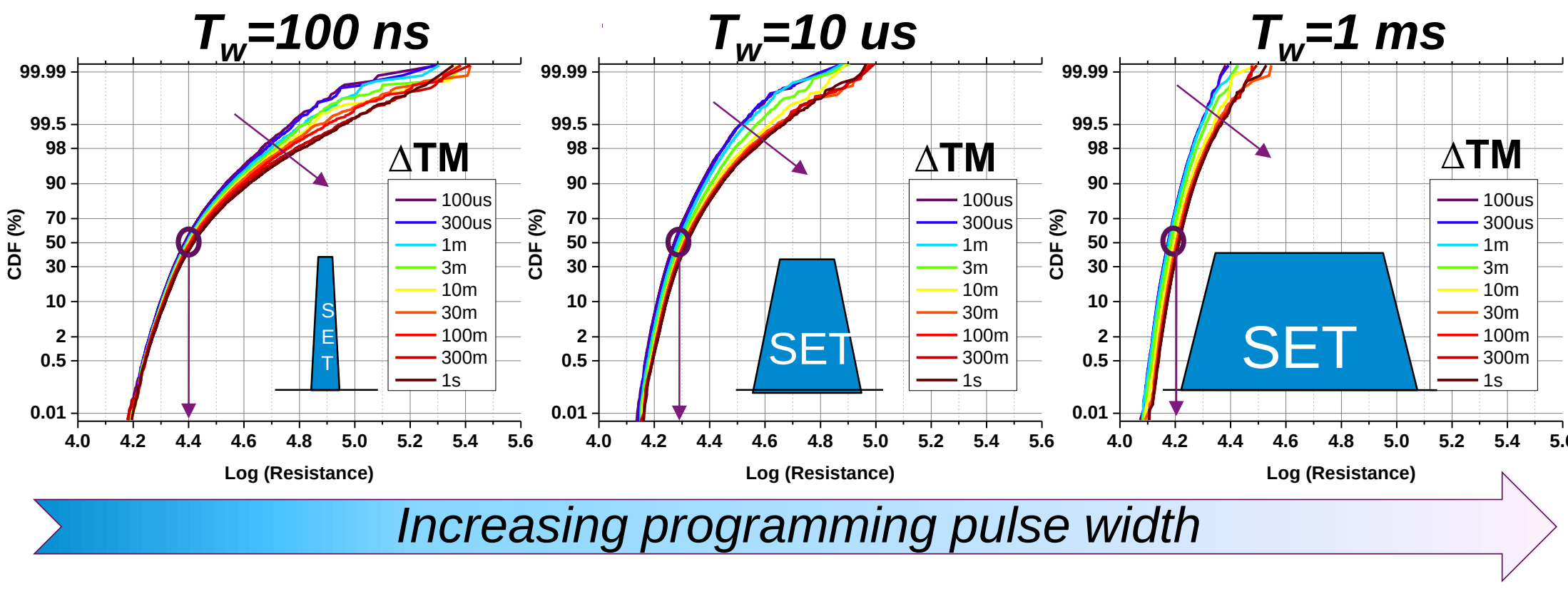

***Figure 3.9: Global Behavior for SET programming Pulse Width sweep. Each plot shows the CDFs for populations sampled at different ΔTM from 100µs to 1s. In all three plots, a considerable tail shift is observed (see arrows).***

Comparing the first plot ($T_w$ = 100 ns) with the next two ($T_w$ = 10 us and 1 ms) gives am intuitive approach to understand the impact of a longer programming pulse width:

- the decreasing initial median resistance ($10^{4.4}$, $10^{4.3}$ , and $10^{4.2}$) with increasing programming pulses width indicating a larger filament formation (Figure 3.10*a*);
- the standard deviation (and hence the distribution dispersion) reduces with increasing pulse width (Figure 3.10*b*), suggesting that larger filaments are also more precisely programmed (the CDF is more similar to a vertical line);

- from a retention point of view, an overall shift, along with the arising of tails, is observed in all the three cases.

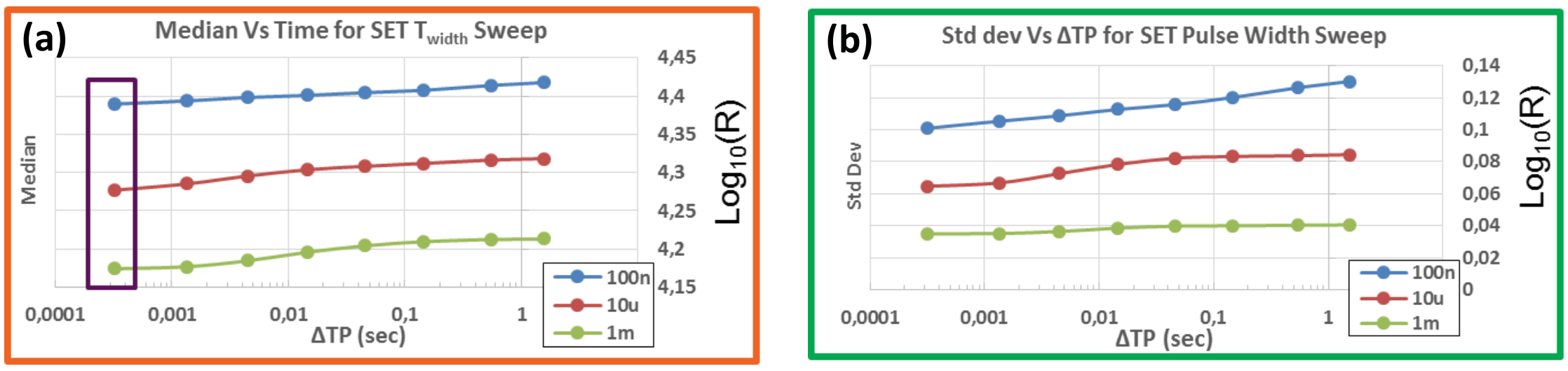

***Figure 3.10: Median (a) and Standard Deviation (b) evolution over time for the three considered experiments.***

The main conclusion that can be extracted from the observation of the global behavior, is that, even though the SET current amplitude was of 50 µA in all the tests performed, a larger pulse width leads to a larger conductive filament, which is slightly more stable. However, the relative increase in the Median value as well as the standard deviation improvement are negligible.

### 3.3.1.2 Metric B (Local Behavior) Results

Figure 3.11 shows the local behavior extracted from the experimental data. Three subpopulations were marked (top, mid, and low 10% of the initial population) as shown in Figure 3.11*a*, and their behavior over time was monitored.

From Figure 3.11b, it is apparent that the overall distribution shifts and its dispersion increases over time (as also understood with the previous metric). In addition, thanks to the tracking of the single subpopulations, it is notable that all the subpopulations tend to mix up. This can be observed even better by plotting the separated CDFs corresponding to each subpopulation. At the First Readout, each CDF have the shape of a truncated curve as shown in Figure 3.11*c*. This is due to the previously enforced selection (corresponding to the dotted lines which delimit the top 10%, mid 10%, and lower 10% of the whole population). The distributions after 1 s are shown in Figure 3.11*d.* It is clear that all the considered subpopulations behave exactly in the same way, since all three subpopulations are moving upon the same shape.

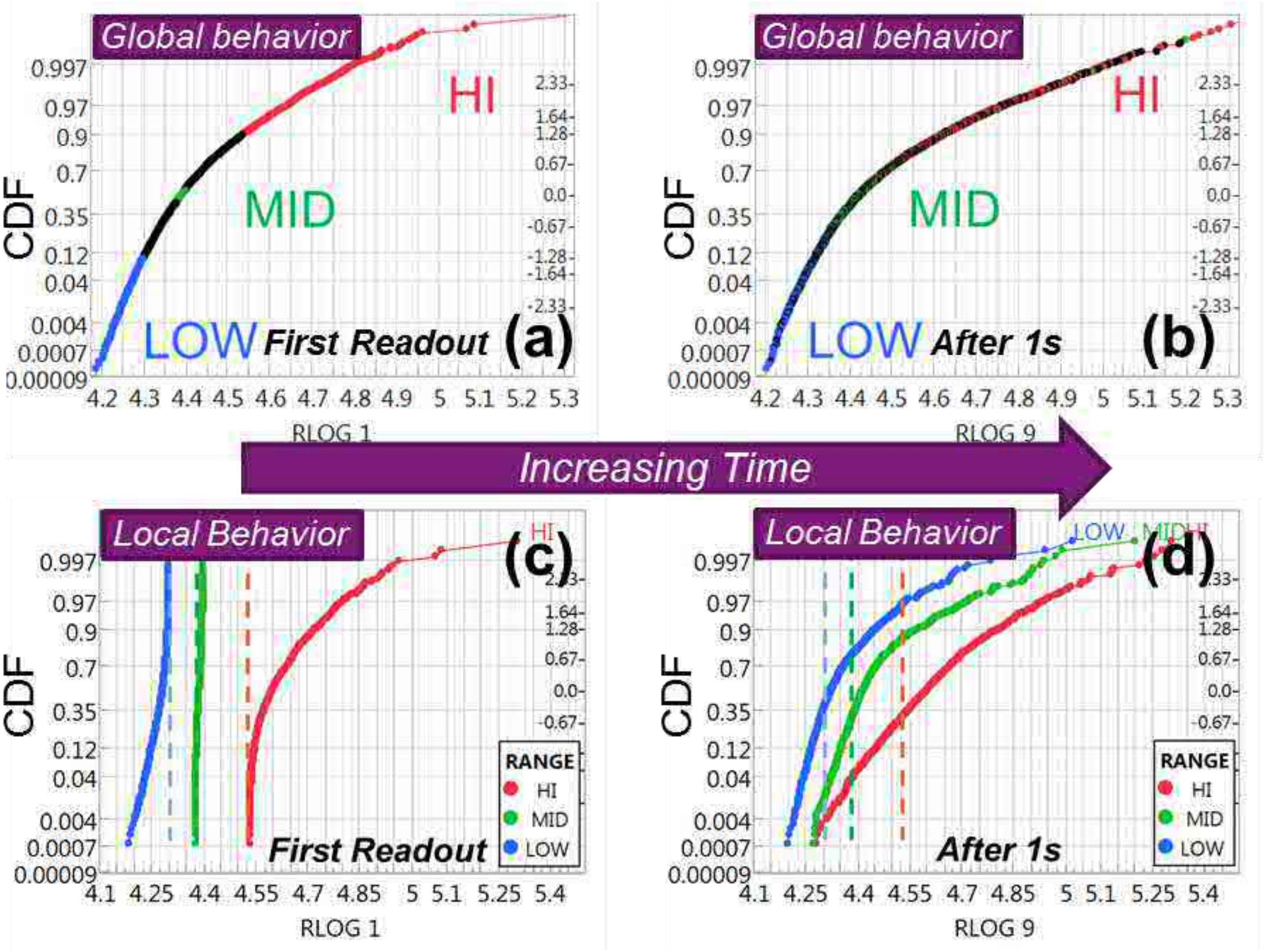

***Figure 3.11: Local Behavior for $T_w$ = 100 ns. CDF at First Readout (top, mid, and low 10% are highlighted) (a) and after 1 s (b), and corresponding separated CDFs: first Read-out (c) and after 1 s (d). It is clear that, with increasing time, all the populations have a tendency to assume the same shape.***

The local behavior was studied also for the other two test conditions ($T_W$ = 100 ns, $T_W$ = 10 ms; results are not reported here for simplicity). In all cases, the observed behavior was found to be identical: regardless from where the subpopulations are extracted, even only after 1 second after the reference readout, the CDFs reassembles the "usual" spread distribution. This behavior indicates a "complete randomness" of the resistance evolution over time: at each successive time instant, the resistance has the same probability to increase and decrease over time. This random behavior is further explained and analyzed in Chapter 4.

### 3.3.1.3 Metric C (Correlation) Results

The correlation evolution over time was extracted using JMP software, thus obtaining the results shown in Figure 3.12. The plots in this figure allows evaluating how the population is drifting. The automatically calculated correlation coefficient was collected for each test condition, as summarized in Figure 3.13.

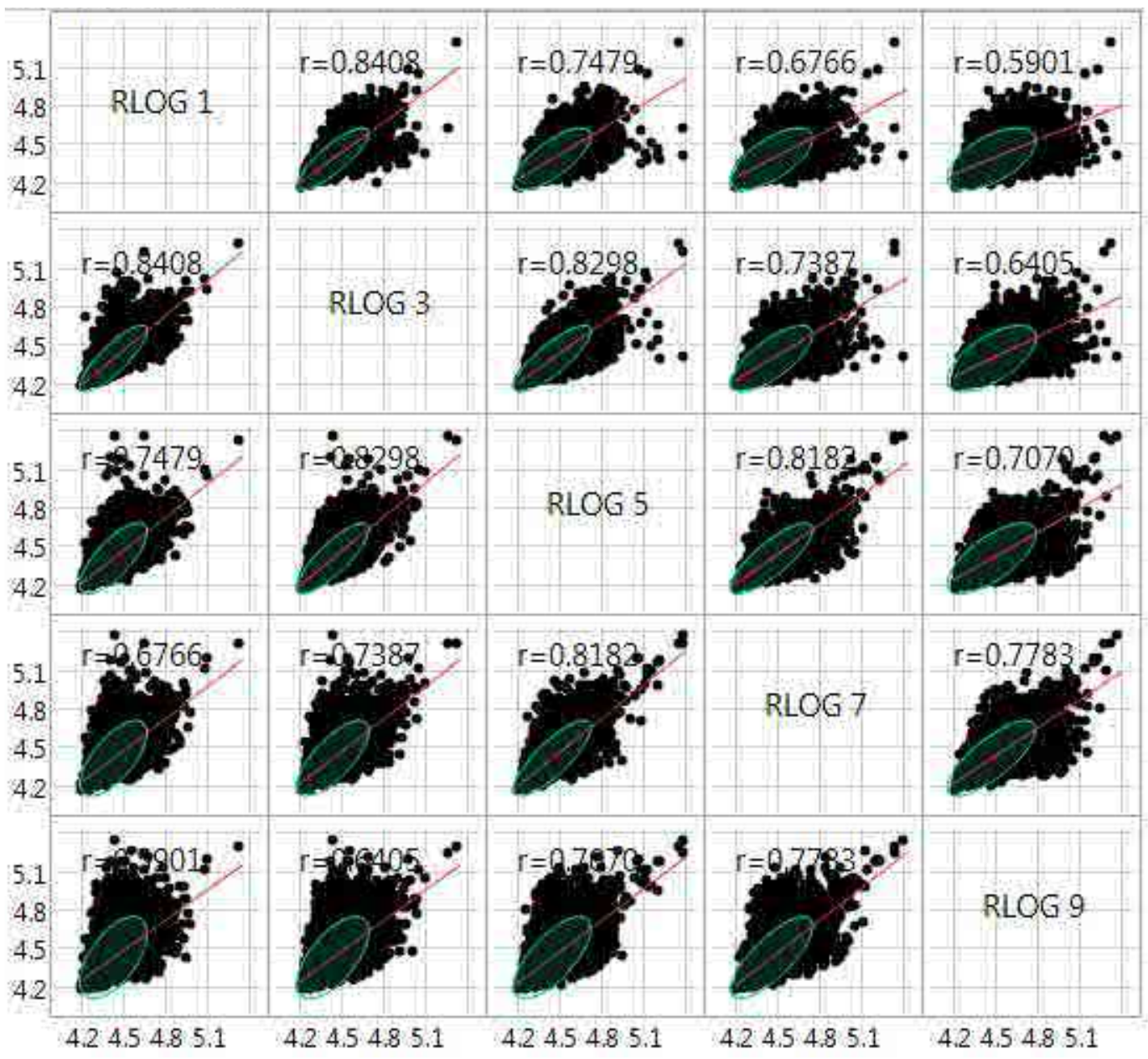

***Figure 3.12: Multivariate plot showing all the scatterplots for populations sampled at 4 consecutive time decades.***

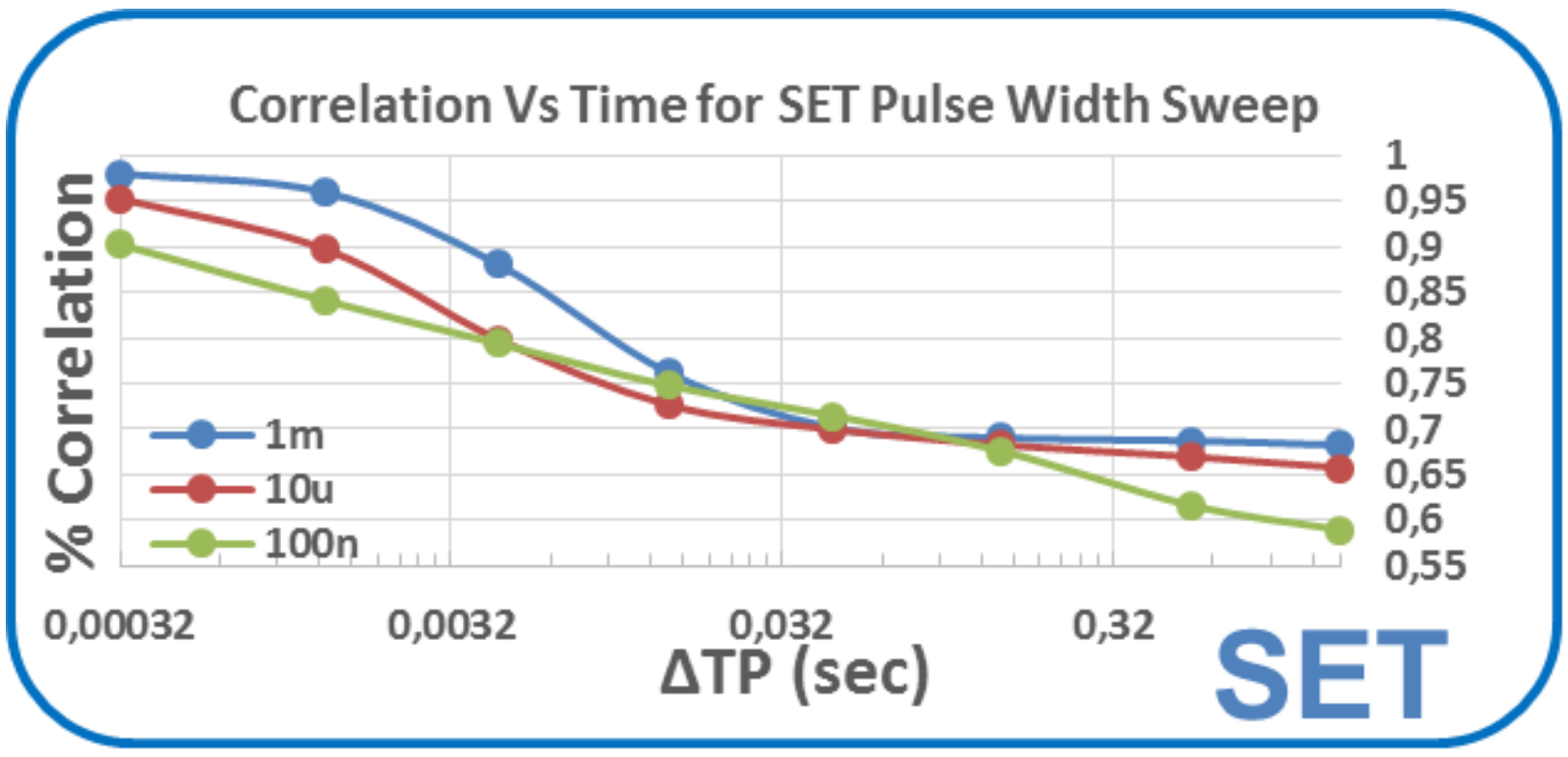

***Figure 3.13: Evolution of the correlation coefficient over time for different values of Program Pulse Width.***

The main result of this analysis is that the programming Pulse Width does not impact significantly on the correlation coefficient. In all three cases, a similar correlation loss is experienced: as graphically represented in Figure 3.12, the “cloud” of points tends to spread and deviate from the straight-line into a less correlated circle.

### *3.3.2 Effect of Fall Time variation*

The second experiment carried out was testing the impact of Fall Time on the program stability. As in the case of the test involving Pulse Width variation, this experiments addresses the hypothesis that a longer programming pulse gives rise to a thicker filament. In the case of Fall Time variation, the idea is to reduce the abrupt current switching through the filament by lowering the voltage over the RME with a more gradual slope.

The measurement parameters of the three performed tests were:

- DC Forming at an operating current $I_{op}$ = 50 µA.
- Single-Pulse programming (following POR conditions):
    - different SET (or RES) pulse fall time $T_{fall}$ (100 ns, 10 us, 1 ms), as shown in Figure 3.14;
    - for opposite polarity restoring pulses, fixed RES (or SET) pulse (RES -1.5V for 100ns and SET 50µA for 100ns);
    - resistance monitoring using the reference readout sequence shown in Figure 3.2.
- Statistics for each test collected on 5 different cells, with 2500 SET/RESET cycles each.

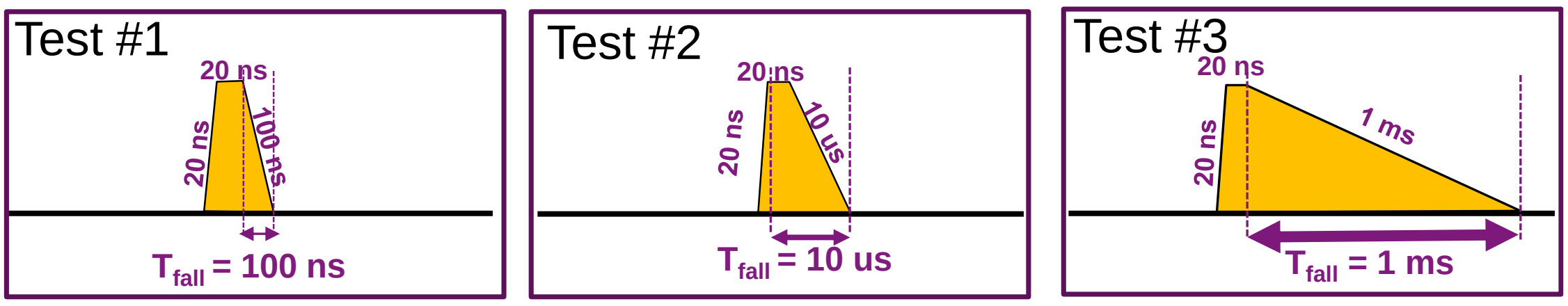

***Figure 3.14: Programming pulse timing, as changed for the three different Fall Time tests.***

### 3.3.2.1 Metric A (Global Behavior)

By observing the Global Behavior shown in Figure 3.15, it is clear that there is no noticeable impact on program stability. In all the three cases, the CDFs shift and spread, thus losing their programmed state.

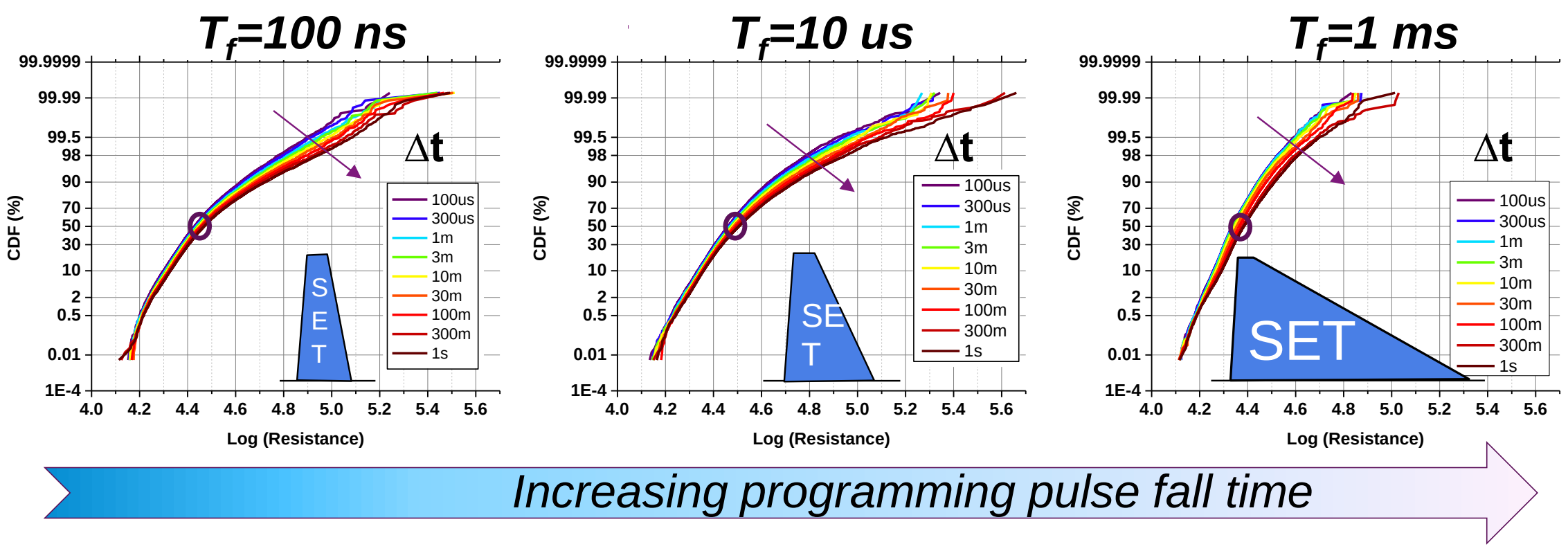

***Figure 3.15: Global Behavior for SET programming Fall Time sweep. Each plot shows the CDF for populations sampled at different ΔTM from 100µs to 1s. In all the three cases, a considerable tail shift is observed (see arrows).***

### 3.3.2.2 Metric B (Local Behavior)

Also the Local Behavior (shown in Figure 3.16) did not show any significant difference from the previous experiment involving Pulse Width variation.

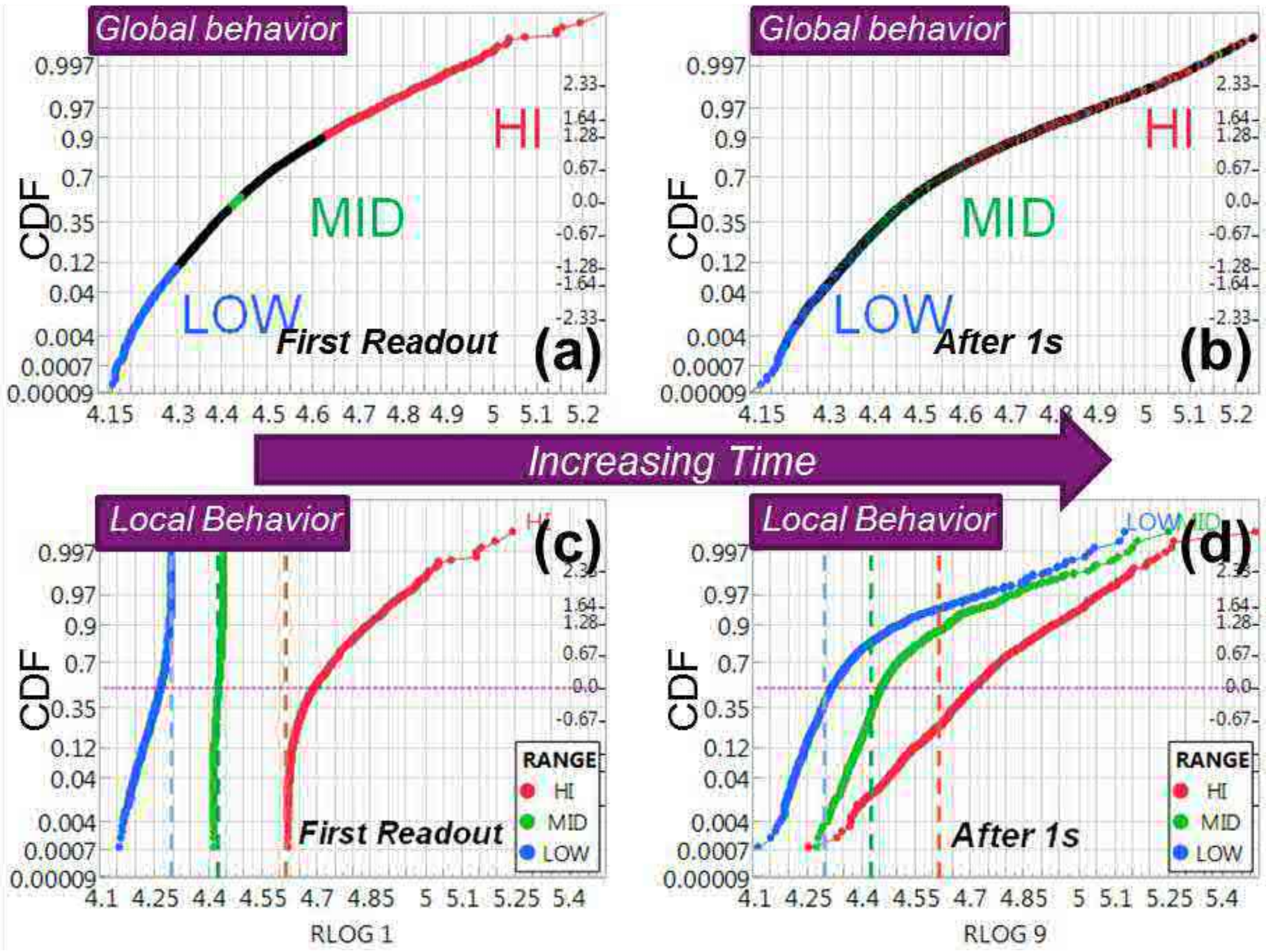

***Figure 3.16: Local Behavior for Tf =100 ns. See caption of Figure 3.11 for details.***

### 3.3.2.3 Metric C (Correlation)

As in the case of the experiment involving Pulse Width variation, the correlation coefficient does not show significant improvements when increasing the Fall Time. (Figure 3.17)

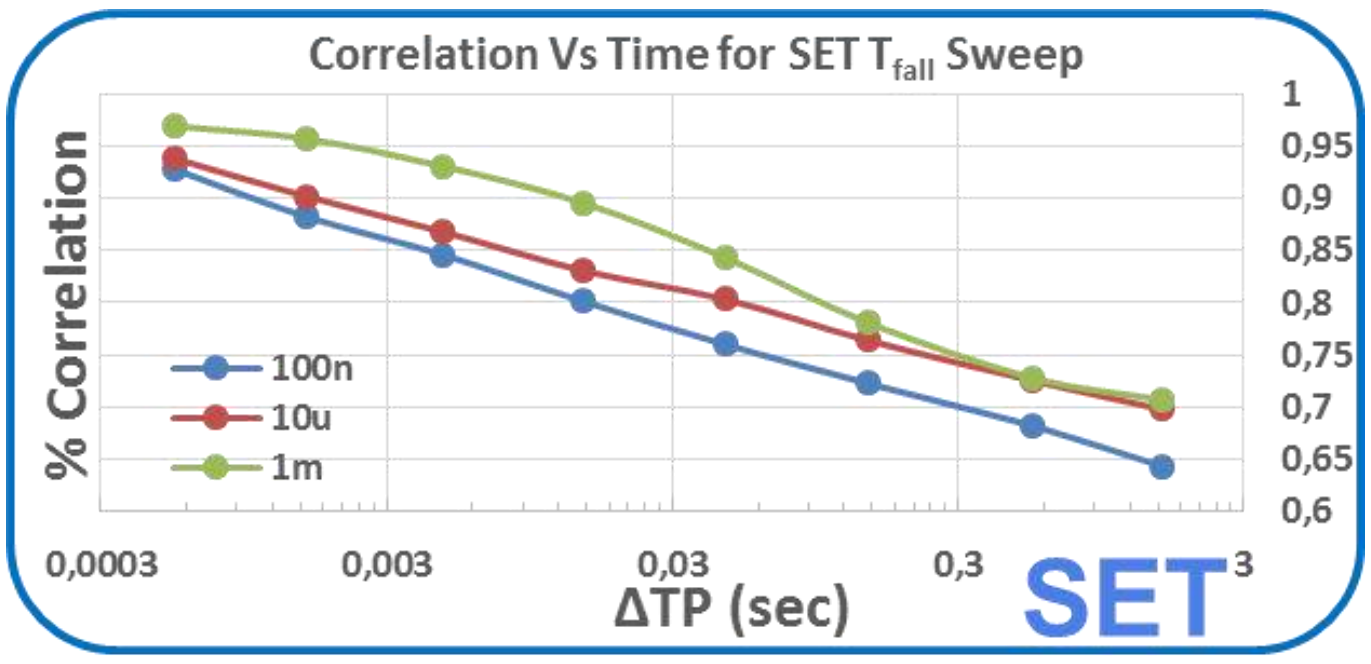

***Figure 3.17: Evolution of the correlation coefficient over time for different values of Program Pulse Fall Time.***

### 3.3.3 Temperature variation

Continuing the investigation on program stability, a third experiment was designed and carried out. In this experiment, the parameter changed was temperature.

For this test, the K4200 Semiconductor Parameter Analyzer was combined with a Cascade Microtech Elite 300. The Elite 300 provides an isolated wafer chamber where temperature can be increased up to 300˚C. As shown in Figure 3.18, only three temperatures were investigated: room temperature (using the data already collected for the previous experiments), 85˚C, and 150˚C.

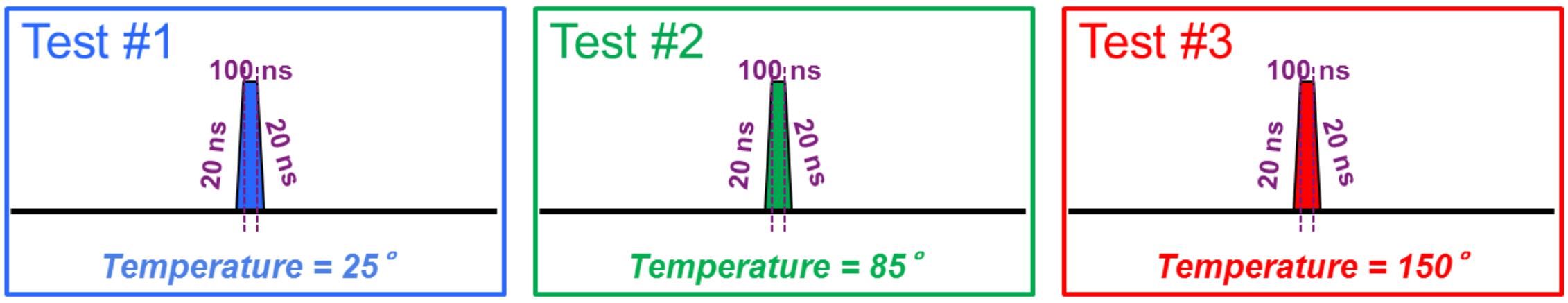

***Figure 3.18: Programming pulse timing (always POR). The Elite 300 probe station has been heated up to 25˚C for test #1, at 85˚C for test #2, and at 150˚C for test #3.***

The measurement parameters of the three performed tests were:

- DC Forming at an operating current of $I_{op}$ = 50 µA.
- Different Temperature for each test (Room Temp, 85˚C, 150 ˚C).
- Single-Pulse programming (following POR conditions):
    - same SET and RES pulse shape ([ $T_r$ ,$T_w$ ,$T_f$ ] = [20,100, 20] ns );
    - different gate voltage during SET to keep the current compliance at 50 µA;
    - different drain voltage during RESET to keep the voltage across the RME at about -1.5V;
    - resistance monitoring using the reference readout sequence shown in Figure 3.2
- Statistics for each test collected on 5 different cells, with 2500 SET/RESET cycles each.

When performing measurements at high temperatures, particular care should be paid in the characterization of the MOS transistor of the 1T1R structure of the tested cells. The high temperature affects the conduction properties of the transistor, which requires a continuous trimming of the gate voltage to keep the current compliance at 50 µA during SET. Furthermore, problems occur also during the SET-to-RESET transition due to high temperature. To induce switching to the High Resistive State, a negative voltage is applied to the Top Electrode of the RME. A voltage partition then occurs between the MOSFET and the RME, which implies that, with increasing MOSFET resistance, the voltage drop across the transistor increases, thus decreasing the effective voltage applied across the RME that is responsible for the SET-to-RESET transition.

#### 3.3.3.1 Metric A (Global Behavior) Results

Figure 3.19 shows the collected data for each considered temperature. As in the case of the experiment involving Pulse Width sweep, a decrease in the Median level is experienced with increasing temperature. This is ascribed to a higher thermal energy of the vacancies inside the filament. However, differently from the case of the first experiment, tails in the distributions arise substantially in the same way for all temperatures with increasing time. This is well highlighted by the vertical lines drawn on the plots. While, after 100 µs, the 10% to 90% percentile at first readout reduces with increasing temperature, in all the three cases, the final 10% to 90% percentile has the same width. Practically, while the initial programming seems to be more stable and to have small standard deviation, after even 1 second, the distributions spread similarly.

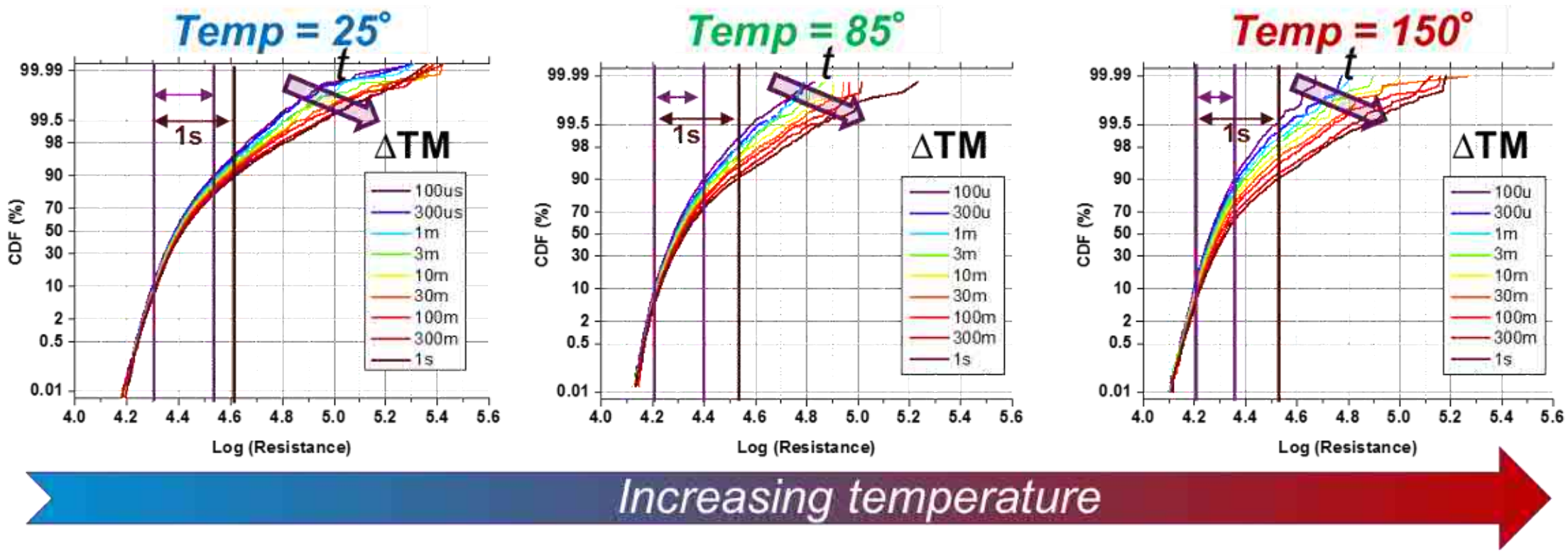

***Figure 3.19: Global Behavior for SET programming at different temperatures. The vertical lines indicate the 10% to 90% percentile at the time of the reference readout, and the same percentile after 1 s. The big arrows show that, in all three cases, the spread in the higher tails increases over time.***

#### 3.3.3.2 Metric B (Local Behavior) Results

The analysis of the local behavior (shown in Figure 3.20) demonstrates a similar behavior as that observed for the case of programming pulse shape variations. At 150˚C, the subpopulations reshape even more for the same distributions.

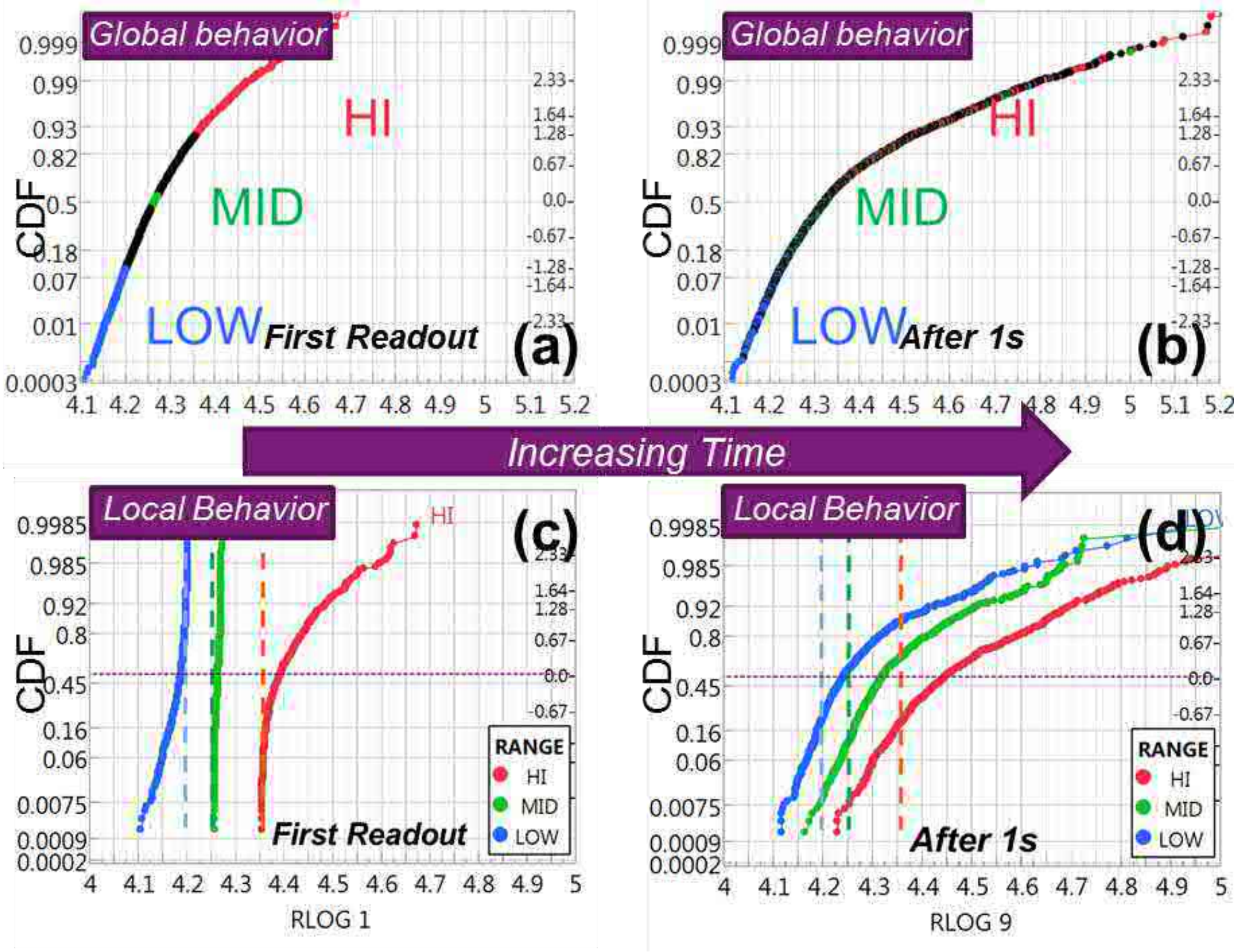

***Figure 3.20: Local Behavior for subpopulations of the test at 150˚C. See caption of Figure 3.11 for details.***

#### 3.3.3.3 Metric C (Correlation) Results

Figure 3.21*a*, shows the normalized ΔMedian evolution over time (ΔMedian calculated as the median value at time ΔTP minus the initial median). The normalization was needed to better compare the three evolutions that occur for three different initial resistive levels. It can be observed that at a higher temperature results in a faster increase in the median after tens of ms. Figure 3.21*b* highlights even more the increase of drift and spreading at higher temperature, showing a faster decrease of the correlation coefficient at 150˚C.

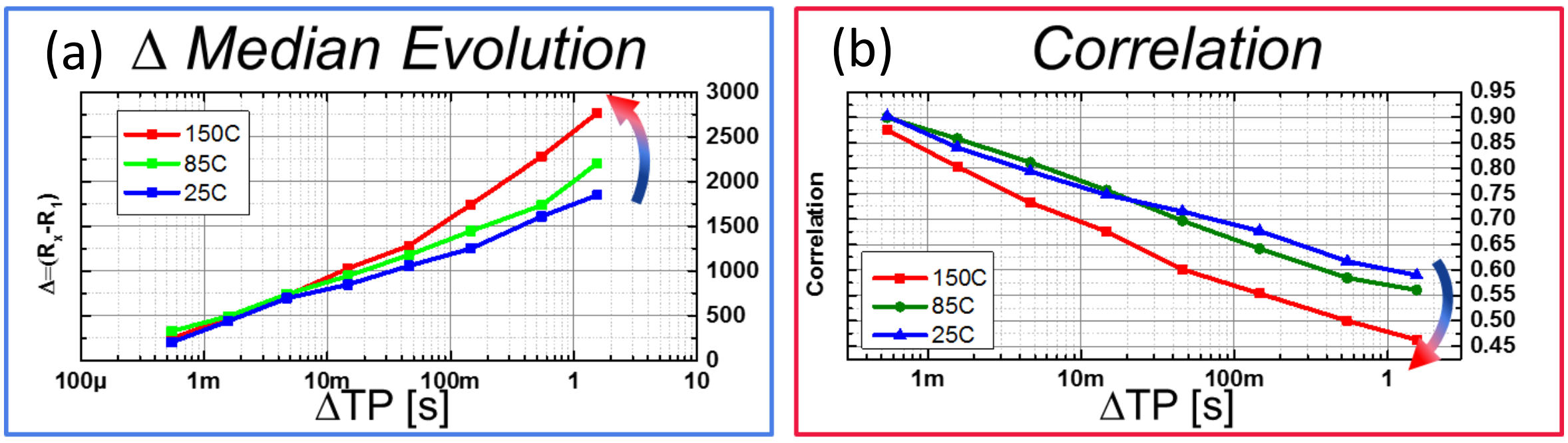

***Figure 3.21: Evolution of the median normalized to the initial median value (a) and evolution of the correlation coefficient over time (b), for three different operating temperatures.***

This temperature test was one of the most important of all performed experiments. The effects of several phenomena, such as drift, are intrinsically temperature dependent, and these measurements are useful to isolate temperature-activated behaviors in order to ease their analysis. Further research on the collected data is being carried out by the modelling team of imec's Memory Device Design group.

### 3.3.4 *Material Stack variation*

The last POR variation analyzed is Material Stack variation. The POR programming conditions were used to perform the Unverified Single-Pulse Algorithm on three different structures, as shown in Figure 3.22:

- the already extensively measured RRAM wafer with Hafnium Oxide based stack;
- a wafer with RRAM cells with Tantalum Oxide based stack. Tantalum, along with Hafnium, oxide are nowadays the most popular oxides used in RRAM technology (as shown in Chapter 1);
- a wafer with Hafnium Aluminate Oxide stack, a variation of the standard Hafnium structure; research in this field aims at finding a solution with higher endurance.

In this experiment, the POR conditions are fundamental to be applied for a valid comparison between different technologies. The different stack oxide introduces many differences in terms of forming and SET/RESET voltage levels, along with different breakdown conditions. Also the current limiting transistors on each wafer had to be carefully changed to ensure an operating current of the POR equal to 50 µA.

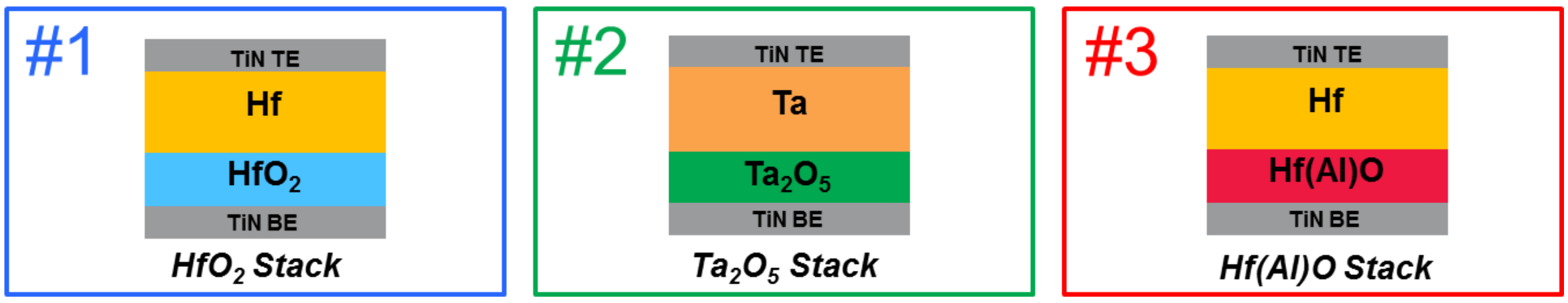

***Figure 3.22: Different Material stacks considered in this experiment. Hafnium Oxide for test #1, Tantalum Oxide stack for #2, and Hafnium Aluminate Oxide for #3.***

The measurement parameters for the three performed tests were:

- DC Forming at an operating current of $I_{op}$ = 50uA.
- Different Temperature for each test (Room Temp, 85˚C, 150 ˚C).
- Single-Pulse programming (following POR conditions):
  - same SET (or RES) pulse shape ( [ $T_r$ ,$T_w$ ,$T_f$ ] = [20,100, 20]ns );
  - different gate voltage on SET to keep the current compliance at 50 µA;
  - for opposite polarity restoring pulses, fixed RES (or SET) pulse (RES -1.5V for 100ns and SET 50µA for 100ns);
  - resistance monitoring using the reference readout sequence shown in Figure 3.2.
- Statistics for each test collected on 5 different cells, with 2500 SET/RESET cycles each.

Figure 3.23 shows the Global Behavior analysis for the three considered material stacks. By inspection, a big difference in the stability between the cells based on different stacks can be noticed. The Tantalum Oxide based stack does not show any program instability: the resistance distribution remains unchanged.

This behavior was expected, being that $Ta_2O_5$ was introduced in the first place as candidate oxide due to its better endurance and retention proprieties (at the cost of a smaller resistive window), related to a different chemical-potential profile along the with respect to $HfO_2$: this is the reason why today much effort is spent on this new type of material. The Hafnium Aluminate Oxide stack, on the contrary, still shows the presence of tails in the resistance distribution as the simple Hafnium Oxide stack, but the initial distribution is much more peaked, showing a lower standard deviation.

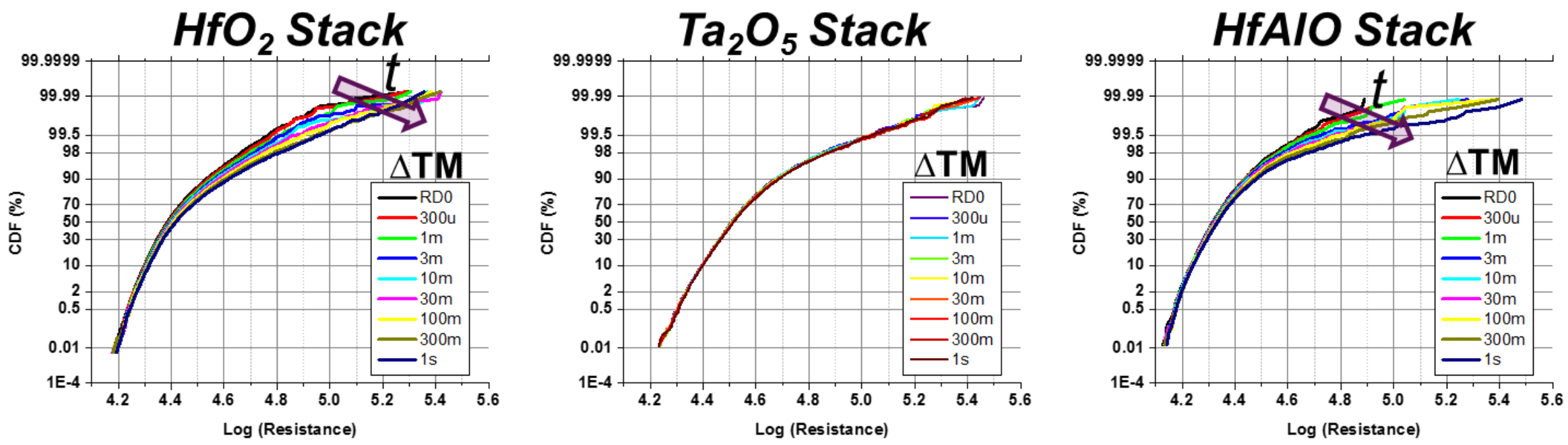

***Figure 3.23: Global Behavior over time for the three different material stacks considered. While for the Hafnium based stacks (in particular for the HfAlO based stack), the initial distributions have a slightly smaller standard deviation than for the $Ta_2O_5$ based stack, and those stacks show a retention loss over time. A very good program stability, practically showing no tails in the distribution even after 1 s, was measured on the $Ta_2O_5$ based stack.***

The above aspects can be seen more in detail by analyzing the ΔMedian evolution over time, which is much higher in the case of the Hafnium Oxide based stack that, thus, is demonstrated to be the stack that displays the higher drift (Figure 3.24*a*). Also the Standard Deviation evolution over time (Figure 3.24*b*) highlights the better stability of the Tantalum Oxide based stack, which has on average a higher standard deviation that, however, remains constant over time.

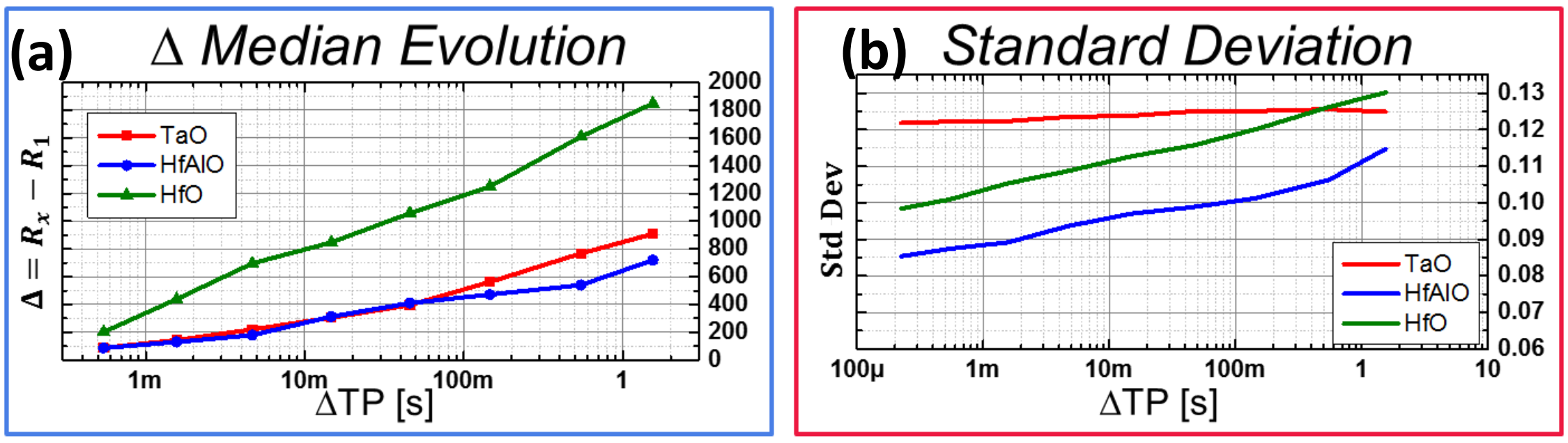

*Figure 3.24: Evolution of the median normalized to the initial median over time (a), and standard deviation (b).*

Also the correlation plots for the different stacks were evaluated, confirming what was already clear when looking at the Global Behavior. As shown in Figure 3.25, the correlation for the Tantalum Oxide based stack is much higher than in the case of the Hafnium based ones.

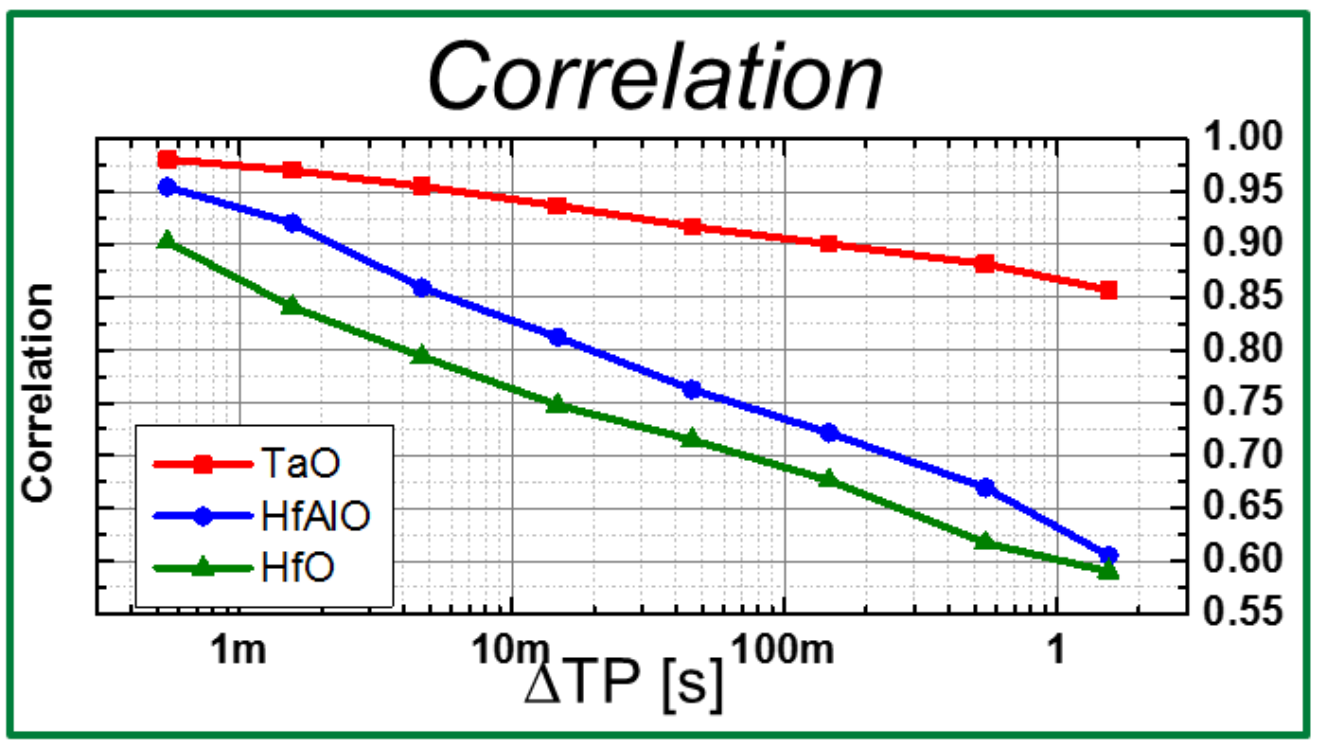

*Figure 3.25: Correlation evolution over time for different material stacks. TaO-based stack shows the best stability (i.e. the least decreasing correlation).*

## 3.4 Program Stability in Verified Programming

Figure 2.6 shows that, when using a standard program-and-verify algorithm, the value of the programmed data is quickly lost after programming for both logic states. The new UTM, as explained in Paragraph 2.5.3, has been developed not only to perform unverified retention measurements, but also with the capability of performing verifying algorithms.

The measurement parameters for the performed tests were:

- DC Forming at an operating current $I_{op}$ = 50uA.
- Pulse timings taken from POR conditions shape ([ $T_r$ ,$T_w$ ,$T_f$ ] = [20, 100, 20] ns ).
- Step Programming with Verify with thresholds fixed at median values:
    - R ≤ 20 kΩ for SET and R ≥ 200 kΩ for RESET.
- Statistics data collected on 5 different cells, with 1000 Verified SET/RESET cycles each.
- Incremental Step Programming (ISP) (see Table 2 in POR conditions):
    - for SET programming, Gate voltage swept [0.8 : 0.025 : 1.3] V;
    - for RESET programming, Drain voltage swept [ -1 ; -0.1 ; -2] V;
    - resistance monitoring using the reference readout sequence shown in Figure 3.2.
- Fixed Step Programming (FSP), max 10 attempts to reach threshold before returning a fail.

### *3.4.1 Incremental Vs. Constant Program-and-Verify Programming*

This paragraph compares the results obtained by using two common Program- Verify algorithms, detailed in Paragraph 2.5.3 and repeated in Figure 3.26 for convenience.

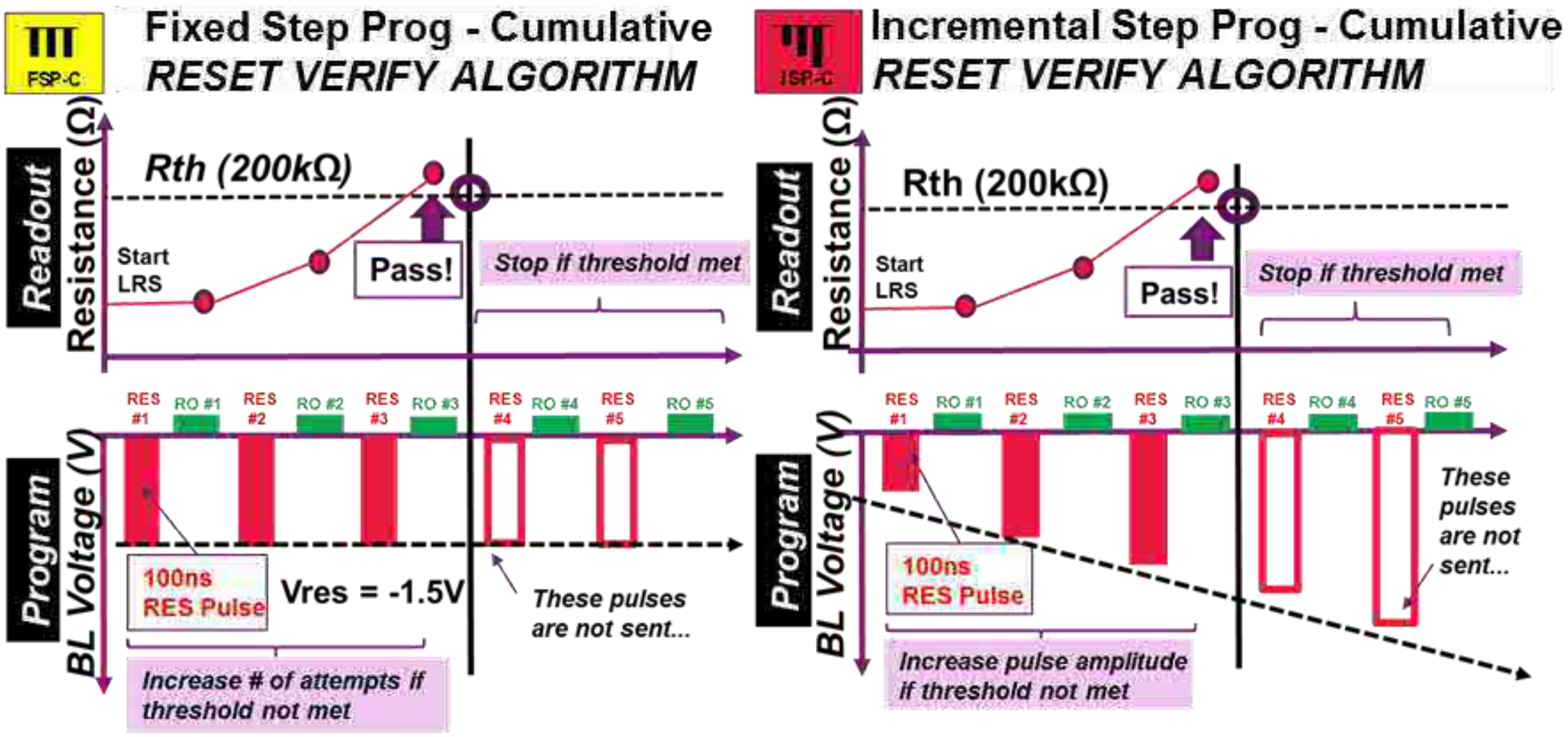

***Figure 3.26: Schemes of the two Verified Programming Algorithms that were tested on imec`s RRAM. See Figure 3.2.6 in Paragraph 2.5.3 for details.***

#### 3.4.1.1 Metric A (Global Behavior) Results

The first analysis was performed by inspecting the Global Behavior. The plots in Figure 3.27 differ from the ones obtained in the case of unverified programming in that at RD0 (first readout), the CDFs follow a straight line, or at least are squeezed and do not cross the verify level (dotted line). In Figure 3.27, both SET and RESET programming data is shown. In all cases, independently from the algorithm in use, big tails are apparent in the distributions even after few microseconds. It is interesting to note that after an initial large shift, the distributions seem to stop drifting (starting from the red curves corresponding to ΔTM = 100ms). This slower relaxation of the programmed state is better seen in the bottom RESET plots (at the bottom of the Figure 3.27, where, after 100ms, the distributions are superimposed.

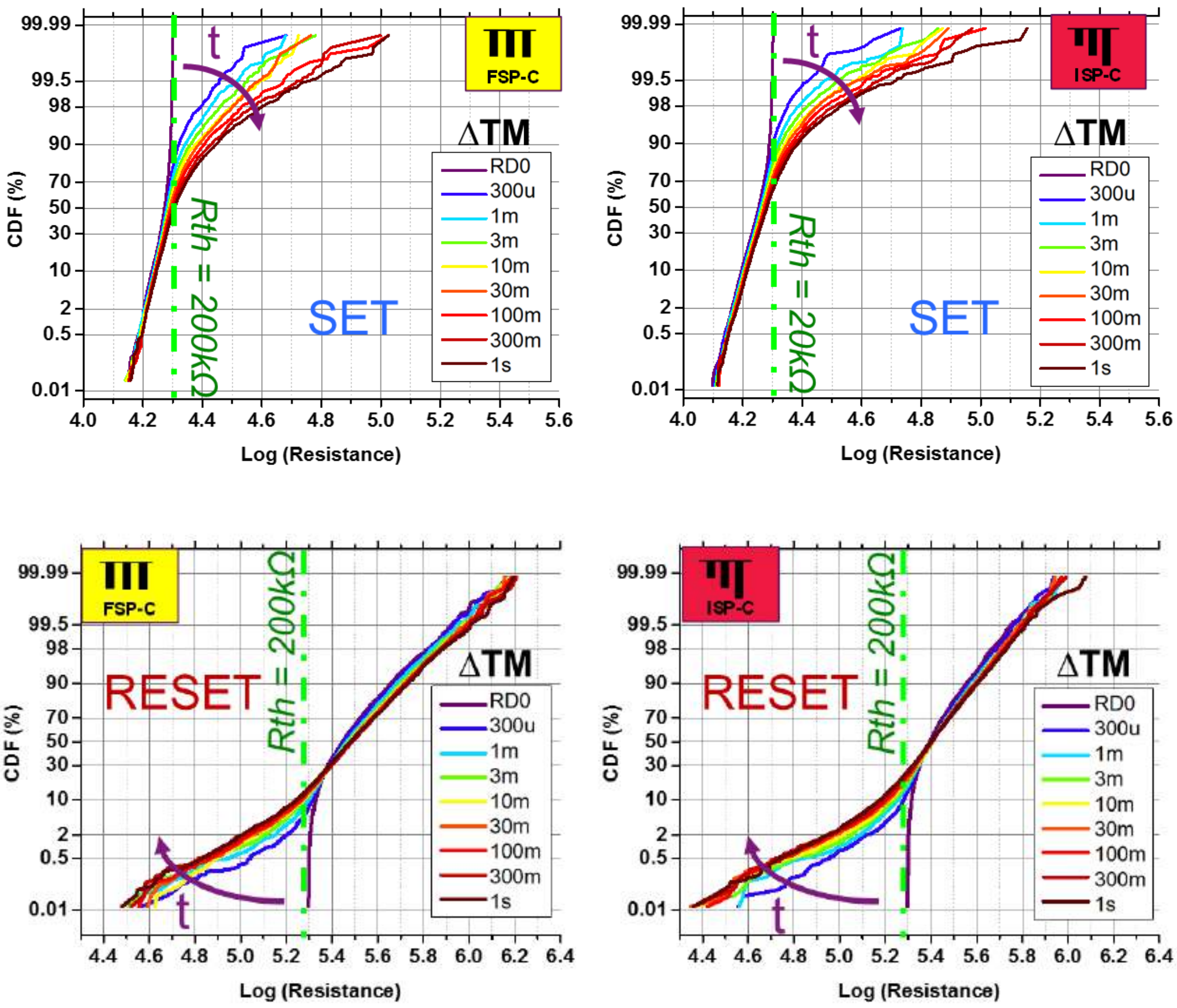

*Figure 3.27: Global Behavior for SET (top) and RESET (bottom) when using a Program-and-Verify algorithm. Dotted lines specify the threshold value at which the verify is performed (programming is verified if, at RD0, R ≤ 20 kΩ for SET and R ≥ 200 kΩ for RESET). In all considered cases, the verified state is not held over time, and tails due to a big relaxation appear, as highlighted by the arrows.*

This reduction in relaxation speed can be also observed by analyzing the Median level, as shown in Figure 3.28 (in particular, a saturation of the median value is detected for the RESET state).

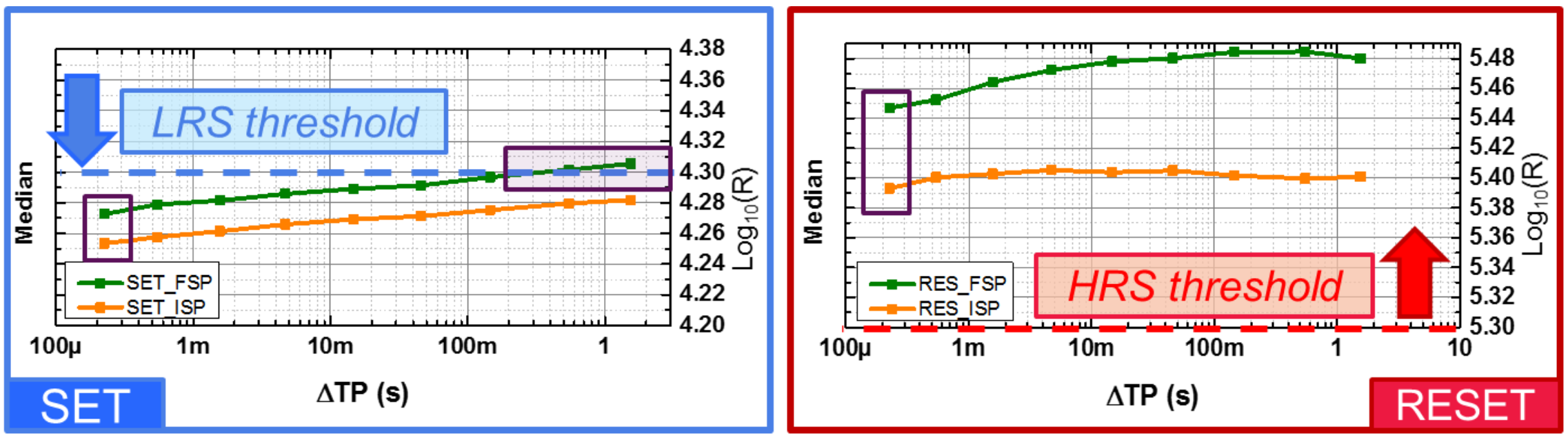

***Figure 3.28: Median evolution over time for the SET (left) and the RESET state (right). For both states, the ISP algorithm returns a lower median level with respect to the FSP algorithm. [OK] This is probably due to a higher average programming current obtained with the incremental step programming algorithm then with the fixed step programming.***

Another way of benchmarking the FSP and the ISP algorithms is shown in Figure 3.29. The plots in this figure show the percentage of the overall programmed states (% of the total 1000 cycles performed over 5 cells), that cross the verify threshold level. It can be noticed that for SET programming, the FSP algorithm experiences up to 55% of failed states after 1 second. This was expected fromFigure 3.28, where the median level for the green curve crosses the LRS verify threshold.

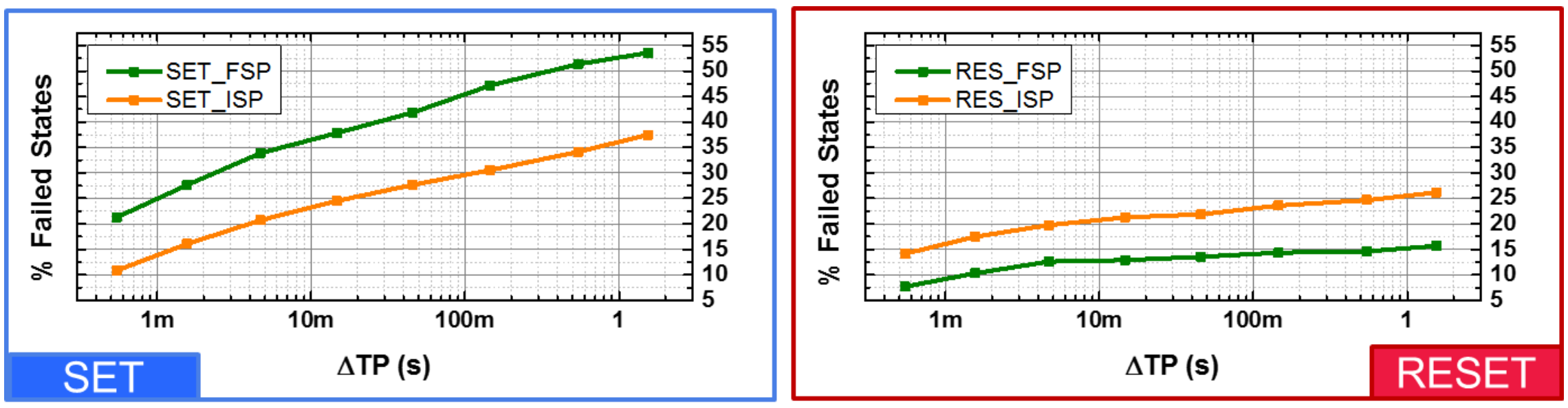

***Figure 3.29: Percentage of failed states, i.e. of states that cross the threshold level after a given time interval.***

### 3.4.1.2 Metric (B) – Local Behavior Results

To understand how the subpopulations evolve over time, a study of the Local Behavior was performed. Figure 3.30 shows the distributions for the Top, the Mid, and the Low 10% of the initial distribution. As for the case of the unverified distributions, it results that as soon as the "verify is released" all the three subpopulations show a very similar drift and relaxation.

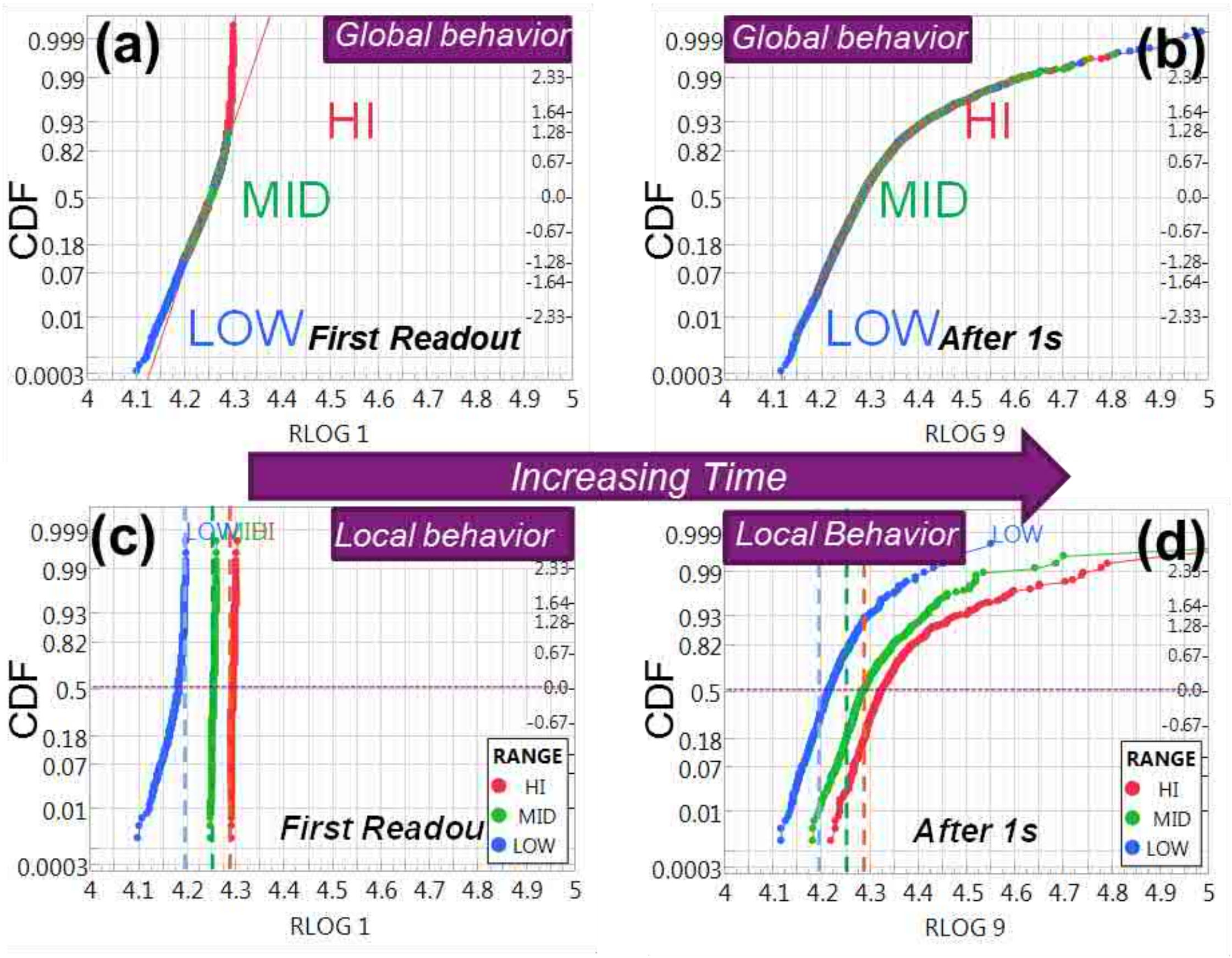

***Figure 3.30: Local Behavior for $T_W$=100 ns. CDF at First Readout with highlighted (top, mid, low 10% are highlighted) (a) and(after 1s in (b), and corresponding separated CDFs (c) (after 1s in (d)). With increasing time, all the population have a tendency to assume the same shape.***

#### 3.4.1.3 Metric (C) – Correlations Results

In the last analysis, the correlation coefficients were calculated (Figure 3.31). From this test, it is clear that ISP programming leads to a much narrower distribution than FSP programming which results in higher correlated states than with FSP. This is particularly true for the case of SET programming.

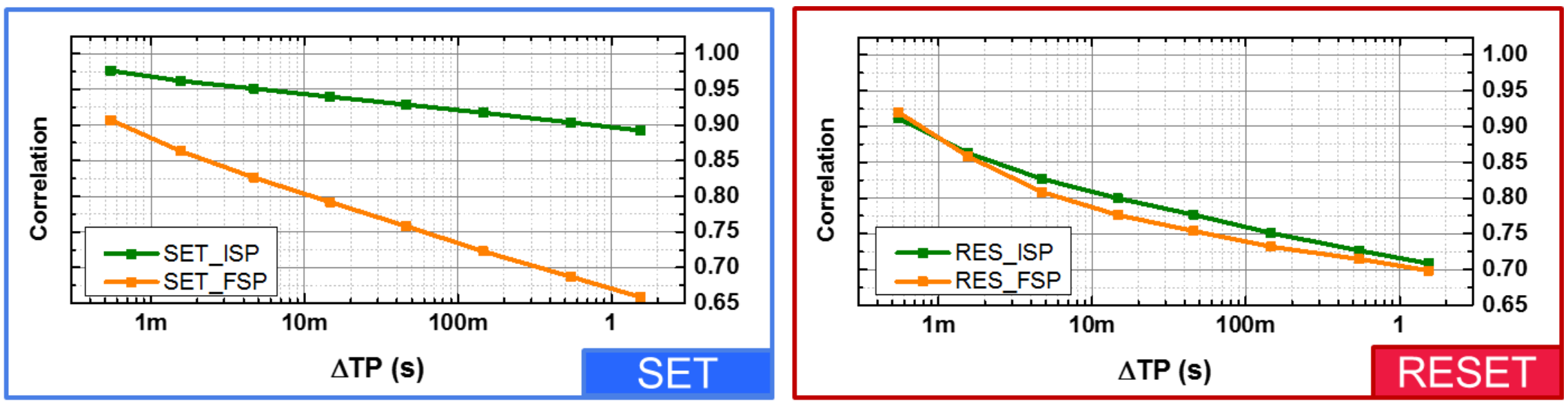

***Figure 3.31: Correlation coefficients calculated for SET (left) and RESET (right) programming with the ISP and FSP algorithm.***

#### *3.4.1.4 Results Summary*

A summary of all the results presented in this paragraph is provided in Table 6. For each considered metric, the better algorithm, between ISP and FSP, for SET and RESET verified programming is shown.

| Metric | | LRS | HRS | COMMENTS |
|---|---|---|---|---|
| (A) | Median | ISP | FSP | ISP always leads to lower median value |
| (A) | Δ Median | = | ISP | Big variation for FSP HRS prog |
| (A) | Relaxation | ISP | FSP | Connected to starting Median value |
| (B) | Local Behavior | = | ISP | Result confirmed by STD DEV |
| (C) | Correlation | ISP | = | |
| | **The winner is:** | **ISP** | **=** | **Incremental SP programming** |

***Table 6: Benchmark of Incremental vs Fixed Step Programming, summarizing the results provided in paragraph 3.4. For LRS programming, the best algorithm is ISP, whereas for HRS programming, the two algorithms give similar performance.***

The main result highlighted by this analysis is that the programming algorithm does not impact significantly on the tails that appear in the resistance distributions with increasing time. In both cases, even few microseconds after the verify, the distribution shifts and reassembles in an equilibrium shape similar those found during unverified programming tests. No matter whether program-and-verify is carried out of not, in all three cases a similar correlation loss is experienced, and, as shown in Figure 3.12, the “cloud” of points tends to spread and deviate from the straight line shape tunring into a lower correlated circle.

## 3.5 Summary

This chapter presented the main experiments conducted during this thesis project, focusing on the algorithms used, explaining the workflow of the benchmarking and providing the obtained results.

First, the data measured with an Unverified-Single Pulse Programming algorithm using different timing parameters, were presented, and the following conclusions were drawn:

- *Pulse Width* does not impact effectively on the relaxation behavior, showing just a small improvement, which is ascribed to a slightly larger conductive filament formation;
- *Fall Time* does not affect the relaxation, confirming that instability is not related to the abrupt quenching of current;
- *Temperature* variation highlighted that relaxation increases with increasing temperature, but further investigation is still needed to exclude other causes;
- The choice of the right *Material Stack* choice does affect relaxation and
    - HfO based stack has faster median resistance increase and correlation loss,
    - HfAlO based stack has limited median drift, but sill loss of tails,
    - TaO based stack has limited median drift and best correlation (no tails).

Secondly, the data measured with Program-and-Verify Step Programming algorithms were presented and the following conclusions were drawn:

- neither ISP nor FSP programming algorithms impact on relaxation;
- Incremental Step Programming algorithm turned out to give better results.

# Chapter 4
# Refined Hourglass Model

As already stated in the introduction, Valence Change Mechanism RRAM demonstrated to be a valid technology to be used in Storage Class Memory thanks to its low power consumption, multi-level cell capability, and 3D crossbar integration. RRAM is also promising in the embedded memory market, thanks to the CMOS process compatibility of the “fab-friendly” oxides used in their production. As shown in Figure 1.25, in the ITRS emerging memory technologies roadmap, the only big drawback of RRAM technology is ***Variability***.

The experimental analysis presented in Chapter 3 shows, through various tests and experiments, that variability, in terms of program instability, is an important problem for this emerging technology. Thus, further research is needed in order to engineer better performing materials and develop more reliable programming algorithms, before making this technology available on the market.

Chapter 4 is organized as follows. First, an introductive explanation of the imec’s Hourglass model to describe RRAM switching is given. Secondly, the Random Walk Model is introduced since its features are used for the development of our provisional model. Thus, our provisional Hourglass refinement model addressing the stability and relaxation issue is detailed and fitted against the actual collected data.

## 4.1 Stochastic Device

RRAM switching is often described as a stochastic process, due to its intrinsic random behavior observed after several measurements.

As reported in the literature [1], no differences have been recorded between measurements performed intra-wafer or intra-device, confirming the intrinsic nature of RRAM variability. Moreover, no correlation was found even from cycle to cycle in the same cell.

In addition to the large variability, the statistical dispersion of the resistance values during switching is still a topic of discussion among researchers. Up to now, there is still not a unique and universally accepted physical model to describe the switching. Many different models of switching are found in the literature (as shown in Chapter 1). In addition, when describing the switching of the same material by using the accepted 1D filament structure, the SET and RESET process can be described in many different ways, without a dominant trusted physical model.

D.Wouters [2] showed that various interpretations and models have been published by the scientific community on this topic and that no dominant model has been selected so far (Figure 4.).

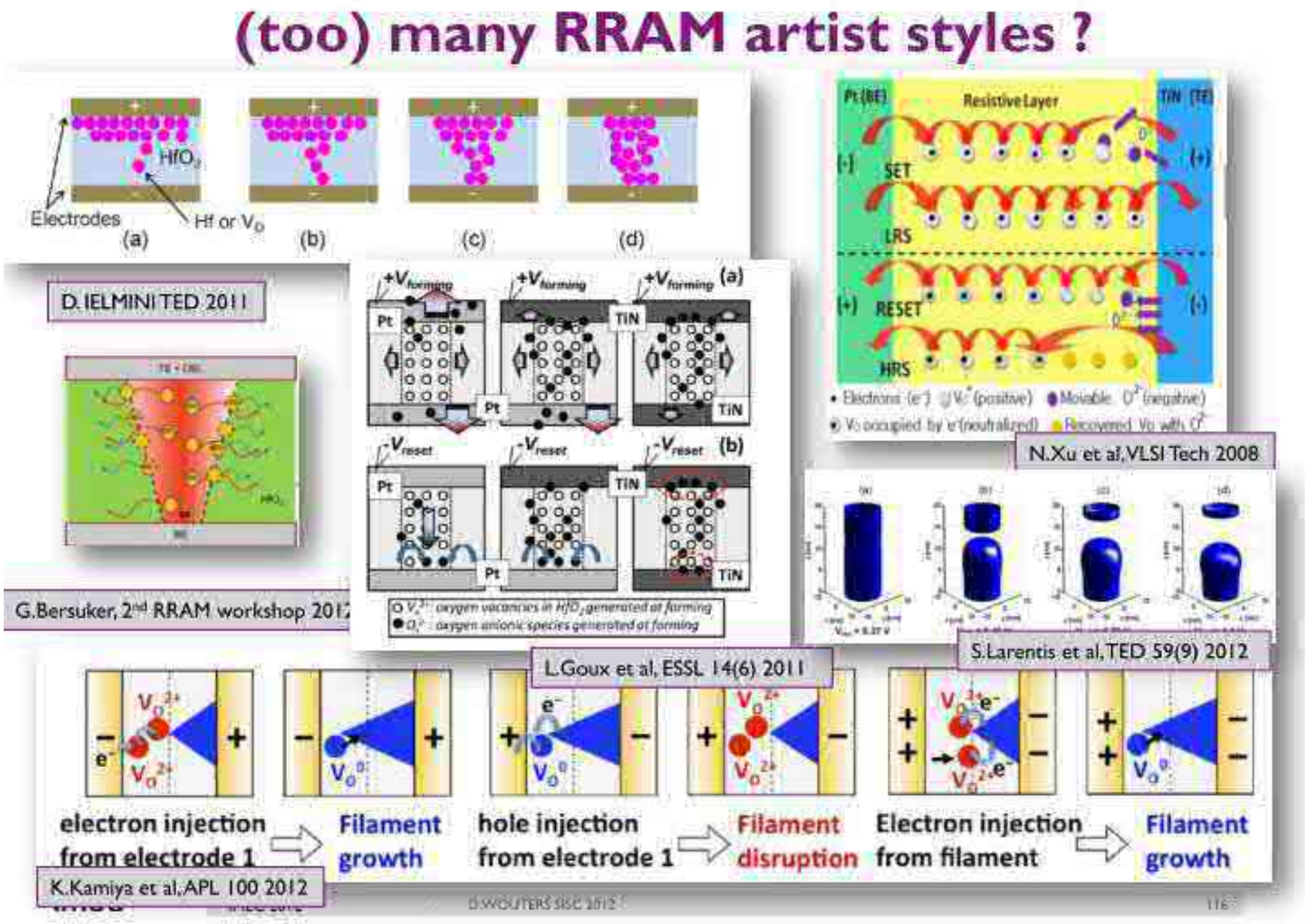

***Figure 4.1: Some of the numerous proposed models for resistive switching in Valence Change memories. Source: [2].***

## 4.2 The Hourglass Model

The Hourglass model, developed in imec by R.Degraeve et al. and extensively explained [3], gives a description of switching in RRAM devices. The name “hourglass” was chosen because of the shape described. In fact, the RRAM structure is modeled by a top reservoir of oxygen vacancies (TR) connected to a bottom reservoir (BR) by a constriction (C) with variable cross section, as shown in Figure 4.2. This shape resembles an hourglass with a variable nozzle size and with sand (= oxygen vacancies) that moves from TR to BR or vice versa depending on the applied polarity.

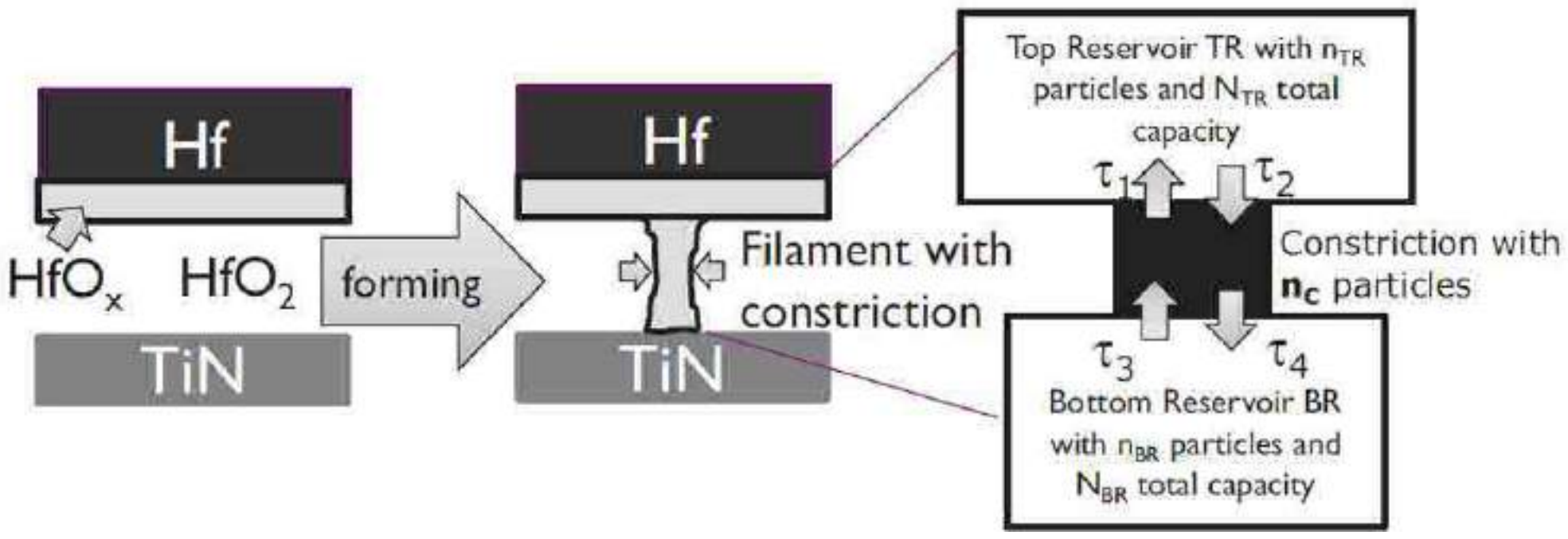

***Figure 4.2: (a) Stoichiometric HfOx is present between the top electrode (Hf) and the dielectric ($HfO_2$). (b) Forming ‘extends’ the HfOx region to the filament. (c) Filament is modeled as a container with a top and a bottom vacancy reservoir (TR and BR, respectively), connected by a constriction C with variable cross section (Source: [3])***

Figure 4.2 also shows the main parameters that are used in this model to fully define the exchange of oxygen vacancies between the two top and bottom reservoirs. τ1, τ2, τ3, and τ4, are indeed the flux variables used in the dynamic modeling of the filament variation, describing the probability of emission of an oxygen vacancy in the four possible directions, as shown in Figure 4.3 by the red arrows.

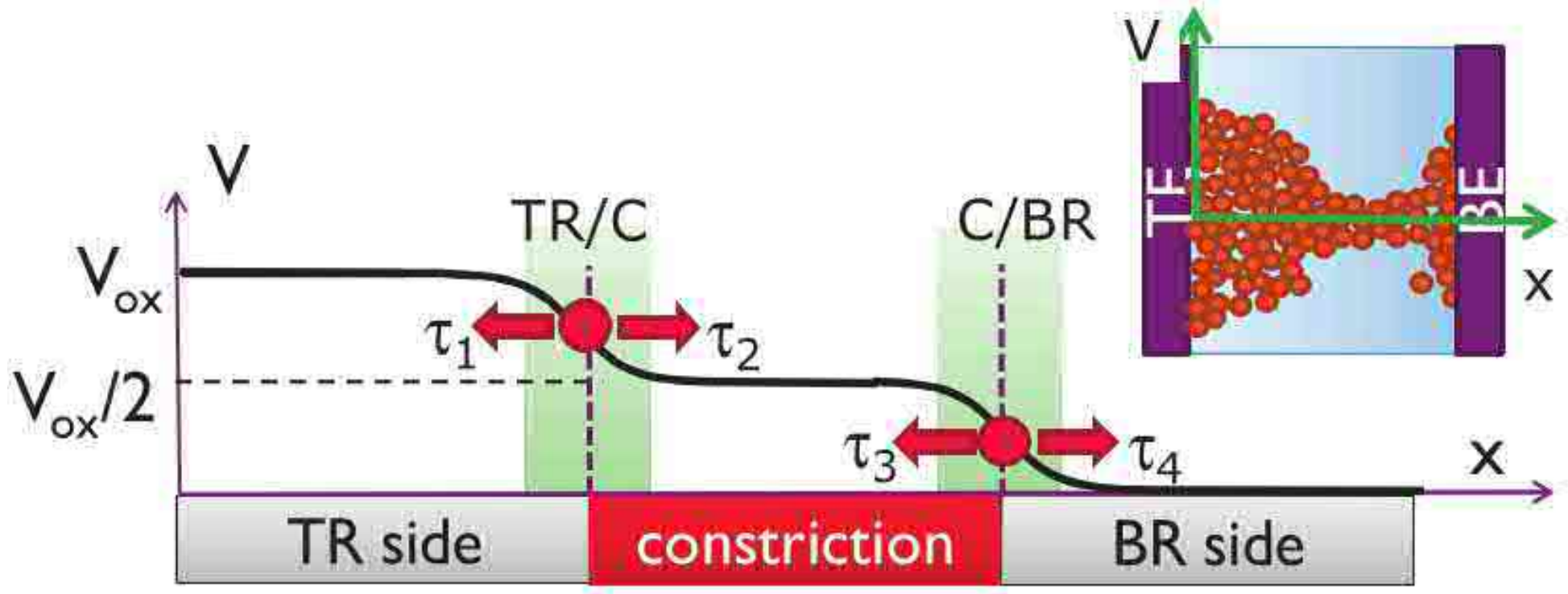

***Figure 4.3: Potential drop in the filament shows high field at the TR/C and C/BR interfaces. The field depends on the cross section of C, increasing for smaller C. [4]***

There are five basic “ingredients” in the hourglass model [3]:

I. an electron conduction model for describing the current-voltage characteristic,

II. a structural model describing the shape of the filament,

III. a kinetic model describing the vacancy movement inside the filament

IV. a thermal model describing the heat generation and its catalyzing effect on switching, and

V. a stochastic model describing the statistical variations in the switching behavior.

The main equations used by the analytic model are shown in Table 4.1 [3].

Current $I = \frac{2e}{h} \int_{-qVox/2}^{qVox/2} T(E)dE$ with $E(x,y) = qV_0 - \frac{1}{2}m\omega_x^2 x^2 + \frac{1}{2}m\omega_y^2 y^2$ (Eqs 1)

and $T$ total transmission probability through all energy levels $E_n = eV_0 + \hbar\omega_y\left(n + \frac{1}{2}\right)$ n=0,1,2,...

No annihilation/generation: $n_{total} = n_{BR} + n_{TR} + n_C$ (Eq. 2)

Set of IV curves corresponding to integer state variable $n_C$ with $\omega_y(n_C) = \omega_{y,min} \cdot \frac{1}{n_C}$ (Eq. 3)

Kinetic model:

$$1/\tau_1 = c.n_C.(1 - \frac{n_{TR}}{N_{TR}}).\exp\left(-\frac{E_a - \alpha qV}{kT}\right) \quad 1/\tau_2 = c.n_C.\frac{n_{TR}}{N_{TR}}\exp\left(-\frac{E_a + \alpha qV}{kT}\right)$$

$$1/\tau_3 = c.n_C.\frac{n_{BR}}{N_{BR}}.\exp\left(-\frac{E_a - \alpha qV}{kT}\right) \quad 1/\tau_4 = c.n_C.(1 - \frac{n_{BR}}{N_{BR}}).\exp\left(-\frac{E_a + \alpha qV}{kT}\right)$$

(Eqs. 4)

Constriction-dependent barrier lowering: $\alpha = \alpha_0 + m_n / n_c$ (Eq. 5)

Thermal model: $T = T_{ambient} + \frac{\alpha VI}{n_c} R_{th}$ (Eq. 6)

***Table 4.7: the equation set of the Hourglass model.***

## 4.3 Random Walk Model

Another (more mathematical) model considered in the analysis is the Random Walk Model.

The Random Walk is defined as a process where the present value of a variable is composed of the past value plus an error term defined as a white noise (a normally distributed variable with zero mean and variance equal to unit).

The original Random Walk was theorized by K. Pearson in a letter to Nature in 1905 [5]. In Pearson's version, a man starts at the origin and walks a fixed distance in any direction. He then walks the same distance again, in some randomly chosen direction, and the process is repeated. For this reason the model was called Random Walk.

Random Walk model is famous among economists and mathematicians that seek to understand and predict the stock price of a company. The advantage of this model is the fact that it provides a prediction interval that helps in forecasting the outcome of a future state.

Algebraically, a random walk is represented as follows:

$$y(t) = y(t-1) + \varepsilon(t) \qquad \text{(Eq. 4.1)}$$

where y(t) is the present state, whose outcome is the previous sate y(t-1) plus a random error ε(t).

The graphical representation of this process is shown in Figure 4. 4*a*. In each time period, going from left to right, the value of the variable takes an independent random step up or down. If up and down movements are equally likely at each intersection, then every possible left-to-right path through the grid is equally likely a priori. An example of random walk solutions starting from the same initial point is shown in Figure 4. 4*b.*

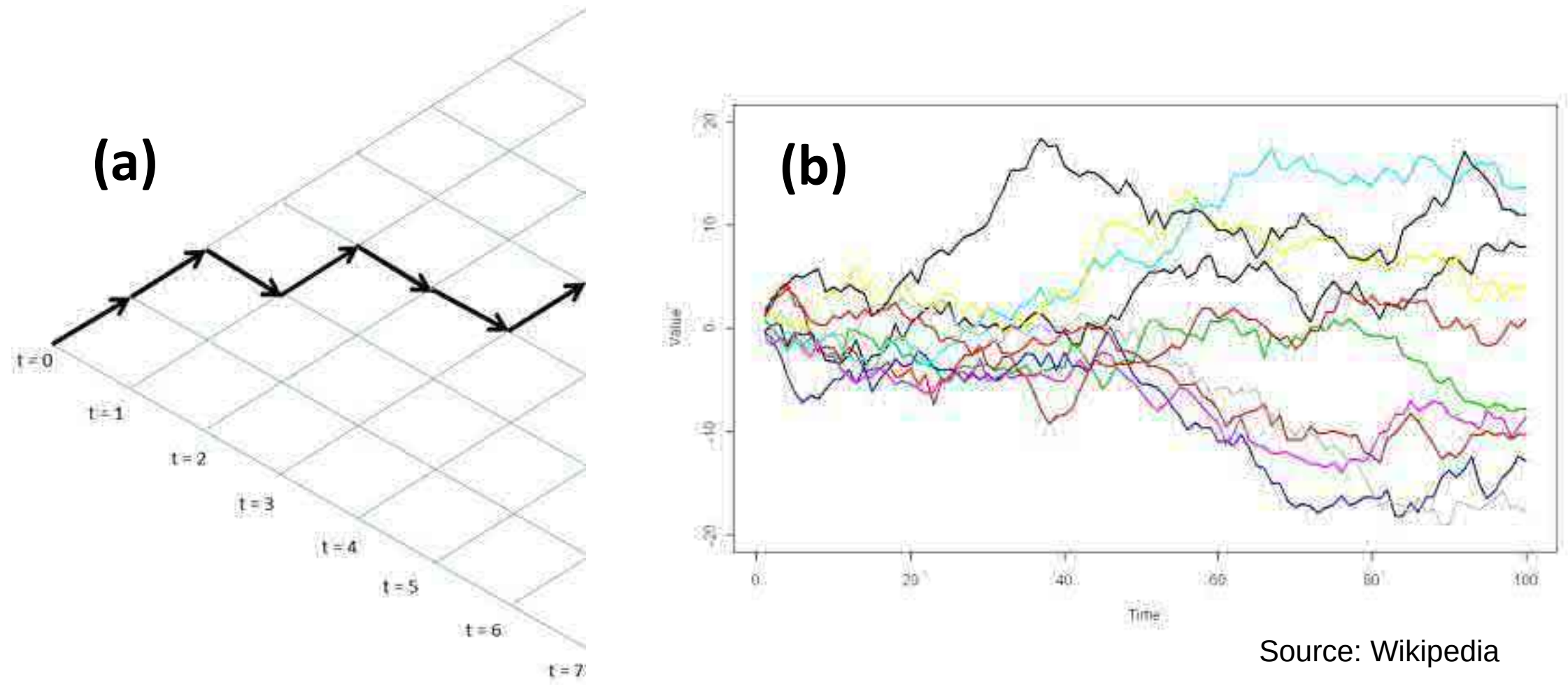

***Figure 4. 4: Random Walk possible outcomes at different time instances (a), and an example of possible Random Walk trajectories starting from the same initial point (b).***

A refinement of the Random Walk model considers also a drift component:

$$y(t) = y(t-1) + \varepsilon(t) + \mu(t) \qquad \text{(Eq. 4.2)}$$

where y(t) and y(t-1) are again the present and past state, respectively, ε(t) is a random noise component, and μ(t) is the added drift component. This process shows both a deterministic trend, associated to the drift component μ(t) and a stochastic trend, associated to the random noise ε(t). This model is often used in economics since economic time series follow a pattern that resembles a random oscillation over a deterministic trend.

Because of the similarities of this model with the measurements presented in Chapter 3,where always an increase in time of the median was observed, the Random Walk with Drift is a good starting point for describing the stochastic variation of the resistance of a programmed RRAM cell.

The first equation of the Hourglass model, shown in Table 4.7, describes the relationship between the filament dimensions ($\omega_x$ and $\omega_y$) and the current that it can carry. The following relationship is thus found:

$$R \propto \exp(\omega_x) \quad \text{(Eq. 4.3)}$$

Which means that the resistance of the filament is proportional to the exponential of the lateral dimensions. Another consideration can be made by taking into account the width of the filament in two consecutive time instances, which is supposed to be equal to a deterministic drift part plus an oscillatory random noise, as stated by the Random Walk model with Drift:

$$\omega_{x_{t2}} - \omega_{x_{t1}} = e(t) + \mu(t) \quad \text{(Eq. 4.4)}$$

By substituting the proportion shown in Eq. 4.3 into Eq. 4.4 (with the intermediate step in Eq. 4.5), an equation very similar to the random walk is obtained:

$$\exp(\omega_{x_{t2}} - \omega_{x_{t1}}) = \exp(e(t) + \mu(t)) \quad \text{(Eq. 4.5)}$$

$$\log(R_{t2}) - \log(R_{t1}) = e'(t) + \mu'(t) \quad \text{(Eq. 4.6)}$$

A comparison between Eq. 4.6 and Eq. 4.2 directly shows that the Random Walk Model is a valid starting point for describing the stochastic variation of imec's RRAM.

The Hourglass and the Random Walk models are the basic hypotheses that were considered when analyzing program stability. The obtained provisional model is presented below.

## 4.4 Time Evolution Description

Thanks to the large data collected, a refinement of the Hourglass model is possible so as to include the observed and measured relaxation in the short time range (from μs to seconds).

As described in Chapter 3, tests show that instability is experienced with increasing time. In addition, we see spreading of the distributions and an overall shift over time. This behavior is depicted in Figure 4.5 where, intuitively, it is possible to note that bits move within different portions of the distributions while the distribution itself only shows a limited change as a function of time.

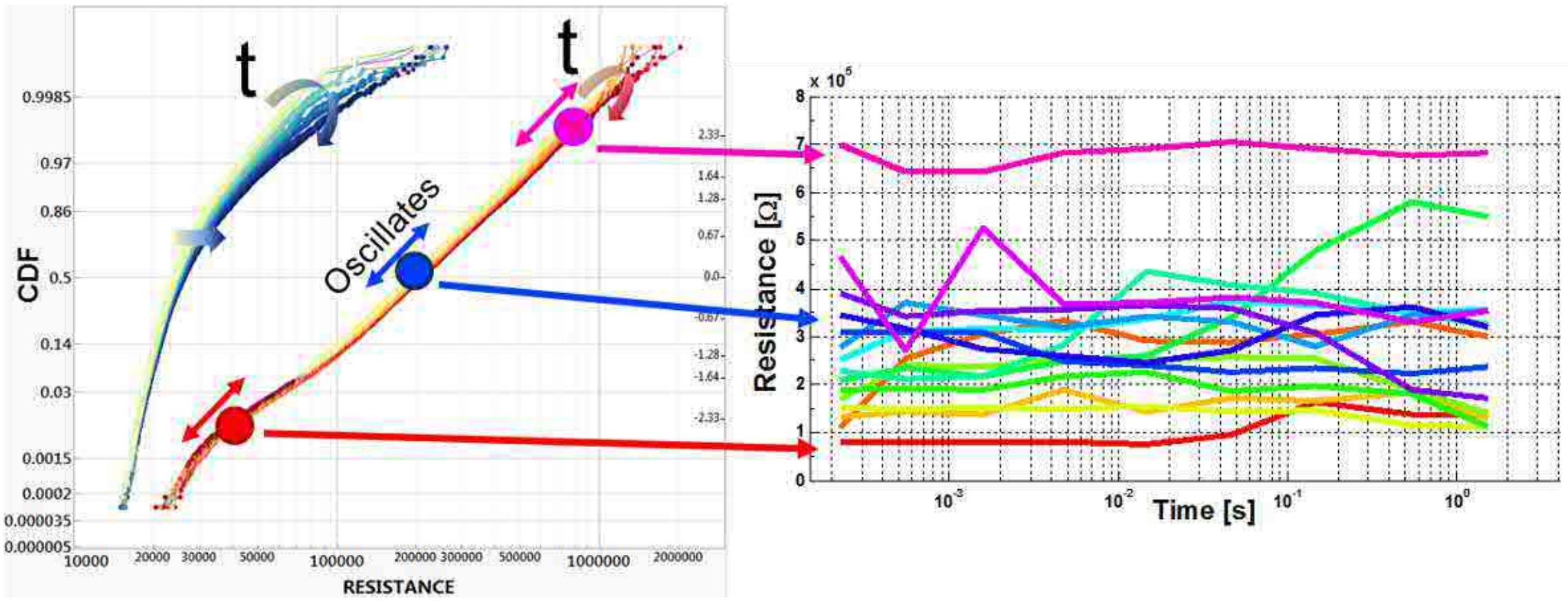

***Figure 4.5: Left: CDFs for the SET (blue) and the RESET (red) state. Right: a subset of 10 randomly chosen cycles and their evolution over time, showing random oscillations.***

Figure 4.5 is the starting point for the interpretation that follows.

As a first-order hypothesis, it is assumed that the resistance evolution over time is described as the super imposition of two distinct physical processes (as synthetized by Eq. 4.2), and reformulated below:

$$R(t) - R_0 = e(t) + \mu(t) \qquad \text{(Eq. 4.7)}$$

where e(t) is described by a random stochastic process with zero mean, whereas μ(t) is a deterministic process with a well-defined function of time and zero variance. This approximation is justified by the fact that, for each programmed state, the change of resistance between two different sampling instants is much greater than the change of median resistances.

Thus, in order to investigate the best physical description of the resistance drift component, a fitting of the median evolution can be found by following the relationship,

$$median(R(t)) =< \mu(t) > \qquad \text{(Eq. 4.8)}$$

where <μ(t)> is the median value obtained through a fitting function.

The most common physical laws were tested in order to find the better fitting for the collected data, by using the Rsquare value (also referred to as the coefficient of determination, evaluated as the square of the sample correlation coefficient between the outcomes and their predicted values) and the root mean square (RMS) error (which is a statistical measure, defined as the square root of the mean of the squares of a errors, defined as the difference between the measured data and the fitted value) as goodness-of-fit discriminant.

The tested laws together with the extractions from the respective LDE (linear differential equation) are shown in Table 4.8 and explained in the following list:

- ***Linear Law:*** even though this law can be discarded a priori by simply inspecting the collected data, it has been considered for the completeness of the analysis. This behavior assumes a linear dependence over time of the resistance. In terms of the LDE, a variation of the resistance is proportional to a variation in time.
- **Exponential Law:** this law is commonly used to describe exponential relationships. In terms of the LDE, this behavior assumes that a relative variation in the resistance value is proportional to a linear variation in time.
- ***Power Law:*** such a behavior in terms of its LDE, assumes a linear dependence, proportional to the drift coefficient, between the relative resistance increase and the relative time increase. This behavior has been largely investigated in many resistive memories, such as CBRAM [6], and PCM [7].
- **Logarithmic Law:** in this fitting, the resistance drift is logarithmically dependent on time. By previous imec knowledge, the logarithmic time dependence was already presumed. For this reason, already during UTM development and POR condition definition, it has been chosen to collect the data with samples logarithmically spaced in time.

| | **Fit Type** | **LDE (rate equation)** | **Solution (Fit law)** |
|---|---|---|---|
| (a) | Linear | $dR = \mu dt$ | $R(t) = R_0 + \mu(t - t_0)$ |
| **(b)** | Exponential | $\frac{dR}{R} = \mu dt$ | $R(t) = R_0 \exp(\mu(t - t_0))$ |
| **(c)** | Power-law | $\frac{dR}{R} = \mu \frac{dt}{t}$ | $R(t) = R_0 \left(\frac{t}{t_0}\right)^{\mu}$ |
| **(d)** | Logarithmic | $dR = \mu \frac{dt}{t}$ | $R(t) = R_0 + \mu \log\left(\frac{t}{t_0}\right)$ |

***Table 4.8: List of the fitted laws, reporting the Linear Differential Equation that, for each case, relates the Resistance to the Drift coefficient.***

The different fits were tested on the available data and are shown in Figure 4.6. It is clear, that both the exponential fit (pink) and the linear fit (light blue) are not capable of fitting the median of the data (black). On the contrary, both the power fit (red) and the logarithmic fit (green) show very good consistency with an Rsquare coefficient close to unit (perfect fit) and small RMS error. Probably, sampling a larger amount of data (for at least 100 seconds) would highlight which of the two above fits is the best. In this case, since the median evolution is observed only over a short time after programming, both logarithmic and power fits are considered to be valid. Accordingly, the measured data were sampled in time on a log-spaced scale. Indeed, as shown in Table 4.8, both LDEs for Power fit and Logarithmic fit, relate the relative resistance evolution to the relative increase of time.

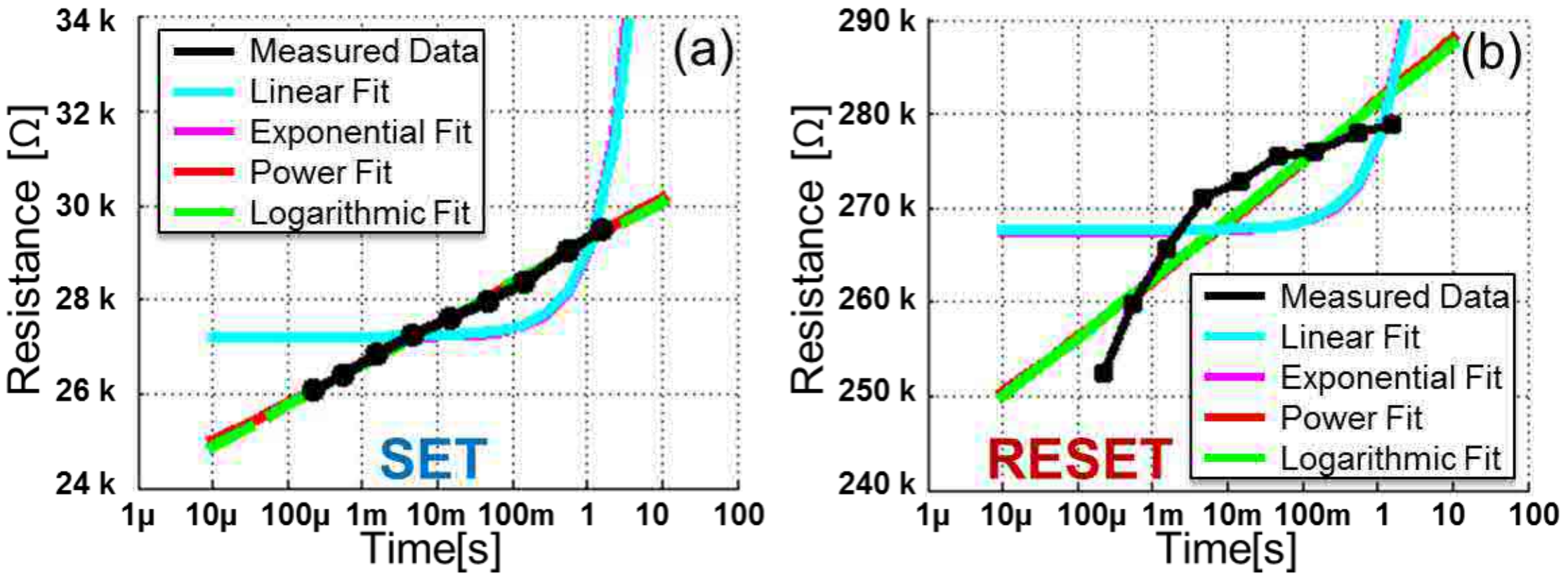

***Figure 4.6: Median of the measured data (black) fitted with the different laws listed in Table 4.8 for LRS (a) and HRS (b).***

To finally verify the assumption made in Eq. 4.2, an extra step is needed to demonstrate that, within the distribution considered, the determinist change of resistance (described by μ) is only a function of time and not (also) of the instantaneous value of resistance R(t). For this purpose, the data sampled at time RD0 (reference readout) were divided into 10 different bins and for each of them a separate fitting was performed by using the law with the best RMS error (the slightly better logarithmic law), as shown in Figure 4.7.

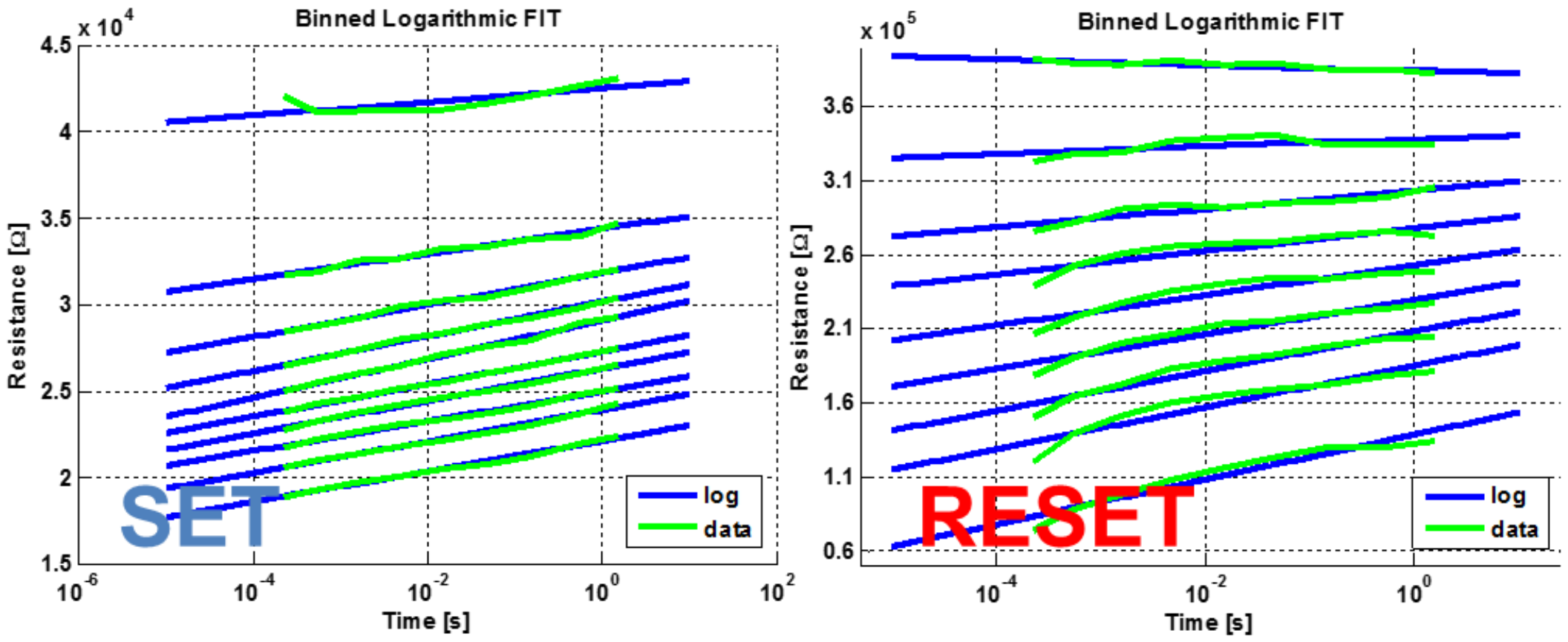

***Figure 4.7: Fit performed on 10 subpopulations extracted from the initial distribution at RD0. For both the SET and the RESET state, the logarithmic law (blue) fits very well the data (green) both when the initial resistance is lower and when it is higher than the median.***

By looking at the data (green), a great reduction of the oscillations is noticed. This is because of the process of averaging over several cycles, which contributes to eliminating the Zero Mean error component e(t) leaving just μ(t).

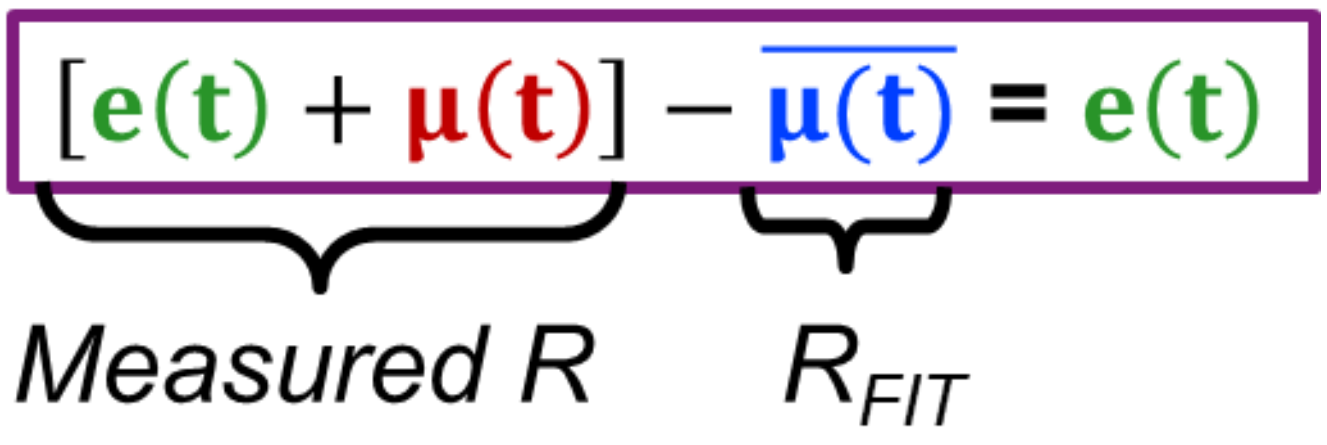

***Figure 4.8: Verifying whether by subtracting the fitted R from the data, the only variable left is the oscillation e(t).***

The next step, as shown in Figure 4.8, is the most crucial one in this analysis. It consists in verifying whether, by subtracting the fitted resistance values from the measured data, the residuals are truly just a random white noise with zero mean.

This calculation was performed by using Matlab, and the resulting residue extraction is shown in Figure 4. 9. The obtained plots are random oscillation centered around zero, which confirms the white noise hypothesis.

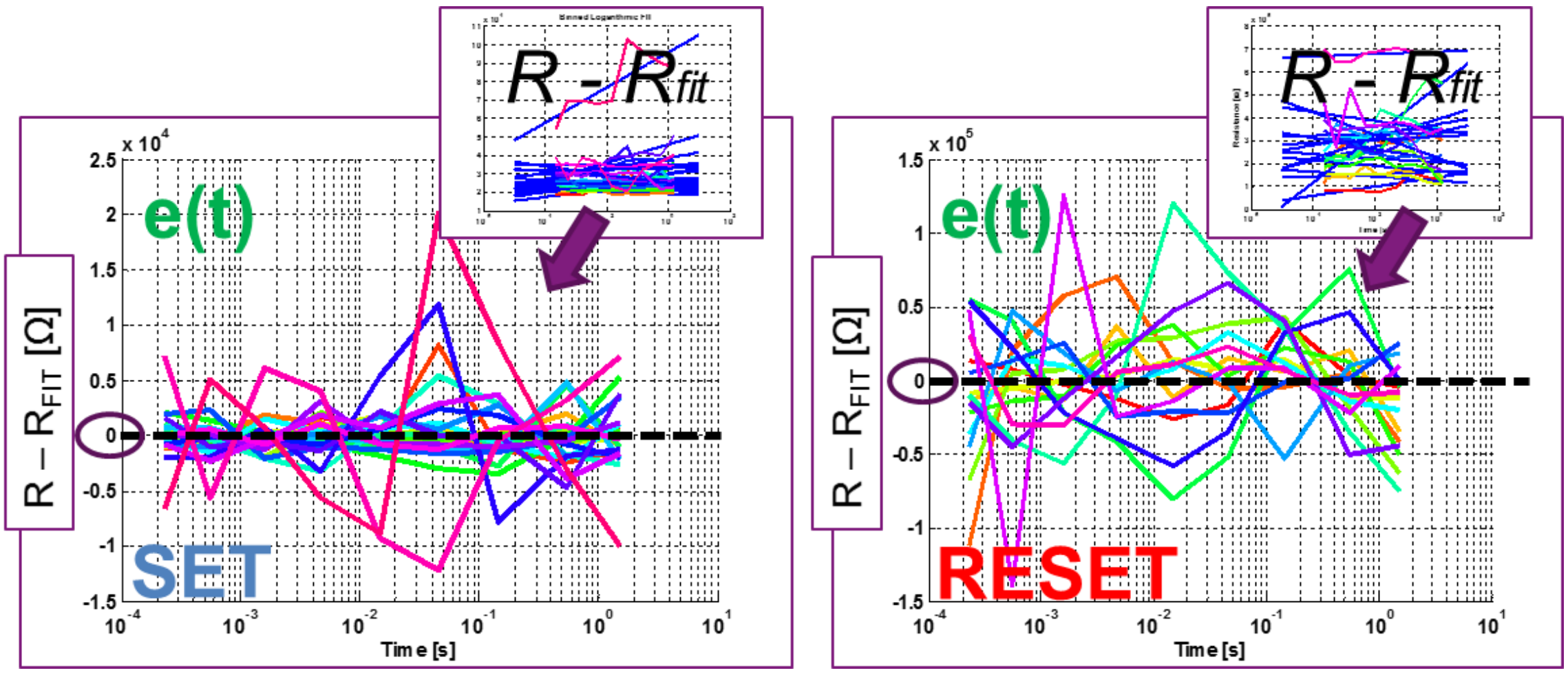

***Figure 4. 9: Isolation of the oscillatory component by subtracting the fitted drift from the measured data. Here 10 random oscillation components are shown.***

As a conclusive step, the assumption of a zero mean oscillation component was verified by analyzing the distribution evolution over time of the previously extracted e(t). The result is shown in Figure 4.10. It is clear that oscillations increase over time, but their mean value is always zero, thus confirming the starting assumption.

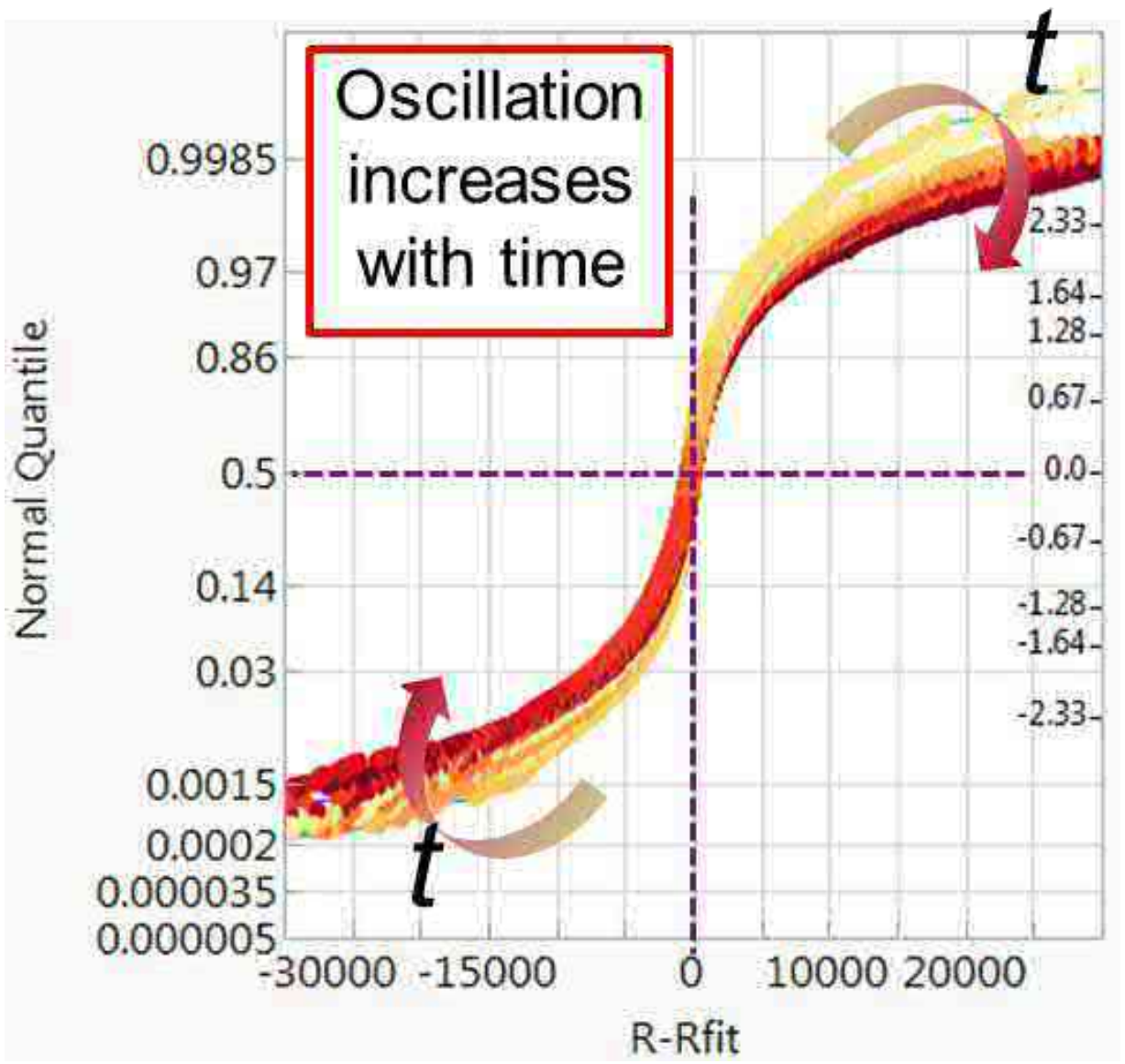

***Figure 4.10: CDF plot of the previously extracted e(t) oscillations in different time instants (time increasing from bright pink to dark red).***

As a conclusive remark, it must be noted that the developed model is a provisional model. Further modeling research is necessary for model refinement through computer simulations. Such research is conducted by imec's Memory Device Design group and does not form part of the research project of this thesis.

# Conclusions

This thesis discussed the experiments and the modeling performed on OxRAM memories during a 6-month internship project at imec.

After introducing the main emerging semiconductor memory technologies in Chapter 1, this thesis addressed the problem of state stability first, in Chapter 2, by presenting equipment and algorithms used to study OxRAM state relaxation, over short time intervals. Secondly, in Chapter 3, the various experiments and tests performed were presented. The impact of the following four parameters upon state stability in the case of the Unverified Single-Pulse programming algorithm was analyzed: programming pulse width, programming pulse fall time, temperature, and composition of the material stack of the cell. The impact of each parameter was investigated according to three metrics: global behavior, to observe the overall impact of the tested parameters, local behavior, to analyze in detail the comportment of subpopulations of the overall distributions, and correlation, very useful to describe with one synthetic value the amount of relaxation occurring.

The results obtained for each experiment were provided, and the following conclusions were drawn:

- *Pulse Width* does not impact effectively on the arising of unwanted relaxation tails, showing just a small improvement, which is ascribed to a slightly larger conductive filament formation;
- *Fall Time* does not affect relaxation, confirming that instability is not related to the abrupt quenching of current;
- *Temperature* variation highlighted that relaxation increases with increasing temperature, but further investigation is still needed to exclude other causes;
- The choice of the right *Material Stack* choice does affect relaxation:
    - HfO based stack has faster median resistance increase and correlation loss,
    - HfAlO based stack has limited median drift, but sill loss of tails,
    - TaO based stack has limited median drift and best correlation (no tails).

Two different Program-and-Verify Step Programming algorithms, namely Incremental Step Programming (ISP) and Fixed Step Programming (FSP), were also tested. The measured data were presented and the following conclusions were drawn:

- neither ISP nor FSP programming algorithms impact on relaxation; but
- Incremental Step Programming algorithm turned out to give better results.

Thanks to the work presented in this thesis, two main conclusions on HfO-based OxRAM memories were formulated:

- resistance evolution in a short timeframe range after programming was confirmed to be log-time dependent and, hence, log-spaced read-out measurements are a must in order to adequately observe data retention in OxRAM memories;
- the trend of any selected subpopulation, independently of its starting median value, (e.g. selecting the tails taking the top (bottom) 10% of the distribution with higher (lower) median) as well as that of any verified distribution, is to regenerate, even after few milliseconds, the “natural” state distribution, with a relatively slow drift on the median, but with large increasing tails.

Finally, in Chapter 4, a provisional extension to imec`s Hourglass model was designed and the resistance relaxation leading to program instability was successfully described by a Random Walk with Drift model, showing that the resistance evolution is a superposition of a random stochastic process, demonstrated to be normally distributed with zero mean value, and a deterministic process, described by a resistance drift over time successfully fitted with a logarithmic or power law.

## Future Work

Thanks to the newly developed software presented in Chapter 2, a whole set of new measurements are now possible to be performed, for example, on other resistive memories (for example CBRAM) or by designing new algorithms.

Moreover, the MDD group has lately purchased a new semiconductor parameter analyzer, an Agilent B1500A (Figure 5.1). This new equipment is an evolution of that used in this thesis work (Keithley K4200) in terms of the Remote Pulse and Measure (RPM) unit settings: while with the K4200, the DAC is initialized once at the beginning of a measurement, the B1500 permits a dynamic initialization of the DAC resolution, which allows using the best resolution for each programmed state (in the case of the OxRAM analyzed in this thesis, where the resistive window is 10x, there is only a factor of 10 between the minimum and the maximum current to be read but, for example, there are CBRAM devices with 1000x resistive window, and it is clear that an accurate measurement over such a wide range becomes non trivial).

The software for the K4200 will be therefore reprogrammed on the new B1500A, thus enabling much more accurate retention measurements.

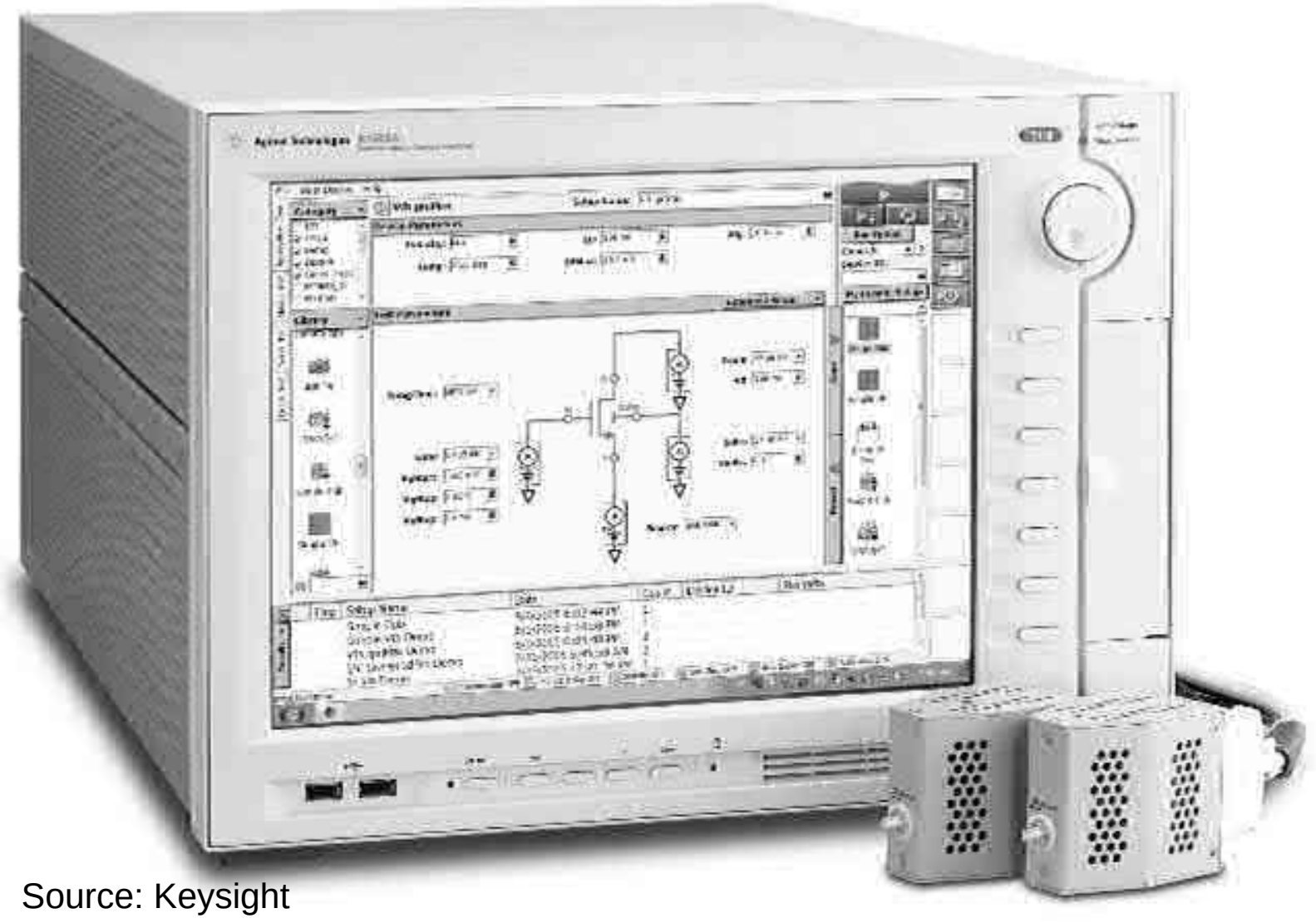

Source: Keysight

***Figure 5.1: Agilent B1500A Semiconductor Parameter Analyzer.***

# Acknowledgments

I would like to show my greatest appreciation to Prof. Guido Torelli whose comments and suggestions were of inestimable value for my study both during classes at University of Pavia, and remotely, through the plentiful conference calls; without your help and guidance, this successful master thesis would have never been possible.

Special thanks also go to Dr. Andrea Fantini who provided technical help and sincere encouragement. Andrea has been a responsible supervisor both professionally and personally. I will never forget the nights together in IMEC 5 or in AMSIMEC, measuring, discussing results, but also laughing and enjoying our time together.

My deepest gratitude goes to all the Memory Device Design team, for their friendship and for always helping me during my 6 months internship. Thanks also to the AMSIMEC staff, for all the trainings they gave and for their valuable help keeping the measurement setups always working.

A huge thanks to all my IMEC and KUL colleagues (Bilal, Nico, Steven, Themis, Lucho, Sid, Oscar, Yi), for making my experience in Leuven remarkable and unforgettable, and for always being with me, even in the most critical moments in Oude Markt. A great high-five to all the wisterians and beachvolley team (Dennis, Vasu, Viet, Mateusz, Jenni, Ilse) with whom we spent all that time in the field, playing with any kind of weather (rain). THANK YOU!

A special hug also to my friends back in Pavia/Milan/Legnano (Bago, Giulia, Laura, Pingu, Potter, Acqua, Vulvix, Ottavia, I LUPINI). You have been and always will be my life pillars.

I would also like to express my gratitude to my parents Sasha and Tania for their moral support and warm encouragements (also thanks to Dusia and Alisa!). Thanks to Nati for your patience as official reviewer; I hope somehow you get to enjoy proof-reading what I write!

I am also deeply grateful to Elisa, for her moral support during all my university studies, and for always cheering for me. Thank you for your approval and faith in me. I will always be grateful to you. Thanks for the hundreds of hours spent on the airplanes (and 616 bus) between Milan and Leuven.

Finally, I would like to express my gratitude to IMEC for their financial support through the Master Thesis Programme. I extremely enjoyed my internship at IMEC, and I can strongly recommend such programme as a very successful and unique learning opportunity.